\documentclass[10pt,a4paper]{revtex4-2}
\usepackage[utf8]{inputenc}
\usepackage[sc,osf]{mathpazo}
\usepackage{amsmath}
\usepackage[T1]{fontenc}
\usepackage{latexsym}
\usepackage{amssymb}
\usepackage[colorlinks=true,citecolor=blue,urlcolor=blue]{hyperref}
\usepackage{color}
\usepackage{graphics,epstopdf}
\usepackage{soul}
\usepackage[demo]{graphicx} 
\usepackage{capt-of}
\usepackage{lipsum}
\usepackage{adjustbox}
\usepackage[normalem]{ulem}
\usepackage[table,xcdraw]{xcolor}

\usepackage{tikz}
\usetikzlibrary{calc}
\newcommand*{\vertchar}[2][0pt]{%
  \tikz[
    inner sep=0pt,
    shorten >=-.15ex,
    shorten <=-.15ex,
    line cap=round,
    baseline=(c.base),
  ]\draw
    (0,0) node (c) {#2}
    ($(c.south)+(#1,0)$) -- ($(c.north)+(#1,0)$);%
}

\usepackage{txfonts}
\usepackage{mathtools}
\usepackage{empheq}
\usepackage[symbol]{footmisc}
\usepackage{mathrsfs}
\usepackage{enumitem}
\usepackage{xpatch}
\usepackage{graphicx}
\usepackage{caption}
\usepackage{secdot}
\usepackage{xurl}
\usepackage[utf8]{inputenc}
\usepackage[english]{babel}

\DeclareMathOperator{\x}{\mathrm{X}}
\DeclareMathOperator{\la}{\mathrm{L}}
\DeclareMathOperator{\ka}{\mathrm{K}}
\DeclareMathOperator{\md}{\mathrm{mod}}

\newtheorem{thm}{Theorem}[section]
\newtheorem{example}{Example}[section]
\newtheorem{definition}{Definition}[section]

\begin{document}





\title{Polynomial representation for multipartite entanglement of resonating valence bond ladders}



\author{Ajit Iqbal Singh\(^1\), Aditi Sen(De)\(^2\), and Ujjwal Sen\(^2\)}
\affiliation{\(^1\)Indian National Science Academy, Bahadur Shah Zafar Marg, New Delhi, 110 002, India\\
\(^2\) 
Harish-Chandra Research Institute,  A CI of Homi Bhabha National Institute, Chhatnag Road, Jhunsi, Allahabad 211  019, India
}


\maketitle

\section*{Abstract}
    A resonating valence bond (RVB) state of a lattice of quantum systems is a potential resource for quantum computing and communicating devices. It is a superposition of   singlet, i.e., dimer, coverings - often restricted to nearest-neighbour ones - of the lattice. We develop a polynomial representation of multipartite quantum states to prove that  RVB states on ladder lattices possess genuine multipartite entanglement.  The multipartite entanglement of doped RVB states and RVB states that are superposed with varying weights for singlet coverings of ladder lattices can both be detected by using this technique. 







\section{Introduction}
\label{section-1}

Genuine multipartite entanglement shared among different, possibly distant, parties has been established as one of the essential resources in quantum information processing tasks~\cite{HHHH}. Hence, there are constant efforts to  
identify physical systems in which such highly multipartite entangled states can be prepared or found, and manipulated. For example, the cluster states created during the dynamics of  Ising spin models, from  suitable initial states, are a notable class of examples of multipartite entangled states, which has been  proved to be suitable for measurement-based quantum computation~\cite{BR}.

Another prominent example is the class of resonating valence bond~(RVB) states. First discovered within the realm of organic chemistry~\cite{Pauling}, 
it has gained significant attention in physics, following a seminal work by P.W. Anderson~\cite{A1} (see also~\cite{A2a, A2, A3,  Ba1, Ba2, PSGC, Sh, Z}),  in which metal-insulator transition was connected to this class of states.
 To define an RVB state on a given lattice
 of  an arbitrary number of sites, pairs of 
  sites - often nearest-neighbour ones - are covered by dimers (i.e., singlets), and the RVB state is obtained by  superposing  all such coverings.  
  The relative contents 
  of bipartite and multipartite entanglements of these states have been analyzed, and their persistence examined also in the presence of defects, for ladder as well as isotropic 
  lattices~\cite{delgado1, delgado2, CKSSV, ravi, DS, DSS, RDSS, SDRSS, SSR}. Either an analytical method based on recursion relations among covering superpositions of different lattice sizes~\cite{delgado1, delgado2, DSS, RDSS, SDRSS, SSR} or numerical simulations are used to detect and quantify multipartite entanglement content, with the quantification being via the  generalized geometric measure~\cite{ggm1} (see also \cite{ggm2, mul}). Notice that a brute force analytical approach is impractical, since the number of covers grows exponentially with the size of the lattice. The goal of this work is to find a fresh analytical approach to deal with RVB states, and in particular, with its multipartite entanglement.



Polynomials have recently been used to characterize entanglement from different perspectives~ (see for instance \cite{F}) \cite{GW,P,Bh,Sk,SAS, ES, JM, Szalay, AR,M,BS}. The method  displays a simple way to use polynomials for understanding multipartite entanglement, in general, and Schmidt rank in particular cases \cite{BS}. For multiqubit pure states, the theory of polynomial representation of quantum entanglement takes  a simple form. Specifically, entanglement can be quantified via polynomial functions which are associated with the coefficients of pure states and remain invariant under stochastic local quantum operations and classical communication~\cite{ES}.

In this work, we provide a polynomial-based technique~\cite{BS} to prove the existence of genuine multipartite entanglement of the RVB states. Going beyond coverings of nearest-neighbour dimers, which has been typically studied until now, this analytical method enables us to  demonstrate multipartite  entanglement content of RVB states with varying coefficients and RVB states which can be constructed with dimer coverings between any two sites of a ladder lattice.  Furthermore, employing the polynomial representation, we  show that doped RVB states on a ladder lattice are genuinely multipartite entangled. 



The rest of the paper is organized as follows. 
In Sec.~\ref{section-2}, we introduce a polynomial representation of quantum entanglement for a multiqubit system. It leads to simple conditions for classification of pure states into  product vectors in a bipartition and genuinely entangled ones, in terms of the degree, support, number of terms or irreducibility  of the corresponding polynomial.   
We provide the main results in Sec.~\ref{section-3}, starting from basics like coverings and resonating valence bond states, nearest-neighbour (in short, NN) coverings, etc., together with a few examples for the sake of completeness. We give their polynomial representations and obtain conditions in terms of them for multipartite entanglement properties of RVB states. This brings us to the notions of decomposability, factorability,  etc. for coverings, together with interrelationships among them. This technique leads to a proof of genuinely multipartite entanglement in  NN and doped RVB states. 
Sec.~\ref{sec-4} deals with RVB states having varying coefficients  and its multipartite entanglement, and we finally include concluding remarks in Sec.~\ref{sec-conclu}.


\section{Polynomial representation of quantum entanglement for multiqubit systems}
\label{section-2}

Let $\mathbb{N}$ be the set of natural numbers $1,2,3, \ldots$. For $m\in \mathbb{N}$, let $\Gamma_{m}= \left\{ j \in \mathbb{N} : j \leq m \right\}$. For any set \(S\), let $P_{S}$ be the set of subsets of S and $\#$ S, the cordinality of S.
We take empty sums to be zero and empty products to be \(1\).
Let $n \in \mathbb{N}$ with $n \geq 2$.

\subsection{Basics of multiqubit systems and polynomials}\label{subsection-2.1}

We begin with qubit systems. Let $l \leq j \leq n.$ 
Let $\mathcal{H}_{j}$ be a qubit system with ordered basis $\left\{ | \uparrow \rangle_{j}, | \downarrow \rangle_{j}\right\}$ or $\left\{ |0\rangle_{j}, |1  \rangle_{j}\right\}$ or $\left\{ \beta^{j}_{1}, \beta^{j}_{2} \right\}$ of orthonormal vectors.
We also consider $\mathcal{H}_{j}$ as the space 
\(\mathbb{C}_{1}[X_{j}]\) of degree at most one,  the constant function $\bold{1}$ and the polynomial $X_{j}$ constituting an orthonormal basis for $\mathcal{H}_{j}$.
In other words, we consider $\mathcal{H}_{j}$ as the subspace spanned by $\bold{1}$ and $X_{j}$ of the Hilbert space $\la^{2}(T)$ of square Lebesgue integrable complex functions $f$ on the unit circle. i.e., torus $T$ with $\langle f, g \rangle = \int \bar{f} g d \lambda$, $\lambda$  being the normalized Lebesgue measure on $T$.


Let $\mathcal{H} = \bigotimes\limits_{\rm d=1}^{\rm n} \mathcal{H}_{j}$ be  tensor product of Hilbert spaces $\mathcal{H}_{j}$, $1 \leq j \leq n$.
Let $\mathcal{I}$ be the set of n-tuples, $\bold{i} = (i_{j})^{n}_{j=1}$ with $i_{j}=0$ or 1 for $1 \leq j \leq n$.  $\mathcal{I}$ is in one-one correspondence with the set $P_{n}$ of subsets of $\Gamma_{n}$ via $\bold{i} \longrightarrow K_{\bold{i}} = \left\{j : i_{j} = 1 \right\}$. We note that for $0 \leq s \leq n $, $\mathcal{I}_{s} = \left\{ \bold{i} \in \mathcal{I} : |\bold{i}| = \sum_{j=1}^{\rm n} i_{j} = s
\right\}$ equals $\left\{\bold{i} \in \mathcal{I} : \# K_{\bold{i}} = s \right\}$ and together, they constitute a decomposition of $\mathcal{I}$.
Further, $\left\{| \bold{i} \rangle = \bigotimes\limits_{j=1}^{\rm n}| i \rangle_{j} : \bold{i} = {({i_{j}})_{j=1}^{n}} \in \mathcal{I} \right\}$  is an orthonormal  basis for $\mathcal{H}$. A generic element $\xi$ of $\mathcal{H}$ has the form $\xi$ = $\sum_{\bold{i}\in \mathcal{I}} a_{\bold{i}} |\bold{i} \rangle$ for a unique $\bold{a} = {(a_{\bold{i}})_{\bold{i}\in \mathcal{I}}} {\in} \mathbb{C}^{\mathcal{I}}$ 


For $\bold{i} \in \mathcal{I}$, let $\boldsymbol{\x}^{\bold{i}} = \prod\limits_{j=1}^{n} \x_{j}^{i_{j}} = \prod \left\{ \x_{j} : j \in K_{\bold{i}} \right\} = \boldsymbol{\x}^{K_{\bold{i}}}$.
Further, we may represent $| \boldsymbol{i} \rangle \in \mathcal{H}$ by $\boldsymbol{\x}^{\bold{i}}$ and accordingly, $\xi~in~\mathcal{H}$, by the polynomial F in several variables $\boldsymbol{\x} = (\x_{j})^{n}_{j=1} = (\x_{1}, \x_{2}, \ldots \x_{n})$ given by ${\rm F} (\boldsymbol{\x}) = \sum_{\bold{i}\in \mathcal{I}} a_{\bold{i}} \boldsymbol{\x}^{\bold{i}}$. This enables us to identify $\mathcal{H}$ with the space 
$\mathbb{C}_{1}[\boldsymbol{\x}]$ of polynomials F in $\boldsymbol{\x}$ with degree of each variable $\x_{j}$ being less than or equal to \(1\). Clearly, the degree of each such F is at most $n$. We may also write F as $\sum\limits_{\rm K \in P_{n}} a_{\rm K} \boldsymbol{\x}^{\ka} = \sum\limits_{\rm K \in P_{n}} a_{\bold{i}_{\ka}} \boldsymbol{\x}^{\bold{i}_{K}}$, where for $\ka \in P_{n}, (\bold{i}_{\ka})_{j} =1$ for $j\in \ka $ and $0$ otherwise.
In other words, we consider $\mathcal{H}$ as the subspace spanned by $\left\{ \boldsymbol{\x}^{\ka}, \ka \in P_{n}\right\}$ of the Hilbert space ${\rm L}^{2} \left(T^{n}\right)$ of square integrable complex functions $f$ on the torus $T^n$ with the normalized product Lebesgue measure $\lambda_{n}$ and take $\langle f, g \rangle = \int\limits_{T^{n}} \overline{f} g d \lambda_{n}$. 
 
 
 Consider any non-empty subset $E$ of $\Gamma_{n}$ with $E \neq \Gamma_{n}$. Let $E' = \Gamma_{n}\smallsetminus E = \{ j \in \Gamma_{n} : j \notin E \}$. Set $\mathcal{H}(E) = \bigotimes\limits_{j \in E} \mathcal{H}_{j}$ and $\mathcal{H}(E') = \bigotimes\limits_{j \in E'}$ $\mathcal{H}_{j}$.
 $\mathcal{H} = \mathcal{H}(E) \bigotimes \mathcal{H}(E')$, a two-fold tensor product in the bipartite cut, $(E, E')$, so as to say.
  Let 
 $\mathbb{C}_{1}[\boldsymbol{\x}_{E}]$ and  
 $\mathbb{C}_{1}[\boldsymbol{\x}_{E'}]$ have their obvious meanings. Further, 
 $\mathbb{C}_{1}[\boldsymbol{\x}] = 
 \mathbb{C}_{1}[\boldsymbol{\x}_{E}] \bigotimes  
 \mathbb{C}_{1}[\boldsymbol{\x}_{E'}]$. We note that for $p \in 
 \mathbb{C}_{1}[\boldsymbol{\x}_{E}]$, $q \in 
 \mathbb{C}_{1}[\boldsymbol{\x}_{E'}]$, $F = p \bigotimes q$ is given by $F(\boldsymbol{\x}) = p (\boldsymbol{\x}_{E}) q (\boldsymbol{\x}_{E'})$.
 We may now  identify $\mathcal{H}(E)$ with 
 $\mathbb{C}_{1}[\boldsymbol{\x}_{E}]$ and $\mathcal{H}(E')$ with 
 $\mathbb{C}_{1}[\boldsymbol{\x}_{E'}]$.
 Caution is needed in the above identification because the context is important. For instance, for $n=4$, $E = \{ 1, 2 \}$, consider the polynomials written in the usual short form $1, \x_{1}-\x_{2}$.
 $1$ means $| 0 \rangle_{1}$ in $\mathcal{H}_{1}$, $|00\rangle_{12} = |0 \rangle_{1} \otimes |0 \rangle_{2}$ in $\mathcal{H}(E)$, $|00 \rangle_{34} = |0\rangle_{3} \otimes |0 \rangle_{4}$ in $\mathcal{H}(E')$ and $|0000\rangle_{1234} = |0 \rangle_{1} \otimes |0 \rangle_{2} \otimes |0 \rangle_{3} \otimes |0 \rangle_{4}$ in $\mathcal{H}$. On the other hand, $X_{1}-X_{2}$ has no meaning in $\mathcal{H}_{1}$ or $\mathcal{H}(E')$ but it stands for $|10\rangle_{12} - |01\rangle_{12} = |1 \rangle_{1} \otimes |0 \rangle_{2} - |0 \rangle_{1} \otimes 1| \rangle_{2}$ in $\mathcal{H}(E)$ and for $|1000 \rangle_{1234} - |0100\rangle_{1234} = |1 \rangle_{1} \otimes |0 \rangle_{2} \otimes |0 \rangle_{3} \otimes |0 \rangle_{4}-|0 \rangle_{1} \otimes |1 \rangle_{2} \otimes |0 \rangle_{3} \otimes |0 \rangle_{4} $ in $\mathcal{H}$.

 
 
 \subsection{Multiqubit entanglement and polynomial representation}\label{subsection-2.2}
 
 Let $0 \neq \xi  \in \mathcal{H}$ and $\digamma (\boldsymbol{\x})$, the corresponding polynomial in 
 $\mathbb{C}_{1}[\boldsymbol{\x}]$. Set $P_{\xi} = \frac{1}{|| \xi||^{2}} | \xi\rangle \langle \xi|$, the corresponding pure state; i.e., the rank one (orthogonal) projection on the linear span of $\xi$. We consider the following concepts related to entanglement of $\xi$ or $P_{\xi}$ to begin with. 
 
 
 
 \begin{enumerate}[label=(\alph*)]
 
 \item $\xi$ is a product vector in $\mathcal{H} = \bigotimes\limits_{j=1}^{\rm n} \mathcal{H}_{j}$ if $\xi =\bigotimes\limits_{j = 1}^{\rm n} \xi_{j}$ with $\xi_{j} \in \mathcal{H}_{j}$ for $1 \leq j \leq n$. Otherwise, $\xi$ is called an entangled vector.
   
 \item Let $\phi \neq E \subset_{\neq} \Gamma_{n}$. $\xi$ is said to be a product vector in the bipartite cut $(E, E')$ if $\xi$ is a product vector in $\mathcal{H}(E)\otimes \mathcal{H}(E')$, i.e., if $\xi = \eta \otimes \zeta$ for some $\eta \in \mathcal{H}(E)$ and $\zeta \in \mathcal{H}(E')$.
 
 \item $\xi$ is said to be  genuinely entangled if it is not a product vector in any bipartite cut.
  
 \end{enumerate}
 Clearly for $n = 2$, $\xi$ is entangled if and only if $\xi$ is genuinely entangled.
 
 The concepts and properties above will be transferred to the corresponding $F(\boldsymbol{\x})$ i.e., $F$.
  We give some easy consequences as a Theorem below.
 We shall freely use the following notation, terminology and facts for that (cf. \cite{F}).
 
 \begin{enumerate}[label=(\alph*)]
 
 \item  Let $S_{F}$, the support of $F =\left\{ j \in \Gamma_{n} \; ; \x_{j}\;\; \text{occurs in}~ F, \text{i.e.,} \,  F~ \text{has degree 1 in} \x_{j} \right\}$ and $m = \# S_{F}$, the number of $\x_{j}$'s that occur in $F$.
 
 \item Let $d$ be the degree of $F$.  $d > 0$ if and only if $m > 0$. Also $m \geq d$ simply because $F$ has degree at most one in each $\x_{j}$.
 
 \item $F$ is said to be homogeneous if each term in $F$ has degree $d$.  
 
 \item Suppose that $F = gh$ with $g$ and $h$ in 
 $\mathbb{C}_{1} [\boldsymbol{\x}]$. Then $S_{g} \cap S_{h} = \emptyset$ because degree of $\x_{j}$ in $F(\boldsymbol{\x})$ is at most one for $1 \leq j \leq n$. If $g$ and $h$ are such that $S_{g} \neq \Gamma_{n} \neq S_{h}$, then we may take any $E$ with $\emptyset \neq E\subset_{\neq} \Gamma_{n}$ with $S_{g} \subset E$ and $S_{h} \subset E'$. This permits us to write $g(\boldsymbol{\x})$ as $g(\boldsymbol{\x}_{E})$ and $h(\boldsymbol{\x})$ as $h(\boldsymbol{\x}_{E'})$ as well and then $F(\boldsymbol{\x}) = g(\boldsymbol{\x}_{E}) h (\boldsymbol{\x}_{E'})$.
 
 \item 
 If $F, g, h$ are as in $(d)$ above and $F$ is homogeneous,  g and h are homogeneous too.
 
 \item Suppose that $F$ is not a constant polynomial. $F$ is said to be irreducible if $F$ has no factors other than itself (up to constants). Clearly, if $F$ has degree $d=1$, then $F$ is irreducible.  If $F$ is not irreducible, there exist $r(\geq 2)$ irreducible polynomials $F_{s}$, $1 \leq s \leq r$ such that $F(\boldsymbol{\x}) = \prod\limits_{s = 1}^{r} F_{s} (\boldsymbol{\x})$.  
 $F_{s}$'s are unique
 to with in order and up to constants.
 In view of $(d)$ above, $S_{F_{s}} \cap S_{F_{s'}} = \emptyset$, for $1 \leq s \neq s' \leq r$.
 \end{enumerate}

 For notational convenience, we may take $r =1$ when $F$ is irreducible and take $F_{1} = F$, and take $r=0$ if $F$ is a non-zero constant function. 
 
 \vspace{.5cm}
 
 \begin{thm}\label{thm-2.1}
  Let $0 \neq F \in 
  \mathbb{C}_{1} [\boldsymbol{\x}]$ and $\xi \in \mathcal{H}$ be the corresponding vector in $\mathcal{H}$.
  
  \begin{enumerate}[label=(\roman*)] 
  \item 
  \begin{enumerate}[label=(\alph*)] 
   \item $\xi$ is a product vector if and only if $F(\boldsymbol{\x}) = \prod\limits_{j=1}^{\rm n} (a_{j} + b_{j} \x_{j})$ with $(0,0) \neq (a_{j}, b_{j}) \in \mathbb{C}^2$ for $1 \leq j \leq n$ if and only if $d=m=r$. 
   \item  In case (a) happens, the number $t$ of terms in $F(\boldsymbol{\x})$ is $2^{u}$ with $u = \# \left\{ j \in \Gamma_{n} : a_{j}\neq  0 \neq b_{j} \right\}$.
  \end{enumerate}
  
  \item Let $\phi \neq E \subset_{\neq} \Gamma_{n}$.
  
  \begin{enumerate}[label=(\alph*)] 
  
  \item $\xi$ is a product vector in the bipartite cut $(E,E')$ if and only if $F(\boldsymbol{\x})=p(\boldsymbol{\x}_E) q(\boldsymbol{\x}_{E'})$ with $p$ and $q$ polynomials in $\boldsymbol{\x}_{E}$ and $\boldsymbol{\x}_{E'}$ respectively.
  
  \item In case (a) happens and neither $p$ nor $q$ has the form $\lambda\boldsymbol{\x}^{K}$ for some scalar $\lambda \neq 0$ and $K$ with $K \subset \Gamma_{n}$, we have that $t$ is a composite number.
  
  \item In case (a) happens and for $1 \leq j \leq n$, neither $\x_{j}$ is missing nor $\x_{j}$ is a factor of $F(\boldsymbol{\x})$, we have that $t$ is a composite number.
  
  \item Suppose that (a) happens. If we know any term of $p(\boldsymbol{\x}_{E})$ or $q(\boldsymbol{\x}_{E'})$ both $p$ and  $q$ can be determined from $F(\boldsymbol{\x})$ and that term.
  
  \end{enumerate}
  \item (a) $\xi$ is genuinely entangled if and only if $m=n$, i.e., all $\x_{j}$s occur in $F(\boldsymbol{\x})$ and $F$ is irreducible. In this case, no $\x_{j}$ is a factor of $F(\boldsymbol{\x})$.\\
  (b) If m=n and d=1, then $\xi$ is genuinely entangled.
  
  \item Suppose that $F$ is homogeneous. Then $t \leq {^n}C_{d}$.
  
  \begin{enumerate}[label=(\alph*)] 
   \item $\xi$ is a product vector if and only if $F(\boldsymbol{\x}) = \lambda\boldsymbol{\x}^{K}$ for some $K\subset \Gamma_{n}$ and scalar $\lambda \neq 0$.
   \item Let $\phi \neq E \subset_{\neq}\Gamma_{n}$, $\alpha = \# E$, $\beta = \#E' = n-\alpha$.   If $\xi$ is a product vector in the bipartite cut $(E,E')$, both $p$ and $q$ in (ii) above are homogeneous, Further, $t \leq \tau_{\alpha}$, where 
   $\tau_{\alpha} = \max \left\{ ^{\alpha}C_{d_{1}} {^{n-\alpha}}C_{d-d_{1}} : \max \{d-\beta, 0\} \leq d_{1} \leq \min \{\alpha, d\}\right\}$. 
   
   \item Let $\tau =  \max \{\tau_\alpha: 1\leq \alpha \leq n-1 \}$. If $t > \tau$,  $\xi$ is genuinely entangled. 
    
    \item $\tau_{\alpha} = \tau_{n-\alpha}$ for $1 \leq \alpha \leq n-1$.
    \end{enumerate}
 \end{enumerate}
 \end{thm}
 
  {\it Proof.}
  (i) and (ii) (a) follow immediately from definitions.
  
  To see (ii) (b) we note that $t = t_{1} t_{2}$ with $t_{1} =$ number of terms in $p (\boldsymbol{\x}_{E})$ and $t_{2}=$ number of terms in $q(\boldsymbol{\x}_{E'})$, and under the given condition, $t_{1} \geq 2$ and $t_{2} \geq 2$.
  
  The condition in (ii) (c) is indeed stronger than that in (ii)(b).
  
  (ii)(d) Suppose we know the term $\lambda\boldsymbol{\x}^{K}$ in $p(\boldsymbol{\x}_{E})$. We figure out the polynomial $q_{K}(\boldsymbol{\x}_{E'})$ so that $\lambda \boldsymbol{\x}^{K} q_{K}(\boldsymbol{\x}_{E'})$ is the part of $F(\boldsymbol{\x})$ in which $\boldsymbol{\x}^{K}$ occurs. Then $q(\boldsymbol{\x}_{E'})$ is precisely $q_{K}(\bold{\x}_{E'})$. Consider any term $\lambda'\boldsymbol{\x}^{K'}$ of $q(\boldsymbol{\x}_{E'})$. We can repeat the arguments above to figure out $p(\boldsymbol{\x}_{E})$
(with no more \(X_j\) with \(j\in E\) with it.)

  (iii)(a) ${\boldsymbol{\Rightarrow {\rm part}}}$. Let $\xi$ be genuinely entangled. 
  
  Let, if possible, $\x_{j}$ not occur in $F(\boldsymbol{\x})$. Set $E=\{j\}$, $p(\boldsymbol{\x}_{E})=$ the constant function 1 in variable $\boldsymbol{\x}_{E}$ and $q(\x_{E'})$ same as $F(\boldsymbol{\x})$ considered as a function in variable $\boldsymbol{\x}_{E'}$ simply because $\x_{j}$ does not occur in $F(\boldsymbol{\x})$. Then $F(\boldsymbol{\x}) = p(\boldsymbol{\x}_{E}) q (\boldsymbol{\x}_{E'})$. 
  So (ii) (a) gives that  $\xi$ is a product vector in the bipartite cut $(E, E')$, a contradiction. 
  
  Now suppose that all $\x_{j}$s occur in $F(\boldsymbol{\x})$. $F$ is not a constant function. If $F$ is not irreducible, then $F(\boldsymbol{\x}) = \prod\limits_{s=1}^{r} F_{s}(\boldsymbol{\x})$ for some $r \geq 2$, irreducible functions $F_{s}(\boldsymbol{\x})$. Let $E= S_{F_1}$, $E' = \Gamma_{n} \backslash E$ then turns out to be $\bigcup\limits_{s=2}^{r} S_{F_{s}}$. Set $p(\boldsymbol{\x}_{E}) = F_{1}(\boldsymbol{\x})$ considered as a function of $\boldsymbol{\x}_{E}$ and $q(\boldsymbol{\x}_{E'}) = \prod\limits_{s=2}^{r} F_{s}(\boldsymbol{\x})$ considered as a function of $\boldsymbol{\x}_{E'}$. As a consequence, $F(\boldsymbol{\x}) = p (\boldsymbol{\x}_{E}) q (\boldsymbol{\x}_{E'})$. This in turn, gives that $\xi$ is a product vector in the bipartite cut $(E, E')$, a contradiction. Hence all $\x_{j}$'s occur in $F(\boldsymbol{\x})$ and $F(\boldsymbol{\x})$ is irreducible.
  
  ${\boldsymbol{\Leftarrow {\rm part}}}$. Suppose that all $\x_{j}$s occur in $F(\boldsymbol{\x})$ and $F$ is irreducible. Let, if possible, $\xi$ be not genuinely entangled. Then by (ii) (a) for some $\phi \neq E \subset_{\neq} \Gamma_{n}$, polynomials $p(\boldsymbol{\x}_{E})$ and $q(\boldsymbol{\x}_{E'})$, we have $F(\boldsymbol{\x}) = p(\boldsymbol{\x}_{E}) q (\boldsymbol{\x}_{E'})$. For $j \in E$, $\x_{j}$ occurs in $F(\boldsymbol{\x})$, and therefore, in $p(\boldsymbol{\x}_{E})$. Similarly for $k \in E'$, $\x_{k}$ occurs in $q(\boldsymbol{\x}_{E'})$. So $p$ and $q$ are not constant polynomials. This forces $F$ to be reducible, a contradiction. Hence $\xi$ is genuinely entangled.
  
  (b) It follows from (a) above because $F$ is irreducible in this case.
  
  (iv) We know that if $F(\boldsymbol{\x})= g(\boldsymbol{\x}) h(\boldsymbol{\x})$ for polynomials $g$ and $h$,  $g$ and $h$ are both homogeneous. To see (a), we have to only note that if $\xi$ a product vector,  for $1 \leq j \leq n$, $(a_{j}, b_{j}) \neq (0,0)$ as in (i), the polynomial $a_{j} + b_{j} \x_{j}$ considered as a polynomial in $\x$ is homogeneous if and only if either $a_{j}$ or $b_{j}$ is zero. We set $K = \left\{ j \in \Gamma_{n} : a_{j} = 0\right\}$. Then $F(\boldsymbol{\x}) = \lambda\boldsymbol{\x}^{K}$ for some scalar $\lambda \neq 0$.
  
  (b) The observations made above give that $p$ and $q$ are homogeneous polynomials in $\boldsymbol{\x}_{E}$ and $\boldsymbol{\x}_{E'}$ respectively. Let their degrees be $d_1$ and $d_2$ respectively.   $0 \leq d_{1} \leq \alpha$, $0 \leq d_{2} \leq \beta $ and number of terms in $p(\boldsymbol{\x}_{E}) \leq {^{\alpha}}C_{d_{1}}$  whereas that in $q(\boldsymbol{\x}_{E'})$ is at most $^{\beta}C_{d_{2}}$. This gives the rest of (b).
  
  (c) It is immediate from the definition and (b) above.
  
  (d) It is obvious from the definition.
 
\begin{figure}[h]
\centering
\includegraphics[scale=.8]{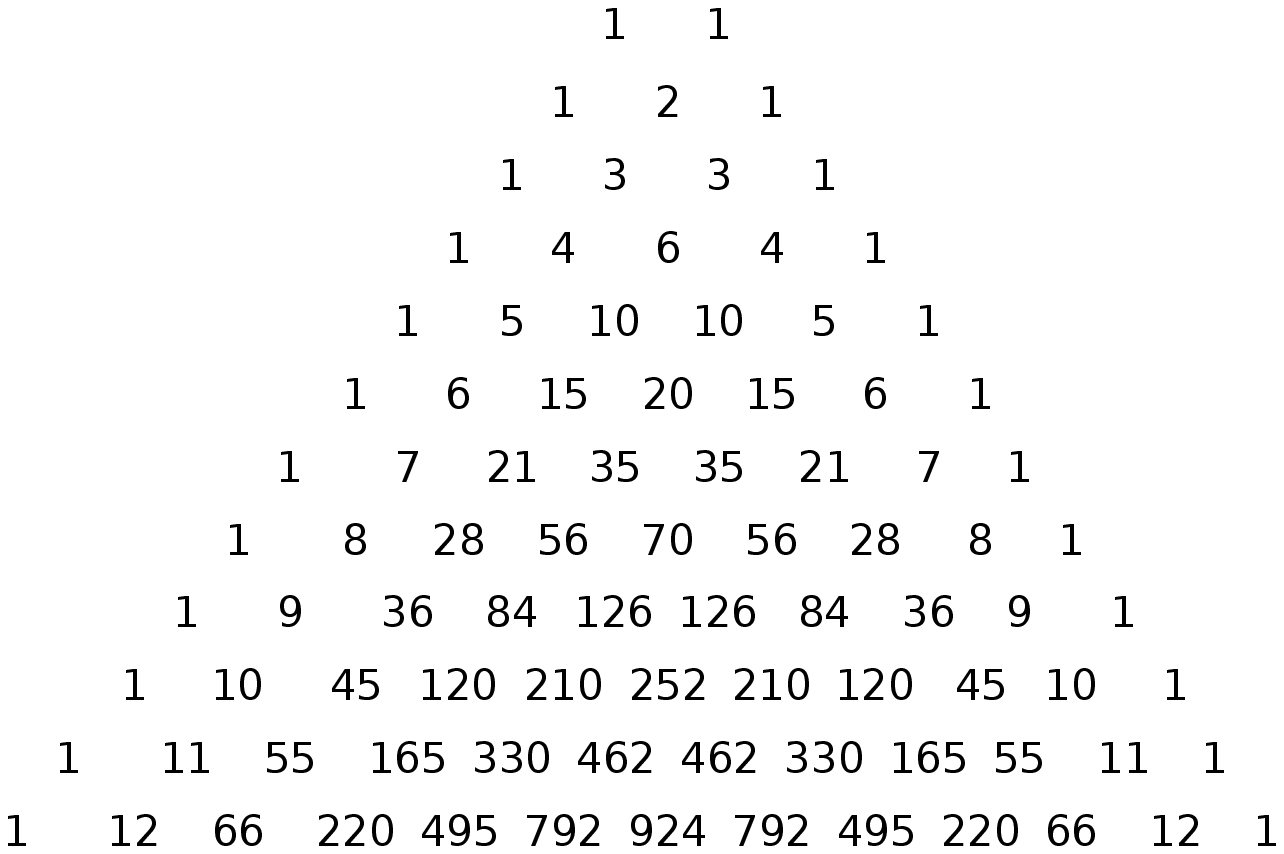}
\caption{ Pascal's triangle.}\label{fig01}
\end{figure}

 \begin{example}\label{example-2.1}
The well-known Pascal's triangle (see Fig.~\ref{fig01}) helps us to give the numbers $\tau_{\alpha}$ occurring in Theorem~\ref{thm-2.1} (iv) above. Counting rows from the top, we have to book at row number $\alpha$ and row number $n-\alpha$ only, consider suitable products and take their maximum.

For $n=6$, $d=3$, we obtain $\tau_{1} = 10$, $\tau_{2} = 12$, $\tau_{3} = 9$ and $\tau = 12$. On the other hand, $t$ can go up to $^6C_{3} = 20$ giving us several genuinely entangled vectors. It is clear how to proceed in general cases.
\end{example}


Above results and discussion give complete information in certain special cases with a little more effort. We record them in the example below.

\begin{example}\label{example-2.2}

 \begin{enumerate}[label=(\roman*)]
 \item $\xi$ is a product vector in the following two cases.
 
   \begin{enumerate}[label=(\alph*)]
   \item $t=1$, i.e., $F(\boldsymbol{\x}) = \lambda \boldsymbol{\x}^{K}$ for some scalar $\lambda \neq 0$ and $K \subset \Gamma_{n}$.
	\item $m=1$ i.e., $F(\boldsymbol{\x}) = a + b\x_{j}$ for some $j$ with $1 \leq j \leq n$ and $(a , b) \in 
 \mathbb{C}^{2}$ with $b \neq 0$.
   \end{enumerate}
   
 \item Let us consider $F$ with $t \geq 2$, $m \geq 2$ and $d =1$. We have the following.
 
 \begin{enumerate}[label=(\alph*)]
 \item $\xi$ is not a product vector because $d < m$. Indeed, $F(\boldsymbol{\x}_{S_{F}})$ is irreducible. Here, we set $F(\boldsymbol{\x}_{S_{F}}) = F(\boldsymbol{\x})$ only.
 \item If $S_{F} = \Gamma_{n}$, i.e., $m=n$,  $\xi$ is genuinely entangled. This does happen when $n=2$.
 \item Consider the case when $n \geq 3$ and $S_{F} \neq \Gamma_{n}$ and any bipartition $(E, E')$. Then $\xi$ is a product vector in this partition if and only if either $E \supset S_{F}$ or $E' \supset S_{F}$.
  \end{enumerate}
 
 \item We now consider the case $n=2 =d = m \leq t$.  $F$ has the form $F(\boldsymbol{\x}) = a_{0} + a_{1} \x_{1} + a_{2} \x_{2} + a_{3}\x_{1} \x_{2}$ with $a_{3} \neq 0$, $(a_{0}, a_{1}, a_{2}) \neq (0,0,0)$.
 
 \begin{enumerate}[label=(\alph*)]
  \item If $t=3$,  $\xi$ is entangled (and therefore, genuinely entangled).
 \item If $t = 2$, the situation gets divided in three forms for $F$, viz., $F(\boldsymbol{\x}) = a_{0} + a_{3}\x_{1} \x_{2}$ with $a_{0} \neq 0$, giving rise to an entangled $\xi$ and $F(\boldsymbol{\x}) = a_{1}\x_{1} + a_{3}\x_{1}\x_{2} = \x_{1}(a_{1} + a_{3}\x_{2})$ with $a_{1}\neq  0$ or $F(\boldsymbol{\x}) = a_{2} \x_{2} + a_{3}\x_{1}\x_{2} = (a_{2} + a_{3}\x_{1})\x_{2}$ with $a_{2}\neq 0$, both giving rise to product vectors.
 
 \item If $t =4$,  each one of $a_{0}, a_{1}, a_{2}, a_{3}$ is non-zero. $\xi$ is a product vector if and only if $a_{0}a_{3}= a_{1}a_{2}$.
   \end{enumerate}
 
 \item Let $t=2$.  $F(\boldsymbol{\x}) = a \boldsymbol{\x}^{K} + b \boldsymbol{\x}^{L}$ for some non-zero scalars $a$ and $b$ and distinct subsets $K$ and $L$ of $\Gamma_{n}$. So $F(\boldsymbol{\x}) = \boldsymbol{\x}^{K \cap L} (a \boldsymbol{\x}^{K \backslash L} + b \boldsymbol{\x}^{L \backslash K})$ and $K \triangle L = (K \backslash L) \cup  (L \backslash K) \neq \phi$. We write $q(\boldsymbol{\x}_{K \triangle L})$ for the second factor. In this case, $q$ is irreducible.
 
 \begin{enumerate}[label=(\alph*)]
 \item $\xi$ is a product vector if and only if $\# K \triangle L = 1$, i.e; either $K \subset L$ or $L \subset K$ and correspondingly $\#(L \backslash K) = 1$ or $\# (K \backslash L) = 1$.
 \item Given a bipartition $(E,E')$, $\xi$ is a product vector in this bipartition if and only if either $K \triangle L \subset E$ or $K \triangle L \subset E'$. This can happen only if $K \triangle L \neq \Gamma_{n}$ 
 \item $\xi$ is genuinely entangled if and only if $K \cap L = \phi$ and $K \cup L = \Gamma_{n}$ i.e., $K \triangle L = \Gamma_{n}$ if and only if for $1 \leq j \leq n$, neither $\x_{j}$ is a factor of $F(\boldsymbol{\x})$ nor it is missing in $F(\boldsymbol{\x})$. \\
 \end{enumerate}
 
 \item Let $t$ be any odd prime.
 
 \begin{enumerate}[label=(\alph*)]
 \item $\xi$ is entangled because $t$ is not of the form $2^{u}$ for any non-negative integer $u$.
 \item $F(\boldsymbol{\x}) = \sum_{j=1}^{t} a_{j}\boldsymbol{\x}^{K_{j}}$ for some non-zero scalars $a_{j}$'s and distinct subsets $K_{j}$ of $\Gamma_{n}$. Let $K =\bigcap\limits_{j=1}^{t} K_{j}$, $L_{j} = K_{j} \backslash K$ for $1 \leq j \leq t$, $L =\bigcup\limits_{j=1}^{t}L_{j}$.  $S_{F} = K \cup L = \bigcup\limits_{j=1}^{t} K _{j}$. Further, $F(\boldsymbol{\x}) = \boldsymbol{\x}^{K}(\sum_{j=1}^{t} a_{j}\boldsymbol{\x}^{L_{j}})$. We write $q(\boldsymbol{\x}_{L})$ for the second factor. We then have $q$ to be irreducible because $t$ is prime and no $\x_{j}$ is a factor of $q(\boldsymbol{\x}_{L})$. Moreover, $S_{q} = L$ and $\# L \geq 2$.
 \item Given a bipartite cut $(E, E')$, $\xi$ is a product vector in this cut if and only if either $L \subset E$ or $L \subset E'$. This can happen only if $L \neq \Gamma_{n}$.
 \item $\xi$ is genuinely entangled if and only if $K = \phi$ and $L = \Gamma_{n}$, i.e., $\bigcap\limits_{j=1}^{t} K_j = \phi$ and $\bigcup\limits_{j=1}^{t} K_j = \Gamma_{n}$ if and only if for $1\leq j \leq n$, neither $X_j$ is a factor of $F(\boldsymbol{\x})$ nor $\x_{j}$ is missing in $F(\boldsymbol{\x})$.
  \end{enumerate}
 
 \end{enumerate}
 This example enables us to narrow our study to cases that satisfy $n \geq 3$, $m \geq 2$, $d \geq 2$ and $t$, a composite number.
\end{example} 

\section{Polynomial Representation of Resonating Valence Bond States  and Their Quantum Entanglement}\label{section-3}

For the sake of self-completeness, convenient notation and terminology, we begin with the simple case of a singlet or a dimer and go up to that of doped resonating valence bond states with variable coefficients. With the right choice made in Sec. \ref{section-2}, their polynomial representation turns out to be a homogeneous polynomial of small degree. This facilitates the study of their quantum entanglement which requires new conditions that we introduce. 

\subsection{Basic definitions and results}\label{subsection-3.1}

\begin{figure}[h]
\centering
\includegraphics[scale=.8]{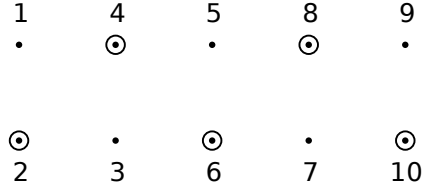}
\caption{Bipartite lattice. The lattice is divided into two sublattices, say $A$ and $B$ such that each point in  sublattice $A$ (marked by ${\bold{\cdot}}$) is surrounded by sublattice $B$ (marked by by $\odot$) and vice versa.}\label{fig02}
\end{figure}

Let $\nu \in \mathbb{N}$ and $n=2\nu$. We divide $\Gamma_{n} = \left\{ j \in N : 1 \leq j \leq n \right\}$ into two equal parts $A$ and $B$, say $A = \left\{ 2t-1 : 1 \leq t \leq \nu \right\}$ and $B = \left\{ 2t : 1 \leq t \leq \nu \right\}$. For pictorial representations of various concepts, we will mark points of $A$ by ${\bold{\cdot}}$ and those of $B$ by $\odot$ and arrange them in different ways. For instance, Fig.  \ref{fig02} is useful in the nearest neighbour (NN) context.

\vspace{.5cm}
\textbf{(1) Coverings.} 

A bijective map $\psi$ of $A$ to $B$ is called a covering, say $C$, of $\Gamma_{n}$. It can be given in terms of a permutation ${\widetilde{\psi}}$ of $\Gamma_{\nu}$ by setting
\begin{equation*}
{\psi}(2t-1) = 2 {\widetilde{\psi}}(t), \tag{3.1}
\end{equation*}
 for $t \in \Gamma_{v}$.
  \begin{enumerate}[label=(\alph*)]
  \item ${\psi}$ is called a nearest neighbour covering  $(NN)$ if $2(t-1) \leq {\psi}(2t-1) \leq 2(t+1)$, i.e., 
  \begin{equation*}
  t-1 \leq {\widetilde{\psi}}(t) \leq t+1, \tag{3.2}
  \end{equation*}
for $t \in \Gamma_{v}$. In other words, the permutation matrix of ${\widetilde{\psi}}$ is banded matrix with band-width $\leq 1$.
   \item Samson and Ezermann~\cite{Sams} study banded matrices and note in Remark 23 (of their paper) that the number of such matrices with bandwidth $\leq 1$ is the Fibonacci number $F_{v}$ with initial conditions $F_{1}=1 =F_{0}$. See also~\cite{RDSS, SSR, RDSS, jome-doi}.
   \item ${\psi}$ is called a periodic $NN$ covering if we permit ${\psi}(2v-1)$ to be 2 and ${\psi}(1)$ to be \(2\nu\) as well. In other words, we work in $\mathbb{Z} \md 2v$ .
   \item The number of all coverings of $\Gamma_{n}$ is $\nu!$, the number of permutations of $\Gamma_{v}$.
    \item We will be considering various sets $\bold{\Psi}$ of coverings with $\# \bold{\Psi} \geq 2$ and $v \geq 2$. For $v=2$, $\# \bold{\Psi}$ can only be 2, since the coverings are only of horizontal or vertical types.

    \item Some easy coverings are depicted in Fig.~\ref{fig03}.
    
    \begin{figure}[h]
\centering
\includegraphics[scale=.8]{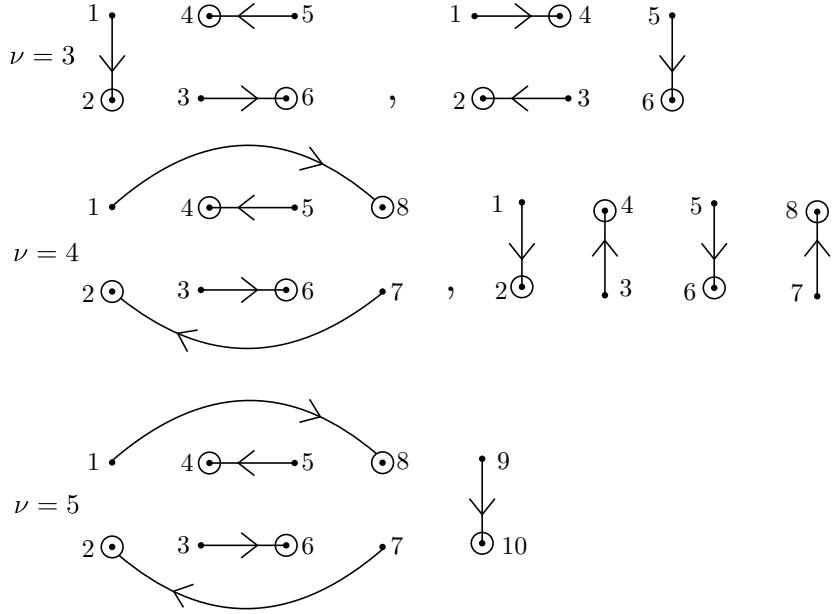}
\caption{Some easy coverings.}\label{fig03}
\end{figure}

  \end{enumerate}


{\bfseries (2) Resonating valence bond (RVB) states and vectors.}

For $a\in A$, $b \in B$, we write $H_{a,b}$ for $\mathcal{H}_{a} \bigotimes \mathcal{H}_{b}$. We follow the polynomial representation as set out in Sec. ~\ref{section-2} above.

\begin{enumerate}[label=(\alph*)]
\item A dimer or a singlet at \(a\), \(b\) is the unit vector
$$
[a, b] = \frac{1}{\sqrt{2}} (| \uparrow \rangle_{a} \otimes | \downarrow \rangle_{b} - | \downarrow \rangle_{a} \otimes | \uparrow \rangle_{b}) = \frac{1}{\sqrt{2}}(| 0\rangle_{a} \otimes | 1 \rangle_{b} - | 1 \rangle_{a} \otimes | 0 \rangle_{b})\quad {\rm in} \quad  \mathcal{H}_{a,b}. 
$$
Its polynomial representation is given by
\begin{equation*}
F_{a_{1}, b}(\x_{a}, \x_{b}) = \frac{1}{\sqrt{2}}(\x_{b}- \x_{a}).\tag{3.3}
\end{equation*}

\item Let  $C$ be a covering of $\Gamma_{n}$ i.e., a bijective map $\psi$ of $A$ to $B$. We let $| C \rangle = | \psi \rangle$ be the vector in $\mathcal{H}$ identified with $\bigotimes\limits_{a \in A} \mathcal{H}_{a, \psi_{(a)}}$ given by $\bigotimes\limits_{a \in A} [a, \psi (a)]$, i.e.,
\begin{gather*}
2^{-v/2} \bigotimes\limits_{a \in A} \left(| \uparrow \rangle_{a} \otimes | \downarrow \rangle_{\psi (a)}) - | \downarrow \rangle_{a} \otimes | \uparrow \rangle_{\psi (a)}\right)\\
= 2^{-v/2} \bigotimes\limits_{a\in A} \left(|0\rangle_{a} \otimes | 1 \rangle_{\psi (a)} - |1 \rangle_{a} \otimes | 0 \rangle_{\psi(a)}\right).
\end{gather*}
It is a unit vector and its polynomial representation is 
\begin{equation}
F_{{\psi}} (\boldsymbol{\x}) = 2^{-v/2} \prod\limits_{a\in A}(\x_{\psi(a)}- \x_{a}).\tag{3.4}\label{eq-3.4}
\end{equation}
It is a non-zero homogeneous polynomial of degree $d=v$ and its support is $S_{F_{\psi}} = \Gamma_{n}$, i.e., $m=n$.

Indeed, 
\begin{align*}
F_{\psi}(\boldsymbol{\x}) &= 2^{-v/2} \sum_{A_{1} \subset A}(-1)^{\# A_{1}} \boldsymbol{\x}^{A_{1}} \boldsymbol{\x}^{B\smallsetminus \psi(A_{1})}\nonumber\\
& = 2^{-v/2} \sum_{\substack{A_{1} \subset A \\ \#A_{1} \leq v/2 \\ 1 \epsilon A_{1} \\ \textrm{if} \, \#A_{1}= v/2}} (-1)^{\# A_{1}} \left(\boldsymbol{\x}^{A_{1}} \boldsymbol{\x}^{B\smallsetminus \psi (A_{1}) } + (-1)^{v} \boldsymbol{\x}^{A \smallsetminus A_{1}} \boldsymbol{\x}^{\psi (A_{1})} \right).\tag{3.5}  
\end{align*}

Clearly, the number $t$ of terms in $F_{\psi}$ is $2^{v}$ and $\boldsymbol{\x}^{K}$ is not a factor of $F_{\psi}(\boldsymbol{\x})$ for any $\phi \neq K \subset \Gamma_{n}$.

\item Let $v \geq 2$ and $\bold{\Psi}$, a set of coverings of $\Gamma_{n}$ with $ \# \bold{\Psi} \geq 2$.  We write $| \bold{\Psi} \rangle $ for the RVB vector $\frac{1}{\# \bold{\Psi}} \sum_{{\psi} \in \bold{\Psi}}  | {\psi} \rangle$ in $\mathcal{H}$. We note that $|| \bold{\Psi} || \leq 1$ and we will come to its normalization later.

\item The polynomial representation of $| \bold{\Psi} \rangle $ is given by
\begin{equation}
F_{\bold{\Psi}}(\boldsymbol{\x}) = \frac{1}{\# \bold{\Psi}} \sum_{\psi \in \bold{\Psi}} F_{{\psi}}(\boldsymbol{\x}) = \frac{1}{\# \bold{\Psi}} 2^{-v/2} \sum_{\substack{{A_{1} \subset A}\\ B_{1}\subset B, \# B_{1}= \# A_{1}}} (-1)^{\# A_{1}} \# \left\{\psi \in \bold{\Psi} : \psi (A_{1}) = B_{1}\right\} \boldsymbol{\x}^{A_{1}} \boldsymbol{\x}^{B\smallsetminus B_{1}}.\tag{3.6}\label{eq-3.6}
\end{equation}
Here $ |\bold{\Psi} \rangle  \neq 0$ simply because $0 \neq 2^{-v/2}(\boldsymbol{\x}^{B} + (-1)^{v} \boldsymbol{\x}^{A})$ is a part of $F_{\bold{\Psi}}(\boldsymbol{\x})$.
Indeed, this also gives that $S_{F_{\bold{\Psi}}} = \Gamma_{n}$, i.e., $m=n$ and $\boldsymbol{\x}^{K}$ is not a factor of $F_{\bold{\Psi}} (\boldsymbol{\x})$ for any $\phi \neq K \subset \Gamma_{n}$.

\item Let $v \geq 2 $, $\bold{\Psi}_{NN} =$ the set of nearest neighbour $(NN)$ coverings of $\Gamma_{n}$ and $\bold{\Psi}_{PNN} =$ the set of periodic nearest neighbour $(PNN)$  coverings of $\Gamma_{n}$.  $|\bold{\Psi}_{NN} \rangle$ and $| \bold{\Psi}_{PNN} \rangle$ can be be called the NN-RVB and the PNN-RVB vectors (in the context of n-qubit system).

We record some immediate applications of Theorem~\ref{thm-2.1}, Examples II.1 and II.2 above in Sec.~\ref{section-2} labelled as Theorem~\ref{thm-3.1} and Example~\ref{example-3.1}.

\end{enumerate}


\begin{thm}\label{thm-3.1}

Considering vectors in their respective spaces, we have the following. 
\begin{enumerate}[label=(\roman*)]

\item A dimer $[a, b]$, $| \psi \rangle$, $| \bold{\Psi} \rangle$ are all entangled.

\item Let $v \geq 2$ and $(E, E')$, a  bipartition of $\Gamma_{n}$.

   \begin{enumerate}[label=(\alph*)]
    \item $| \psi \rangle$ is a product vector in 
    $(E, E')$ if and only if 
			\begin{align*}
			\psi(E \cap A) &= E \cap B \quad \text{if and only if}\\
			\psi(E' \cap A) &= E'\cap B. \tag{3.7}\label{eq-3.7}
			\end{align*}
			
	 In this case, we have that $\# E$ and $\# E'$ are both even.
	 \item If $| \bold{\Psi} \rangle$ is a product vector in the bipartition $(E, E')$,  
	 then
\begin{equation}\tag{3.8}\label{eq-3.8}  
	 		 \psi(E \cap A) = E \cap B \quad ({\text {equivalently}}, \psi (E' \cap A) = E' \cap B)\; \text{for each}\; \psi \;\text{in}\; \bold{\Psi}.
	 		\end{equation} 		
	 	In this case, $\# E$ and $\# E'$ are both even.	
    \end{enumerate}
    
   \item For $v=2$, $| \bold{\Psi} \rangle$ is genuinely entangled.
   
   \item For $\bold{\Psi} $ NN or PNN, $|\bold{\Psi} \rangle $ is genuinely entangled.
   
   \item (a) The converse of (ii) (b) holds for $v=3$ but not, in general.
   
   (b) The converse of (ii) (b) holds for all $v \geq 3$ if we consider special cuts with $\min \left\{ \# E, \# E' \right\} = 2$.
\end{enumerate}
\end{thm}

{\it Proof.}
\begin{enumerate}[label=(\roman*)]
 
 \item  It follows from Theorem~\ref{thm-2.1} (i) (a) because $d=m/2$ for the corresponding polynomials. 

\item (a) The second equivalence is trivial. For the first one, the `if' part is immediate. For the  `only if' part, let, $p(\boldsymbol{\x}_{E}), q (\boldsymbol{\x}_{E'})$ be polynomials that satisfy $F_\psi (\boldsymbol{\x}) = p(\boldsymbol{\x}_{E}) q(\boldsymbol{\x}_{E'})$. Then by Theorem~\ref{thm-2.1} (iv) (b), $p(\boldsymbol{\x}_{E})$ and $q(\boldsymbol{\x}_{E'})$ are both homogeneous, say, of degree $v_{1}$ and $v_{2}$ respectively with $v_{1} \geq 1$, $v_{2} \geq 1$, $v_{1} + v_{2} = v$, $S_{p} =E$ and $S_{q} = E'$.  Now by Eq.~(3.5), $\boldsymbol{\x}^{A}$ and $\boldsymbol{\x}^{B}$ both occur in $F_{\psi}(\boldsymbol{\x})$. The one and only one way it can happen is the $\boldsymbol{\x}^{E \cap A}$, $\boldsymbol{\x}^{E \cap B}$ occur in $p(\boldsymbol{\x}_{E})$, and $\boldsymbol{\x}^{E' \cap A}$ and $\boldsymbol{\x}^{E' \cap B}$ occur in $q(\boldsymbol{\x}_{E'})$. This gives that $\# E \cap A = \# E \cap B = v_{1}$ and $\# E' \cap A = \# E' \cap B =v_{2}$. Now $\boldsymbol{\x}^{E \cap A}$ $\boldsymbol{\x}^{E' \cap B}$ occurs in $p(\boldsymbol{\x}_{E})q(\boldsymbol{\x}_{E'})$. But this can happen for $F_{\psi} (\boldsymbol{\x})$ if and only if $B \backslash \psi(E \cap A) = E' \cap B$, i.e., $\psi(E \cap A) = E \cap B$. As a consequence $\psi (E' \cap A) = E' \cap B$ as well. This confirms the assertion already made about the relationships amongst cardinalities of the sets, $v_{1}$, $v_{2}$. Clearly $\# E = \# (E \cap A) + \# (E \cap B) = 2 v_{1}$.

(b) As noted before,
$\boldsymbol{\x}^{A}$ and $\boldsymbol{\x}^{B}$ both occur in $F_{\bold{\Psi}}(\boldsymbol{\x})$. So arguments in the proof of part (ii) (a) above by replacing $F_{\psi}$ by $F_{\bold{\Psi}}$ with the addition of ``for some $\psi \in \bold{\Psi}$'' in the sentence beginning with `But'. Now consider any $\varphi \in \bold{\Psi}$. Then $\boldsymbol{\x}^{E \cap A} \boldsymbol{\x}^{B \smallsetminus \varphi (E \cap A)}$ occurs in $F_{\bold{\Psi}}(\boldsymbol{\x})$ and, therefore, in $p(\boldsymbol{\x}_{E}) q(\boldsymbol{\x}_{E'})$. The only way this can happen is that $B \smallsetminus \varphi (E \cap A) = E' \cap B$, which gives, in turn, $\varphi(E \cap A) = E \cap B$. This completes the proof of the first assertion. The second statement follows immediately.

\item For $v=2$, the condition~\eqref{eq-3.8}   in (ii) (b) is not satisfied for any $\phi \neq E \subsetneq \Gamma_{4}$ as is clear from previous discussion. 


\item In view of (iii), it is enough to consider the case $v \geq 3$.  Let if possible, the condition in (ii)(b) be satisfied for some $E$ with $\phi \neq E \subsetneq \Gamma_{n}$ and $\bold{\Psi}=\bold{\Psi}_{NN}$.

Consider any $2s-1 \in E \cap A$ with $1 \leq s \leq v$. If $s < v$, then there exist $\psi_{1}, \psi_{2} \in \bold{\Psi}$ with $\psi_{1} (2s-1) = 2(s + 1)$ and $\psi_{2} (2s+1) = 2(s+ 1)$. So $2(s+ 1) \in E \cap B$, which in turn, gives that $2s+1 \in E \cap A$. On the other hand, if $s > 1$, then there exist $\psi_{3}, \psi_{4} \in \bold{\Psi}$ satisfying $\psi_{3}(2s-1)= 2(s-1)$ and $\psi_{4}(2s-3) =2 (s-1)$. So $2(s-1) \in E \cap B$ which leads to $2s-3 \in E \cap A$. Repeating the arguments iteratively we arrive at the conclusion that $\left\{2t-1 : 1 \leq t \leq v \right\} \subset E \cap A$. But that is not so. Hence $| \bold{\Psi}_{NN} \rangle $ is not a product vector in any bipartite cut and is, therefore, genuinely entangled.  

Now we come to $|\bold{\Psi}_{PNN} \rangle$. Since $\bold{\Psi}_{PNN} \supset \bold{\Psi}_{NN}$, it cannot satisfy \eqref{eq-3.8} in (ii)(b) for any $E$ with $\phi \neq E \subsetneq \Gamma_{n}$. Hence $|\bold{\Psi}_{PNN} \rangle$ is genuinely entangled.

\item 
\begin{enumerate}[label=(\alph*)]
\item For any bipartite cut $(E, E')$ of $\Gamma_{6}$ with $\#E$ and $\#E'$ both even, we obtain $\min\left\{\# E, \# E'\right\}=2$.  Suppose $\#E'=2$ and for any $\psi \in \bold{\Psi}$, $\psi (E' \cap A) = E' \cap B$, say, $\psi(a) = b$ where $E'\cap A = \{a\}$, $E'\cap B=\{b\}$. So $\x_{b} - \x_{a}$ is a factor of $F_{\bold{\Psi}} (\boldsymbol{\x})$. Similarly for the case of $\# E=2$. Hence $| \bold{\Psi} \rangle $ is a product vector in the cut $(E, E')$ by Theorem~\ref{thm-2.1}(ii)(a).
We shall give an example in Example~\ref{example-3.1} below to show that the converse of (ii)(b) in Theorem here does not hold in general.

\item We can formulate the proof on the lines of that of (a) above.

\end{enumerate}

\end{enumerate}


\begin{definition}\label{def-3.1}
(i) We call $\bold{\Psi}$ $\text{{\bf decomposable via}} ~E$ if it satisfies the condition~\eqref{eq-3.8} in Theorem~\ref{thm-3.1}(ii)(b) above, i.e., $\psi(E\cap A) = E\cap B$ for $\psi \in \bold{\Psi}$. (ii) $\bold{\Psi}$ is called $\boldsymbol{\rm decomposable}$ if it is decomposable via $E$ for some bipartition $(E, E')$. 
\end{definition}

\begin{example}\label{example-3.1}

\begin{enumerate}
\item Let $v=3$ and $\bold{\Psi}$ consist of two coverings given in Fig. \ref{fig03}. Then $\bold{\Psi}$ is not decomposable which implies that  $ | \bold{\Psi} \rangle $ is genuinely entangled.

\item Let $v=4$ and $\bold{\Psi}$ consist of the two coverings in Fig. \ref{fig03}. Then  $\bold{\Psi}$ is decomposable with $E= \{3,4,5,6\}$ serving the desired purpose. However, $ | {\bf \Psi}\rangle$ is not a product vector in the bipartite cut $(E, E')$  as can be seen by computing $F_{\bold{\Psi}}$. In fact,
\begin{equation*}
\begin{split}
F_{\bold{\Psi}}(\boldsymbol{\x}) = &\left(\x_{2} \x_{8} + \x_{1}\x_{7} - \x_{1}\x_{2} -\x_{7}\x_{8}\right)\left(\x_{4}\x_{6} + \x_{3}\x_{5}- \x_{3}\x_{4} - \x_{5}\x_{6} \right)\\
&+ \left(\x_{2}\x_{8} + \x_{1}\x_{7} - \x_{1}\x_{8}-\x_{2}\x_{7}\right) \left(\x_{4} \x_{6} + \x_{3} \x_{5} - \x_{3}\x_{6}-\x_{4} \x_{5}\right).
\end{split}
\end{equation*}
This can not be factored  as $p(\boldsymbol{\x}_{E})~q(\boldsymbol{\x}_{E'})$ for any polynomials $p$ and $q$. Moreover, apart from $E$, there is no other subset $G$ with $\phi \neq G \subsetneq \Gamma_{8}$ that satisfies the condition~\eqref{eq-3.8} of decomposability. This shows that the converse of Theorem~\ref{thm-3.1}(ii)(b) does not hold in  general.
\end{enumerate}

\end{example}

${{\text{\bf (3) Concepts and properties of}}~ \mathbf{\bold{\Psi}}~ \text{\bf related to decomposability.}}$

Let $v\geq 3$ and $\# \bold{\Psi} \geq 2$.

\begin{enumerate}[label=(\alph*)]

\item  Let $\mathcal{A}_{1} = \left\{A_{1} \subset A : \# \left\{\psi (A_{1}) : \psi \in \bold{\Psi} \right\}  =1\right\}$. For $A_{1} \in \mathcal{A}_{1}$, we write $B_{1}$ for $\psi(A_{1})$ for any (and thus for all) $\psi \in \bold{\Psi}$. Then $\phi, A \in \mathcal{A}_{1}$ and $\mathcal{A}_{1}$ is an algebra of subsets of $A$. As a consequence, $\mathcal{B}_{1} = \left\{B_{1}= \psi(A_{1}) : A_{1} \in \mathcal{A}_{1} \right\}$ is the same for all \(\psi \in \bold{\Psi}\) and  is an algebra of subsets of $B$. In particular, \# $\mathcal{A}_{1} \geq 3$ if and only if \# $\mathcal{B}_{1} \geq 3$. In this case, let $\mathcal{A}_{1}^{\textrm{min}}$ be the subset of $\mathcal{A}_{1}\backslash \{\phi\}$ consisting of its minimal subsets.

\item Now suppose \# $\mathcal{A}_{1} \geq 3$ and consider any $A_{1} \in \mathcal{A}_{1}$ with $\phi \neq A_{1}\neq A$. We set $E = A_{1} \cup B_{1}$. Then  $ \phi \neq E \neq \Gamma_{n}$. Also $\psi(E \cap A) = E \cap B$ for $\psi \in \bold{\Psi}$. In other words, $\bold{\Psi}$ is decomposable via $E$. On the other hand, if $\bold{\Psi}$ is decomposable via $E$,  $\phi \neq E \cap A \neq A$ and $E \cap A \in \mathcal{A}_{1}$. This, in turn, gives that \# $\mathcal{A}_{1} \geq 3$. 

\item Since \# $\bold{\Psi} \geq 2$, there exist $\psi_{1}, \psi_{2} \in \bold{\Psi}$, $a_1 \in A$, $b_{1} \neq b_{2}$ in $B$ with $\psi_{1}(a_{1}) = b_{1}$, $\psi_{2}(a_{1}) = b_{2}$. So $\{a_{1}\} \notin \mathcal{A}_{1}$. Hence $\mathcal{A}_{1} \neq P_{A}$. 

\item Consider $a_{1}$ as in (c) above and $a_{2}\neq a_{1}$ in $A$. If $\left\{a_{1}, a_{2} \right\} \in \mathcal{A}_{1}$,  $\psi_{1}\left\{a_{1}, a_{2} \right\} = \left\{b_{1},\psi_{1} (a_{2})\right\}$ has to be $\psi_{2}\left\{a_{1}, a_{2} \right\} = \left\{b_{2},\psi_{2} (a_{2})\right\}$, so that $\psi_{1}(a_{2}) = b_{2}$ and $\psi_{2}(a_{2}) = b_{1}$ and, thus, $\left\{a_{2}\right\} \notin \mathcal{A}_{1}$. On the other hand, if $\left\{a_{2}\right\} \in \mathcal{A}_{1}$, then $\psi_{1}\left\{a_{1}, a_{2}\right\} = \left\{b_{1}, \psi_{1}(a_{2})\right\} = \left\{b_{1}, \psi_{2}(a_{2})\right\} \neq \left\{b_{2},  \psi_{2}(a_{2})\right\}$ and, thus $\left\{a_{1}, a_{2} \right\} \notin \mathcal{A}_{1}$. Hence\, at most one of $\left\{a_{2}\right\}$, $\left\{a_{1}, a_{2}\right\}$ is in $\mathcal{A}_{1}$.

\item Let $\mathcal{A}_{2} = \left\{A_{2} \subset A : \# \{\psi (A_{2}) : \psi \in \bold{\Psi}\} \geq 2\right\}$. Then $\mathcal{A}_{1} \cap \mathcal{A}_{2} = \phi$, $\mathcal{A}_{1} \cup \mathcal{A}_{2} = P_{A}$ and for $A_{2} \in \mathcal{A}_{2}$, $A \smallsetminus A_{2} \in \mathcal{A}_{2}$. By (c) above $\{a_{1}\} \in \mathcal{A}_{2}$ and therefore, $A\smallsetminus \{a_{1}\} \in \mathcal{A}_2$. Now by (d), for $a_{2}\in A\smallsetminus\{a_{1}\}$, at least one of $\{a_{2}\}$
and $\{a_{1}, a_{2}\}$ is in $\mathcal{A}_{2}$ and neither of them or their complements in $A$ equal $\{a_{1}\}$ or $A\smallsetminus\{a_{1}\}$ because $v\geq 3$. Hence $\# \mathcal{A}_{2}\geq 4$.

\item $\phi$ and $A$ are in $\mathcal{A}_{1}$ and therefore, not in $\mathcal{A}_{2}$. So for $A_{2} \in \mathcal{A}_{2}$ $v-1 \geq \#$ $A_{2}\geq 1$ and, therefore, $^{v}C_{\# A_{2}} \geq v$.

\item 
\begin{align*}
F_{\bold{\Psi}}(\boldsymbol{\x}) &= 2^{-v/2} \sum_{A_{1} \in \mathcal{A}_{1}}(-1)^{\# A_{1}}  \boldsymbol{\x}^{A_{1}} \boldsymbol{\x}^{B\smallsetminus B_{1}} + \\
   & 2^{-v/2} \frac{1}{\# \bold{\Psi}} \sum_{A_{2} \in \mathcal{A}_{2}} (-1)^{\# A_{2}} \boldsymbol{\x}^{A_2} \left(\sum_{B_{2} \subset B} \# \left\{\psi \in \bold{\Psi} : \psi(A_{2}) = B_{2}\right\}\boldsymbol{\x}^{B\smallsetminus B_{2}}\right).\tag{3.9}
\end{align*}

\item Let $A_{2} \in \mathcal{A}_{2}$. We write, for $B_{2} \subset B$, $C_{A_{2}, B_{2}}$ for $\frac{1}{\#\bold{\Psi}} \# \{\psi \in \bold{\Psi} : \psi (A_{2}) = B_{2} \}$, which is $\geq 0$.

We note that $C_{A_{2}B_{2}} \neq 0$ if and only if $B_{2} \in \{ \psi (A_{2}) : \psi \in \bold{\Psi}\}$.

In particular, $C_{A_{2}B_{2}} = 0$ for $\# A_{2} \neq \# B_{2}$. Moreover $\sum_{B_{2 \subset B}}$ $C_{A_{2}, B_{2}} =1$. So $^{v}C_{\# A_{2}} \geq \# \{B_{2} \subset B : C_{A_{2}, B_{2}} \neq 0 \} \geq 2$.

This, in turn, gives that $C_{A_{2}, B_{2}}>0$ for at last two distinct subsets $B_{2}$ of $B$. As a consequence, $C_{A_{2}, B_{2}} < 1$ for all $B_{2} \subset B$.

\item Let 
\begin{align*}
F^{1}_{\bold{\Psi}}(\boldsymbol{\x})  &= 2^{-v/2} \sum_{A_{1} \in \mathcal{A}_{1}}(-1)^{\# A_{1}} \boldsymbol{\x}^{A_{1}} \boldsymbol{\x}^{B\smallsetminus B_{1}} \\
F^{2}_{\bold{\Psi}}(\boldsymbol{\x})  &= 2^{-v/2} \sum_{A_{2} \in \mathcal{A}_{2}} \sum_{B_{2} \subset B} C_{A_{2}, B_{2}}(-1)^{\# A_{2}} \boldsymbol{\x}^{A_{2}} \boldsymbol{\x}^{B\smallsetminus B_{2}}. \tag{3.10}\label{eq-3.10}\\
\text{So}\quad\quad F_{\bold{\Psi}} (\boldsymbol{\x}) &= F^{1}_{\bold{\Psi}} (\boldsymbol{\x}) + F^{2}_{\bold{\Psi}} (\boldsymbol{\x}).
\end{align*}

Let $t$, $t^{(1)}$, $t^{(2)}$ be the number of terms in $F_{\bold{\Psi}}(\boldsymbol{\x})$, $F^{(1)}_{\bold{\Psi}}(\boldsymbol{\x})$ and $F^{(2)}_{\bold{\Psi}}(\boldsymbol{\x})$ respectively. Then $t^{(1)} = \# \mathcal{A}_{1}$.

Using (h), we have $\sum_{A_{2} \in \mathcal{A}_{2}} \quad ^{v}C_{\# A_{2}} \geq t^{(2)} \geq 2 \# \mathcal{A}_{2}$. But by (e), $\# \mathcal{A}_{2} \geq 4$  and therefore, $t^{(2)} \geq 8$. Hence 
\begin{align*}
2^ {v} + \sum_{A_{2} \in \mathcal{A}_{2}}(^{v}C_{\# A_{2}}-1) &= \# \mathcal{A}_{1} + \sum_{A_{2} \in \mathcal{A}_{2}}\, ^{v}C_{\# A_{2}} \geq t^{(1)} + t^{(2)} = t \geq \# \mathcal{A}_{1} + 2 \# \mathcal{A}_{2}  \\[-.1cm]
 & = \#P_{A} + \# \mathcal{A}_{2}\\
& = 2^{v} + \# \mathcal{A}_{2} \geq 2^{v} + 4. \tag{3.11}  
\end{align*}

\item If $\{a\} \in \mathcal{A}_{1}$ for some $a\in A$,  by (b) above and Theorem~\ref{thm-3.1} (v)(b), $|\bold{\Psi} \rangle $ is a product vector in the bipartite cut $(E, E')$ with $E= \{a, b\}$ for $b =\psi(a)$ for some $\psi  \in \boldsymbol{{\Psi}}$.

\item We confine our attention to the case that $\{a\} \notin \mathcal{A}_{1}$, $a\in A$. In that case, $\{a\} \in \mathcal{A}_{2}$ for $a\in A$. So $\# \mathcal{A}_{2} \geq 2v$ because $\{A \smallsetminus \{a\}, \{a\} : a \in A \}$ are all different sets in view of $v \geq 3$. This refines the second part of (3.11) viz; 
\begin{equation*}
t \geq 2^{v} + \# \mathcal{A}_{2} \geq 2^{v} + 2 \nu = 2(2^{v-1} + v).\tag{3.12} 
\end{equation*} 

\end{enumerate}

We record a few tests of decomposability which are implicit 
above.


\begin{thm}\label{3.2}
{\bf (Tests of decomposability.)} Let $v\geq 3$ and $\# \bold{\Psi} \geq 2$.
\begin{enumerate}[label=(\roman*)]
\item The following are equivalent.
\begin{enumerate}[label=(\alph*)]
\item $\bold{\Psi}$ is decomposable.

\item $\# \mathcal{A}_{1} \geq 3$.

\item $\# \mathcal{A}_{1} \geq 4$.

\item The number $t^{\hat{}}$ of terms in $F_{\bold{\Psi}}(\boldsymbol{\x})$ with $|$co-efficient$|= 2^{-v/2}$ is $\geq 3$.

\item The number $t^{\hat{}}$  as in (d) above is $\geq 4$.

\end{enumerate}

\item If $t < 2^{v} + 2v$ then $\bold{\Psi}$ is decomposable via  some $E=\left\{ a, b \right\}$ with some $a \in A$ and $b \in B$. Further, in this case $| \bold{\Psi} \rangle$ is a product vector in the bipartite cut $(E, E')$.
\end{enumerate}

\end{thm}


{\bf (4) Normalization of $| \bold{\Psi} \rangle$}. 

Let $v \geq 2$ and $\# \bold{\Psi} \geq 2 $.

 (a) As discussed before, 
 for $v=2$, there is only one $\bold{\Psi}$. 

The polynomial representation of $| \bold{\Psi} \rangle$ is given by
\begin{align*}
F_{\bold{\Psi}} (\boldsymbol{\x}) &= \frac{1}{2^{2}}\left((\x_{2} - \x_{1}) (\x_{4} - \x_{3}) + (\x_{4}-\x_{1}) (\x_{2}-\x_{3})\right)\\
 &= \frac{1}{4} \left(2(\x_{1} \x_{3} + \x_{2} \x_{4}) - \x_{1} \x_{4}- \x_{2}\x_{3} -\x_{1}\x_{2}-\x_{3}\x_{4}\right).\\
\text{Hence} \quad || | \bold{\Psi} \rangle || &= \frac{1}{4} \sqrt{\left(4+4+1+1+1+1\right)} = \frac{\sqrt{3}}{2}.
\tag{3.13}
\end{align*}

Hence the corresponding RVB state is $| \widetilde{\bold{\Psi}} \rangle  = \frac{2\sqrt{3}}{3} | \bold{\Psi} \rangle$ and its polynomial representation is given by 
\begin{equation*}
\widetilde{F}_{\bold{\Psi}}(\boldsymbol{\x}) = \frac{\sqrt{3}}{6} \left(2(\x_{1}\x_{3} + \x_{2}\x_{4}) - \x_{1}\x_{4} - \x_{2}\x_{3}- \x_{1}\x_{4} -\x_{3}\x_{4}\right).\tag{3.14}
\end{equation*}

Let $v\geq 3$.
The terms of the polynomial representation of $\widetilde{F}_{\bold{\Psi}}(\boldsymbol{\x})$ coming from non-zero values of $C_{A_{2}, B_{2}}$s are mutually orthogonal in $\text{L}^{2}\left(T^n\right)$, and hence we have
\begin{equation*}
|| | \bold{\Psi} \rangle || = ||\widetilde{F}_{\bold{\Psi}}(\boldsymbol{\x})||_{2} = 2^{-v/2}\sqrt{\left(\# \mathcal{A}_{1}  + \sum_{A_{2} \in \mathcal{A}_{2}} \left(\sum_{B_{2} \subset B} C_{A_{2}, B_{2}}^{2}\right) \right)}.
\tag{3.15}
\end{equation*}

But for each $A_{2} \in \mathcal{A}_{2}$, $\# \{B_{2} \subset B  : C_{A_{2}, B_{2}} \neq 0\} \geq 2$ and $\sum_{B_{2}\subset B} C_{A_{2}, B_{2}} = 1$;
Therefore, $\sum_{B_{2} \epsilon B} C_{A_{2}, B_{2}}^{2} < 1$ for $A_{2} \in \mathcal{A}_{2}$.
Hence
\begin{equation*}
|| | \bold{\Psi} \rangle || < 2^{-v/2} \sqrt{( \#\mathcal{A}_{1} +  \#\mathcal{A}_{2}) }=1.
\tag{3.16}
\end{equation*}

The corresponding RVB state, say $\widetilde{| \bold{\Psi} \rangle} = \frac{1}{|| | \bold{\Psi} \rangle||} | \bold{\Psi} \rangle$ and its polynomial representation is 
\begin{equation*}
\widetilde{F}_{\bold{\Psi}} (\boldsymbol{\x}) = \frac{1}{|| | \bold{\Psi} \rangle||} F_{\bold{\Psi}}(\boldsymbol{\x}).
\tag{3.17}
\end{equation*}


{\bf (5) Motivation for factorability of sets of coverings}.

Theorem~\ref{thm-3.1} and previous discussion 
permit us to confine out attention to $v\geq 3$ and decomposable coverings $\bold{\Psi}$ with $\# \bold{\Psi} \geq 2$ and $\{a\} \notin \mathcal{A}_{1}$, for any $a \in A$ on the one hand and motivate the new concept of factorability of $\bold{\Psi}$ on the other hand.

\begin{enumerate}[label=(\alph*)]
\item Suppose $\bold{\Psi}$ is decomposable via $E$. Let $v_{1}= \# E \cap A$, $v_{2} = \# E' \cap A$.

Let $\bold{\Psi}_{E} = \left\{ \psi /E : \psi \in \bold{\Psi}\right\}$ and $\bold{\Psi}_{E'} = \left\{\psi/E': \psi \in \bold{\Psi}\right\}$.

We may set up labellings $\left\{\psi_{j}^{E} : 1 \leq j \leq j' \right\}$ and $\left\{\psi_{k}^{E'} : 1 \leq k \leq k'\right\}$ for $\bold{\Psi}_{E}$ and $\bold{\Psi}_{E'}$ respectively. Then $\max\left\{j', k'\right\} \geq 2$ simply because $\# \bold{\Psi} \geq 2$. 

\item For any bijective  maps $\psi_{1} : E \cap A \longrightarrow E \cap B$ and $\psi_{2} : E' \cap A \longrightarrow E' \cap B$,\\
we write $\psi_{1} \times \psi_{2}$ for the bijective map of $A$ to $B$ given by
\[
(\psi_{1} \times \psi_{2})~(a) =  
\begin{cases*}
\psi_{1}(a),~~ \text{if}~~ a\in E\tag{3.18} \\
\psi_{2}(a),~~ \text{if}~~a \in E'. 
\end{cases*}
\]

For instance, $\psi \in \bold{\Psi}$ can be written as $\psi_{E} \times \psi_{E'}$.

\item Let $\overline{\bold\Phi}_{1}$ and $\overline{\bold{\Phi}}_{2}$ be any two sets of bijective maps $\phi_{1}$ and $\phi_{2}$  on $E \cap A$ to $E \cap B$ and on $E' \cap A \rightarrow E' \cap B$  respectively. Then we denote $\{ \phi_{1} \times \phi_{2} : \phi_{1} \in \overline{\bold\Phi}_{1}, \phi_{2} \in \overline{\bold\Phi}_{2}\}$ by $\overline{\bold\Phi}_{1} \times \overline{\bold\Phi}_{2}$.

\item For the sake of convenience, we call $\psi_{1}$ and $\psi_{2}$ as in (b) \textbf{coverings of $\mathbf{E}$ and $\mathbf{E'}$} respectively. We note 
that

 \begin{equation} \tag{3.19}\label{eq-3.19}
 \bold{\Psi}=\bigcup_{ \psi \in \bold{\Psi}} \left(\{\psi/E\} \times \{\psi/E' \} \right) \subset \bold{\Psi}_{E} \times \bold{\Psi}_{E'} = \left\{\psi_{j}^{E} \times \psi_{k}^{E'} : 1 \leq j \leq j', 1 \leq k \leq k'\right\}.
\end{equation}

\item We may picture $\bold{\Psi}_{E} \times \bold{\Psi}_{E'}$ as a rectangular array $\mathcal{M}$ of dots, or, for that matter the $k' \times j'$ matrix $M$ with all entries 1. Then $\bold{\Psi}$ is a subset of $\mathcal{M}$ having at least one dot in each row and at least one dot in each column:
We call it  the \textbf{grid} of \(\boldsymbol{\Psi}\). The corresponding matrix for $\bold{\Psi}$ is a matrix of $0$s or $1$s with sum of each row and sum of each column bigger than or equal to one. 
\end{enumerate}

\begin{definition}\label{definition-3.2}
Let $v \geq 3$ and $\bold{\Psi}$, a set of coverings of $\Gamma_{n}$ with $\# \bold{\Psi} \geq 2$. Let $(E, E')$ be a bipartite cut. We define certain types for $\bold{\Psi}$. 
\end{definition}

\begin{enumerate}[label=(\roman*)]

\item $\bold{\Psi}$ is \textbf{factorable via} $E$ if for some sets $\overline{\boldsymbol{\phi}}_{1}$ and $\overline{\boldsymbol{\phi}}_{2}$ of coverings of $E$ and $E'$ respectively, $\bold{\Psi} = \overline{\boldsymbol{\phi}}_{1} \times \overline{\boldsymbol{\phi}}_{2}$. In this case $\bold{\Psi}$ is decomposable via $E$, $\overline{\boldsymbol{\phi}}_{1}= {\boldsymbol\Psi}_{E}$, $\overline{\boldsymbol{\phi}}_{2}= {\boldsymbol\Psi}_{E'}$, so that ${\boldsymbol\Psi}= {\boldsymbol\Psi}_{E} \times {\boldsymbol\Psi}_{E'}$ and $\# {\boldsymbol\Psi}=\# {\boldsymbol\Psi}_{E} \cdot \# {\boldsymbol\Psi}_{E'}$. These are indeed equivalent conditions in place of factorability and we may use the names \textbf{a grid of full-size} as well.

\item Let $\bold{\Psi}$ be decomposable via $E$. We say that $\bold{\Psi}$ is \textbf{flat} or \textbf{a pole via} $E$ if $\# \bold{\Psi}_{E'} =1$ or $\# \bold{\Psi}_{E}=1$ respectively. In both these cases, $\bold{\Psi} = \bold{\Psi}_{E} \times \bold{\Psi}_{E'}$, so that $\bold{\Psi}$ is factorable. This does happen if $\#E'=2$ or $\# E=2$ respectively.

\item Let  $\bold{\Psi}$ be decomposable via $E$. $\bold{\Psi}$ is \textbf{steep} or \textbf{diagonal via }  $E$ if $\psi_{E}$'s are all distinct and $\psi_{E'}$'s are all distinct as $\psi$ varies in $\bold{\Psi}$, in other words, $j'=k' = \# \psi$, or equivalently, $\psi \longrightarrow \psi/E$ and $\psi \longrightarrow \psi/E'$ are injective maps on $\bold{\Psi}$ to the sets of coverings of $E$ and $E'$ respectively.

\item $\bold{\Psi}$ is \textbf{hilly via} $E$ if $\bold{\Psi}$ is decomposable via $E$ but neither factorable nor steep via $E$.

\item $\bold{\Psi}$ is \textbf{factorable} \textbf{(flat, a pole)} if it is factorable (flat, a pole) via $E$ for some $\phi \neq E \subset_{\neq} \Gamma_{n}$.

\end{enumerate}

\begin{example}\label{example-3.2}
The pictures in Figs.~4-8 serve as examples for the concepts defined above.

\begin{figure}[h]
\centering
\includegraphics[scale=1]{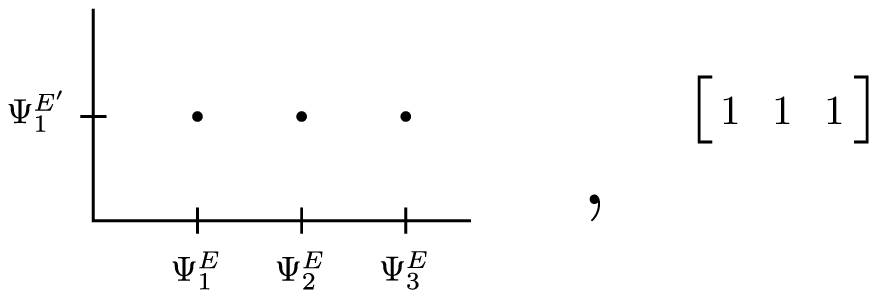}
\caption{Flat.}\label{fig04}
\end{figure}


\begin{figure}[h]
\centering
\includegraphics[scale=.80]{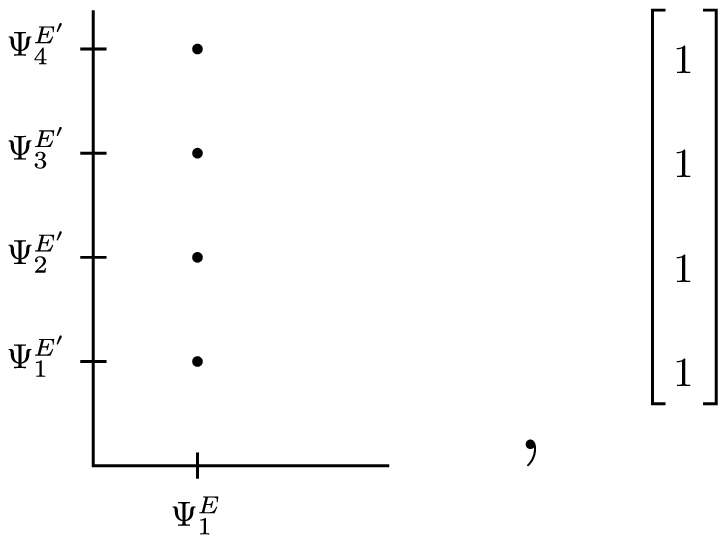}
\caption{ Pole.}\label{fig05}
\end{figure}

\begin{figure}[h]
\centering
\includegraphics[scale=1]{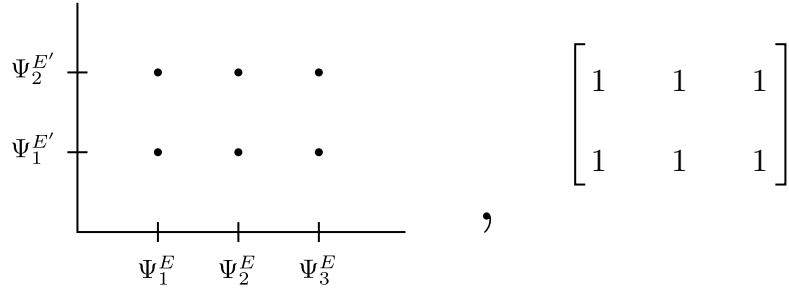}
\caption{ Grid of full size, factorable.}\label{fig06}
\end{figure}


\begin{figure}[h]
\centering
\includegraphics[scale=1]{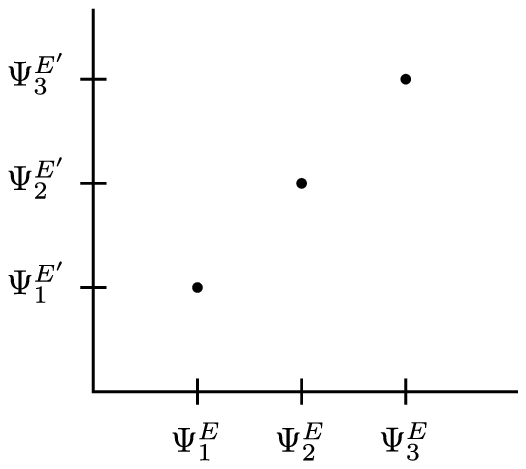}
\caption{Steep.}\label{fig07}
\end{figure}

\begin{figure}[h]
\centering
\includegraphics[scale=1]{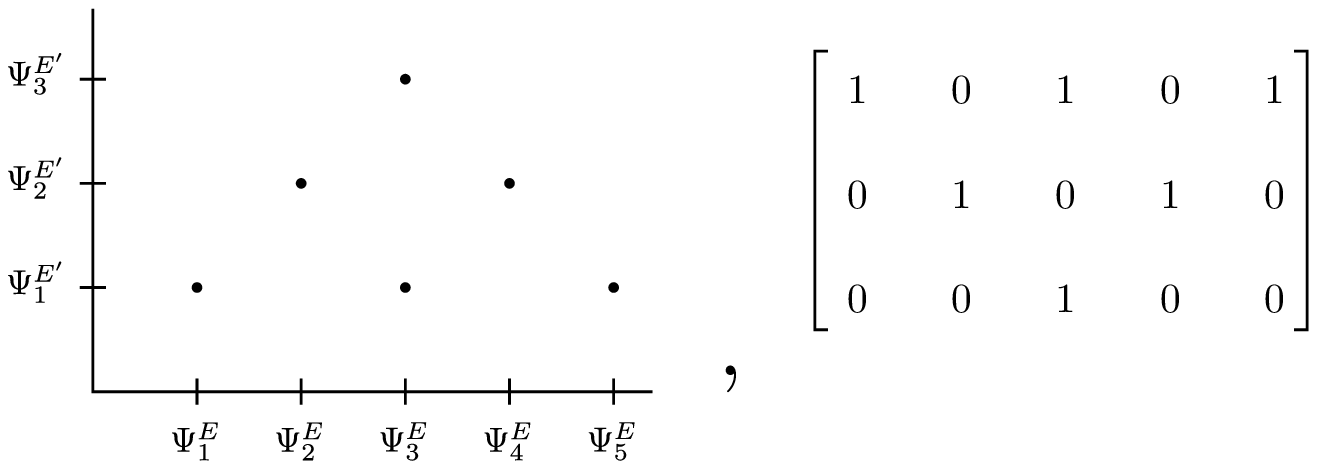}
\caption{ Hilly.}\label{fig08}
\end{figure}
\end{example}

\begin{thm}\label{thm-3.3}
If $\bold{\Psi}$ is factorizable via $E$,  then $| \bold{\Psi} \rangle$ is a product vector in the bipartite cut $(E, E')$. In particular, it is so if $\bold{\Psi}$ is flat or a pole via $E$.
\end{thm}

{\it Proof.}
It is immediate from definition that 
\begin{equation}
| \bold{\Psi} \rangle = \frac{1}{\# \bold{\Psi}} \sum_{\substack{1 \leq j \leq j'},\\ 1 \leq k \leq k'} | \psi^{E}_{j} \rangle \bigotimes | \psi_{k}^{E'} \rangle = \frac{1}{\# \bold{\Psi}}\left(\sum_{j=1}^{j'} | \psi_{j}^{E} \rangle\right) \bigotimes \left(\sum_{k=1}^{k'} | \psi_{k}^{E'} \rangle\right).
\tag{3.20}\label{eq-3.20}
\end{equation}
Hence $| \bold{\Psi}\rangle $ is a product vector in the cut $(E, E')$.



\vspace{0.5cm}
\textbf{(6) Entanglement properties of $\mathbf{| \bold{\Psi} \rangle}$ via polynomial representation.} 

In view of Theorems~\ref{thm-3.1}, \ref{thm-3.3} and intermediary discussions,  we can confine our attention to the case when $v \geq 4$, $\bold{\Psi}$ is decomposable via some  $E$ with $\phi \neq E \neq  \Gamma_{n}$ and $ \bold{\Psi}$ is not factorable. We will do a little reparation here in connection with polynomial representation of $| \bold{\Psi} \rangle$ to streamline the proof of our next Theorem. 

\begin{enumerate}[label=(\alph*)]

\item The polynomial representation of the unit vectors $| \psi_{1} \rangle$, $| \psi_{2} \rangle$, $| \psi_{1} \times \psi_{2} \rangle$  in $\mathcal{H}_{E}$, $\mathcal{H}_{E'}$, $\mathcal{H}$ respectively are given by 
\begin{align*}
F_{\psi_{1}} (\boldsymbol{\x}_{E}) & =  2^{-v_{1}/2} \prod_{a\in E \cap A}(\x_{\psi_{1}(a)}- \x_{a}),~~\\
 \quad F_{\psi_{2}} (\boldsymbol{\x}_{E'})&= 2^{-v_{2}/2} \prod_{a \in E' \cap A}(\x_{\psi_{2}(a)}-\x_{a}),\\
F_{\psi_{1} \times \psi_{2}} (\boldsymbol{\x}) &= 2^{-v/2} \prod_{a \in A} (\x_{\psi_{1} \times \psi_{2}(a)} - \x_{a}).~\\
\text{So} \, \, F_{\psi_{1} \times \psi_{2}} (\boldsymbol{\x}) &= F_{\psi_{1}} (\boldsymbol{\x}_{E})~~ F_{\psi_{2}}(\boldsymbol{\x}_{E'}).
\tag{3.21}
\end{align*}
This justifies the rotation $\psi_{1} \times \psi_{2}$ set up before. 

\item For $1 \leq j \leq j'$, let $T_{j} = \left\{k : 1 \leq k \leq k', \psi_{j}^{E} \times \psi_{k}^{E'} \in \bold{\Psi}\right\}$, and for $1 \leq k \leq k'$, let $S_{k} = \left\{j : 1 \leq j \leq j', \psi_{j}^{E} \times \psi_{k}^{E'} \in  \bold{\Psi}\right\}$. Then all these sets are non-empty  and $\bold{\Psi} = \underset{j=1}{\overset{j'}{\bigcup}} ~~ \underset{k \in T_{j}}{\overset{}\bigcup} \left\{\psi_{j}^{E} \times \psi_{k}^{E'}\right\} = \underset{k=1}{\overset{k'}{\bigcup}}~~ \underset{j\in S_{k}}{\overset{}{\bigcup}}\left\{\psi_{j}^{E} \times \psi_{k}^{E'} \right\}$.

The unions are all disjoint and $s = \# \bold{\Psi} = \underset{j=1}{\overset{j'}{\sum}} \# T_{j} = \underset{k=1}{\overset{k'}{\sum}} \# S_{k}$.

We may express concepts in Definition~\ref{definition-3.2} in terms of $\# T_{j}$ , $\# S_{k}$ for $1 \leq j \leq j',\, 1\leq k \leq k'.$ For instance, $\bold{\Psi}$ is factorable if and only if $\# T_{j} = k'$ for $1 \leq j \leq j'$, if and only if $\# S_{k} = j'$ for $1 \leq k \leq k'$, if and only if $ s = \# \bold{\Psi} = j' k'$.  

\item For $\psi \in {\boldsymbol{\Psi}}$, let $p_{\psi} ({\boldsymbol{\x}}_{E}) = p_{\psi_{E}}({\boldsymbol{\x}_{E}})= 2^{\nu_{1}/2}F_{\psi/E}({\boldsymbol{\x}}_{E})= \prod\limits_{\substack{a\in E \cap A}}(\x_{\psi(a)}- \x_{a})$ and $q_{\psi}({\boldsymbol{\x}}_{E'}) = q_{\psi/E'}(\boldsymbol{\x}_{E'}) = 2^{v_{2}/2} F_{\psi/E'} ({\boldsymbol{\x}}_{E'}) = \prod\limits_{\substack{a \in E' \cap A}} (\x_{\psi(a)} - \x_{a})$.

Set 
\begin{equation}
p^{0}_{\psi}(\boldsymbol{\x}_{E}) = p^{0}_{\psi/E}(\boldsymbol{\x}_{E}) = \sum_{\phi \neq A_{1} \subset_{\neq} E \cap A}(-1)^{\# A_{1}} {\boldsymbol{\x}^{A_{1}}} {\boldsymbol{\x}^{E \cap B \smallsetminus \psi(A_{1})}},
\tag{3.22}\label{eq-3.22}
\end{equation}
and 
\begin{equation}
q_{\psi}^{0} (\boldsymbol{\x}_{E'}) = q^{0}_{\psi/E'}(\boldsymbol{\x}_{E'})= \sum_{\phi \neq A_{2} \subset_{\neq} E' \cap A}(-1)^{\# A_{2}} {\boldsymbol{\x}^{A_{2}}} {\boldsymbol{\x}^{E' \cap B \smallsetminus \psi(A_{2})}}.
\tag{3.23}\label{eq-3.23}
\end{equation}
Then
\begin{equation}
p_{\psi}(\boldsymbol{\x}_{E}) = \boldsymbol{\x}^{E \cap B} + (-1)^{v_{1}} \boldsymbol{\x}^{E \cap A} + p^{0}_{\psi}(\boldsymbol{\x}_{E}),
\tag{3.24}
\end{equation}
and
\begin{equation}
q_{\psi}(\boldsymbol{\x}_{E'}) = \boldsymbol{\x}^{E' \cap B} + (-1)^{v_{2}} \boldsymbol{\x}^{E' \cap A} + q^{0}_{\psi}(\boldsymbol{\x}_{E'}).
\tag{3.25}
\end{equation}

\item Let $s= \# {\bold{\Psi}}$. We set $| \widetilde{\bold{\Psi}} \rangle = 2^{v/2} \# \bold{\Psi} |\bold{\Psi} \rangle$. Its entanglement properties are same as that for $| \bold{\Psi} \rangle$.

Its polynomial representation has coefficients as integers and that is convenient to work with.
Indeed,
\begin{align*}
\tilde{F}_{\bold{\Psi}}(\boldsymbol{\x}) &= \sum_{\psi \in \bold{\Psi}} p_{\psi} (\boldsymbol{\x}_{E})~~ q_{\psi} (\boldsymbol{\x}_{E'})\\
&= \sum_{\psi \in \bold{\Psi}} \Bigg((\boldsymbol{\x}^{B}) + (-1)^{v}~~\boldsymbol{\x}^{A} + (-1)^{v_{1}}~~ \boldsymbol{\x}^{E \cap A} ~~\boldsymbol{\x}^{E'\cap B} + (-1)^{v_{2}}~~ \boldsymbol{\x}^{E \cap B}~~ \boldsymbol{\x}^{E' \cap A})\\ 
 &  + \left(\boldsymbol{\x}^{E \cap B} + (-1)^{v_{1}}~~ \boldsymbol{\x}^{E \cap A}\right)~~q_{\psi}^{0} (\boldsymbol{\x}_{E'})  + p_{\psi}^{0}(\boldsymbol{\x}_{E})~~ \left(\boldsymbol{\x}^{E' \cap B} + (-1)^{v_{2}}~~ \boldsymbol{\x}^{E' \cap A} \right)\\
 & + p_{\psi}^{0}~~ (\boldsymbol{\x}_{E})~~ q_{\psi}^{0}(\boldsymbol{\x}_{E'}) \Bigg)\\
 & = \# \bold{\Psi} \left(\boldsymbol{\x}^{B} + (-1)^{v} \boldsymbol{\x}^{A} + (-1)^{v_{1}} \boldsymbol{\x}^{E \cap A} \boldsymbol{\x}^{E' \cap B} + (-1)^{v_{2}} \boldsymbol{\x}^{E \cap B} \boldsymbol{\x}^{E' \cap A} \right)\\
 &\quad + \left(\boldsymbol{\x}^{E \cap B} + (-1)^{v_{1}} \boldsymbol{\x}^{E \cap A}\right) \sum_{\psi \in \bold{\Psi}} q_{\psi}^{0}(\boldsymbol{\x}_{E'}) + \sum_{\psi \in \bold{\Psi}} p_{\psi}^{0}(\boldsymbol{\x}_{E}) 
 \left(\boldsymbol{\x}^{E' \cap B} + (-1)^{v_{2}} 
 \boldsymbol{\x}^{E' \cap A} \right)\\
 & \quad + \sum_{\psi \in \bold{\Psi}} p_{\psi}^{0} (\boldsymbol{\x}_{E}) q^{0}_{\psi} (\boldsymbol{\x}_{E'}).
 \tag{3.26}\label{eq-3.26}
\end{align*}

\item Consider homogeneous polynomials $p(\boldsymbol{\x}_{E})$ and $q(\boldsymbol{\x}_{E'})$ of degrees $v_{1}$ and $v_{2}$ respectively. They have the form
$p(\boldsymbol{\x}_{E}) = \delta \boldsymbol{\x}^{E \cap A} + \beta\boldsymbol{\x}^{E \cap B} + p_{0}(\boldsymbol{\x}_{E})$ with
\begin{align*}
p_{0}(\boldsymbol{\x_{E}}) &= \sum\left\{\delta_{A_{1}, B_{1}} \boldsymbol{\x}^{A_{1}} \boldsymbol{\x}^{E \cap B \smallsetminus B_{1}} : \phi \neq A_{1} \subsetneq  E \cap A, \phi \neq B_{1} \subsetneq E \cap B,\right.\\
&\left.\hspace{4cm} \# A_{1}= \# B_{1}\right\},\tag{3.27}\label{eq-3.27} 
\end{align*}
and $q(\boldsymbol{\x}_{E'}) = \delta' \boldsymbol{\x}^{E' \cap A} + \beta' \boldsymbol{\x}^{E' \cap B} + q_{0}(\boldsymbol{\x}_{E'})$ with
\begin{align*}
q_{0}(\boldsymbol{\x}_{E'}) &= \sum \left\{\delta'_{A_{2}, B_{2}} \boldsymbol{\x}^{A_{2}} \boldsymbol{\x}^{E' \cap B \smallsetminus B_{2}} : \phi \neq A_{2 } \subsetneq E' \cap A, \phi \neq B_{2} \subsetneq  E' \cap B,\right.\\ 
 &\left.\hspace{4cm}\qquad\qquad\# A_{2} = \# B_{2}\right\},\tag{3.28}
\end{align*}
where $\delta, \beta, \delta',  \beta', \delta_{A_1, B_1}, \delta_{A_{2}, B_{2}}'$ are all scalars.
Hence
\begin{align*}
p(\boldsymbol{\x}_{E}) q (\boldsymbol{\x}_{E'}) &= \delta \delta' \boldsymbol{\x}^{A} + \beta \beta' \boldsymbol{\x}^{B} + \delta \beta' \boldsymbol{\x^{E \cap A}} \boldsymbol{\x}^{E' \cap B} + \delta'\beta \x^{E' \cap A} \boldsymbol{\x}^{E \cap B} +\\
&\quad(\delta \boldsymbol{\x}^{E \cap A} + \beta \boldsymbol{\x}^{E \cap B}) q_{0}(\boldsymbol{\x}_{E'}) + p_{0}(\x_{E}) (\delta' \boldsymbol{\x}^{E' \cap A} +\beta' \boldsymbol{\x}^{E' \cap B}) +\\
&\quad p_{0}(\boldsymbol{\x}_{E}) q_{0}(\boldsymbol{\x}_{E'}).
\tag{3.29}
\label{eq-3.29}
\end{align*}
 
\item We compare \eqref{eq-3.26} in (d) and \eqref{eq-3.29} in (e) above. 

We obtain that $\widetilde{F}_{\boldsymbol\Psi}(\boldsymbol{\x}) = p(\boldsymbol{\x}_{E}) q(\boldsymbol{\x}_{E'})$ if and only if
\begin{align*}
\beta \beta' &= s = (-1)^{v} \delta \delta' = (-1)^{v_{1}} \delta \beta' = (-1)^{v_{2}} \delta' \beta,
\tag{3.30}\label{eq-3.30}\\
\beta q_{0}(\boldsymbol{\x}_{E'}) &= \sum_{\psi \in \boldsymbol{\Psi}} q_{\psi}^{0}(\boldsymbol{\x}_{E'}) = (-1)^{v_{1}} \delta q_{0} (\boldsymbol{\x}_{E'}),
\tag{3.31}\label{eq-3.31}\\
\beta' p_{0}(\boldsymbol{\x}_{E}) &= \sum_{\psi \in \boldsymbol{\Psi}} p^{0}_{\psi} (\boldsymbol{\x}_{E}) = (-1)^{v_{2}} \delta' p_{0} (\boldsymbol{\x}_E), 
\tag{3.32}\label{eq-3.32}\\
 \text{and}\quad p_{0}(\boldsymbol{\x}_{E}) q_{0}(\boldsymbol{\x}_{E'}) &= \sum_{\psi \in \boldsymbol{\Psi}} p^{0}_{\psi} (\boldsymbol{\x}_{E}) q^{0}_{\psi} (\boldsymbol{\x}_{E'}). \tag{3.33}\label{eq-3.33}
\end{align*}
 
 \item Let us first consider the case when this does happen. By (3.30), $\delta = (-1)^{v_{1}} \beta$, $\delta' = (-1)^{v_{2}} \beta'$ and $\delta, \beta, \delta', \beta'$ are all non-zero.
 
 Therefore,  we may take $\beta = s$, $\delta = (-1)^{v_{1}} s$, $\beta' = 1$, $\delta' = (-1)^{v_{2}}$ and multiply $p_{0}(\boldsymbol{\x}_{E})$ and $q_{0}(\boldsymbol{\x}_{E'})$ by suitable scalars if the need be and retain the same notation for them. Then~\eqref{eq-3.31} and \eqref{eq-3.32} give $s q_{0}(\boldsymbol{\x}_{E'}) = \sum\limits_{\psi \in \boldsymbol{\Psi}} q_{\psi}^{0} (\boldsymbol{\x}_{E'})$ and $p_{0}(\boldsymbol{\x}_{E}) = \sum\limits_{\psi \in \boldsymbol{\Psi}} p^{0}_{\psi}(\boldsymbol{\x}_{E})$. This turns~\eqref{eq-3.33} into
 \begin{equation}
 \left(\sum_{\psi \in \boldsymbol{\Psi}} p^{0}_{\psi}(\boldsymbol{\x}_{E})\right) \left(\sum_{\psi \in \boldsymbol{\Psi}} q^{0}_{\psi} (\boldsymbol{\x}_{E'})\right) = s \sum_{\psi \in \boldsymbol{\Psi}} p^{0}_{\psi} (\boldsymbol{\x}_{E}) q^{0}_{\psi}(\boldsymbol{\x}_{E'}).\tag{3.34}\label{eq-3.34} 
  \end{equation}

For notational convenience, we write $p_{\psi_j^{E}}^{0} (\boldsymbol{\x}_{E})$ as $p^{0}_{j}(\boldsymbol{\x}_{E})$ and $q^{0}_{\psi_{k}^{E'}} (\boldsymbol{\x}_{E'})$ as $q_{k}^{0}(\boldsymbol{\x}_{E'})$ for $1\leq j \leq j'$, $1\leq k \leq k'$. 
We expand both sides of \eqref{eq-3.34} using (b) above in parts.

\begin{align*}
\text{L.H.S}~~\text{of}~~{\eqref{eq-3.34}} &= \left(\sum_{j=1}^{j'}  \# T_{j} p_{j}^{0}(\boldsymbol{\x}_{E})\right)\left(\sum_{k=1}^{k'} \#S_{k}~ q_{k}^{0}(\boldsymbol{\x}_{E'})\right)\\
&= \sum_{j=1}^{j'} \sum_{k=1}^{k'} \# T_{j} \#S_{k} p_{j}^{0}(\boldsymbol{\x}_{E}) q_{k}^{0}(\boldsymbol{\x}_{E'}).\tag{3.35}\label{eq-3.35}
\end{align*}

Now by (b) above, $s= \sum_{k=1}^{k'} \# S_{k} = \sum_{j=1}^{j'} \# T_{j}$.

R.H.S of \eqref{eq-3.34}
\begin{equation}
= s\sum_{j=1}^{j'} p_{j}^{0} (\boldsymbol{\x}_{E}) \left(\sum_{l \in T_{j}} q_{l}^{0}(\boldsymbol{\x}_{E'})\right) = s\sum_{k=1}^{k'}  \left(\sum_{t \in S_{k}}  p_{t}^{0} (\boldsymbol{\x}_{E})\right) q_{k}^{0} (\boldsymbol{\x}_{E'}).\tag{3.36}\label{eq-3.36}
\end{equation}

Equating L.H.S  and R.H.S of~\eqref{eq-3.34}, we obtain
\begin{equation}
\sum_{j=1}^{j'} p_{j}^{0}(\boldsymbol{\x}_{E}) \left(\sum_{l\in T_{j}} (\# T_{j} \# S_{l} -s) q_{l}^{0}(\boldsymbol{\x}_{E'}) + \sum_{\substack{1 \leq k \leq k' \\ k \notin T_{j}}} \# T_{j} \# S_{k} q_{k}^{0} (\boldsymbol{\x}_{E'}) \right) = 0, \tag{3.37}\label{eq-3.37}
\end{equation}
and
\begin{equation}
\sum_{k=1}^{k'} \left(\sum_{t \in S_{k}} (\# T_{t} \# S_{k} -s) p_{t}^{0}(\boldsymbol{\x}_{E}) +  \sum_{\substack{1 \leq j \leq j' \\ j \notin S_{k}}} \# T_{j} \# S_{k}  p_{j}^{0}(\boldsymbol{\x}_{E})\right) q_{k}^{0}(\boldsymbol{\x}_{E'}) = 0.
\tag{3.38}\label{eq-3.38}
\end{equation}

\item Arguments in (g) above can be reversed and, therefore, we call \eqref{eq-3.34}, \eqref{eq-3.37}, \eqref{eq-3.38} as Master Equations for $\widetilde{F}_{\boldsymbol{\Psi}}(\boldsymbol{\x})$ to be expressible as product of some polynomials $p(\boldsymbol{\x}_{E})$ and $q(\boldsymbol{\x}_{E'})$. We emphasize that the Master Equations do not involve $p(\boldsymbol{\x}_{E})$ or $q(\boldsymbol{\x}_{E'})$ and, in  fact, they are determined in an explicit way as explained in (g) above. Moreover, they are equivalent conditions for $| \bold{\Psi} \rangle$ to be a product vector in the bipartite cut $(E, E')$.
\end{enumerate}


We give easy consequences of the Master equations before going to generalities in the spirit of the converse or partial converses of Theorem~\ref{thm-3.3}.

\begin{thm}\label{thm-3.4}
\textbf{Dichotomy}. Consider the following two conditions for $\boldsymbol{\Psi}$ with $\# \boldsymbol{\Psi} \geq 2$.
\begin{enumerate}[label=(\alph*)]

\item Either $\boldsymbol{\Psi}$ is flat or a pole.

\item $| \boldsymbol{\Psi} \rangle$ is genuinely entangled.

\end{enumerate}

If $s=\# \boldsymbol{\Psi}$ is a prime number, one and only one of (a) and (b) holds.

\end{thm}

{\it Proof.}
If (a) holds then by Theorem~\ref{thm-3.3}, $| \boldsymbol{\Psi} \rangle$ is not genuinely entangled. So at most one of (a) or (b) can hold.

Now consider the case when (a) does not hold. Let, if possible, (b) not hold. Then $|\boldsymbol{\Psi} \rangle$ is a product vector in some bipartite cut $(E, E')$. By Theorem~\ref{thm-3.1} (ii), and its proof, $\boldsymbol{\Psi}$ is decomposable via $E$ and $F_{\boldsymbol{\Psi}}(\boldsymbol{\x}) = p(\boldsymbol{\x}_{E}) q (\boldsymbol{\x}_{E'})$ for some polynomials $p$ and $q$.

By Master Eq.~\eqref{eq-3.34}, we have
\begin{equation}
\left(\sum_{\psi \in \boldsymbol{\Psi}} p_{\psi}^{0}(\boldsymbol{\x}_{E})\right) \left(\sum_{\psi \in \boldsymbol{\Psi}} q_{\psi}^{0} (\boldsymbol{\x}_{E'}) \right) = s \sum_{\psi \in \boldsymbol{\Psi}} p_{\psi}^{0}(\boldsymbol{\x}_{E}) q_{\psi}^{0}(\boldsymbol{\x}_{E'}).\tag{ME}\label{eq-ME}
\end{equation}
 
Consider any $a \in E \cap A$, $a' \in E' \cap A$ and any $\phi \in \boldsymbol{\Psi}$. Put $b = \phi(a)$, $b' = \phi(a')$. Let $U = \left\{ \psi \in \boldsymbol{\Psi} : \psi(a) = b \right\}$, $V= \left\{\psi \in \boldsymbol{\Psi} : \psi(a')= b' \right\}$, $W = \left\{\psi \in \boldsymbol{\Psi} : \psi(a) = b, \psi(a') = b'\right\}$,  $u =\# U$, $ v = \# V$, $w = \# W$. Then $\phi \in W \subset U \cap V$. So $1 \leq w \leq \min \{u, v \}$. Since $\boldsymbol{\Psi}$ is neither flat nor a pole, we have $V \subsetneq \boldsymbol{\Psi}$, $U \subset_{\ne} \boldsymbol{\Psi}$. So $u < s$, $v < s$.
 
The co-efficient of $\boldsymbol{\x}^{\{a\}}$ $\boldsymbol{\x}^{E \cap B \smallsetminus \{b\}}$ in $\sum_{\psi \in \boldsymbol{\Psi}} p_{\psi}^{0}(\boldsymbol{\x}_{E})$ is -u and that of $\boldsymbol{\x}^{\{a'\}}$ $\boldsymbol{\x}^{E' \cap B \smallsetminus \{b'\}}$ in $\sum_{\psi \in \boldsymbol{\Psi}} q_{\psi}^{0}(\x_{E'})$ is $-v$. So the co-efficient of $\boldsymbol{\x}^{\{a, a'\}}$ $\boldsymbol{\x}^{B \smallsetminus \{b, b'\}}$ in the L.H.S of \eqref{eq-ME} is $uv$. On the other hand, the coefficient of $\boldsymbol{\x}^{\{a, a'\}}$ $\boldsymbol{\x}^{B \smallsetminus \{b, b'\}}$ on  The R.H.S of \eqref{eq-ME} is $sw$. So we have the equation 
\begin{equation}
uv=sw \tag{3.39}\label{eq-3.39}
\end{equation}

Now $1 \leq u < s$ and $1 \leq v < s$. So $s$ cannot be a factor of $u$ or $v$. So if $s$ is a prime, then $s$ cannot be a factor of $uv$. This contradicts~\eqref{eq-3.39}. Hence $| \boldsymbol{\Psi} \rangle$ is genuinely entangled, i.e., (b) holds.
 

\begin{thm}\label{thm-3.5}
Let $\boldsymbol{\Psi}$ be decomposable via $E$ and $\{p_{j}^{0}(\boldsymbol{\x}_{E}) : 1 \leq j \leq j'\}$ and $\{q_{k}^{0} (\boldsymbol{\x}_{E'}) : 1 \leq k \leq k'\}$ etc. as above. 
Suppose that both these systems are linearly independent. If $| \boldsymbol{\Psi} \rangle$ is a product vector in the bipartite cut $(E, E')$, $\boldsymbol{\Psi}$ is factorable via $E$.
\end{thm}

{\it Proof.}
By the Master Equation \eqref{eq-3.37}, the linear independence of $\left\{p_{j}^{0} (\boldsymbol{\x}_{E}) : 1 \leq j \leq j'\right\}$ forces the following. For $1 \leq j \leq j'$,  
\begin{equation}
\sum\limits_{l\in T_{j}} (\# T_{j} \# S_{l}-s) q_{l}^{0}(\boldsymbol{\x}_{E'}) + \sum_{\substack{1 \leq k \leq k'\\ k \notin T_{j}}} \# T_{j} \# S_{k} q_{k}^{0}(\boldsymbol{\x}_{E'}) = 0.\tag{3.40}\label{eq-3.40} 
\end{equation}

However,  $\left\{q_{k}^{0}(\boldsymbol{\x}_{E}): 1 \leq k \leq k' \right\}$ is linearly independent and $\# T_{j} \neq 0 \neq \# S_{k}$ for $1 \leq j \leq j'$, $1 \leq k \leq k'$. So $T_{j} = \left\{k : 1 \leq k \leq k'\right\}$ for $1 \leq j \leq j'$.
This gives that $\boldsymbol{\Psi}$ is factorable via $E$.

\begin{example}\label{example-3.3}
We give instance of $\boldsymbol{\Psi}$ for which the conditions of linear independence as in Theorem~\ref{thm-3.5} above hold. 

(a) Let $\nu=5$, $E = \{1,2,3,4,5,6\}$, $E' = \{7,8,9,10\}$.

Let $\boldsymbol{\Psi}$ consist of the four coverings $\psi (j), j = 1,2,3,4$ as in Fig.~\ref{fig09}.

\begin{figure}[h]
\centering
\includegraphics[scale=1.1]{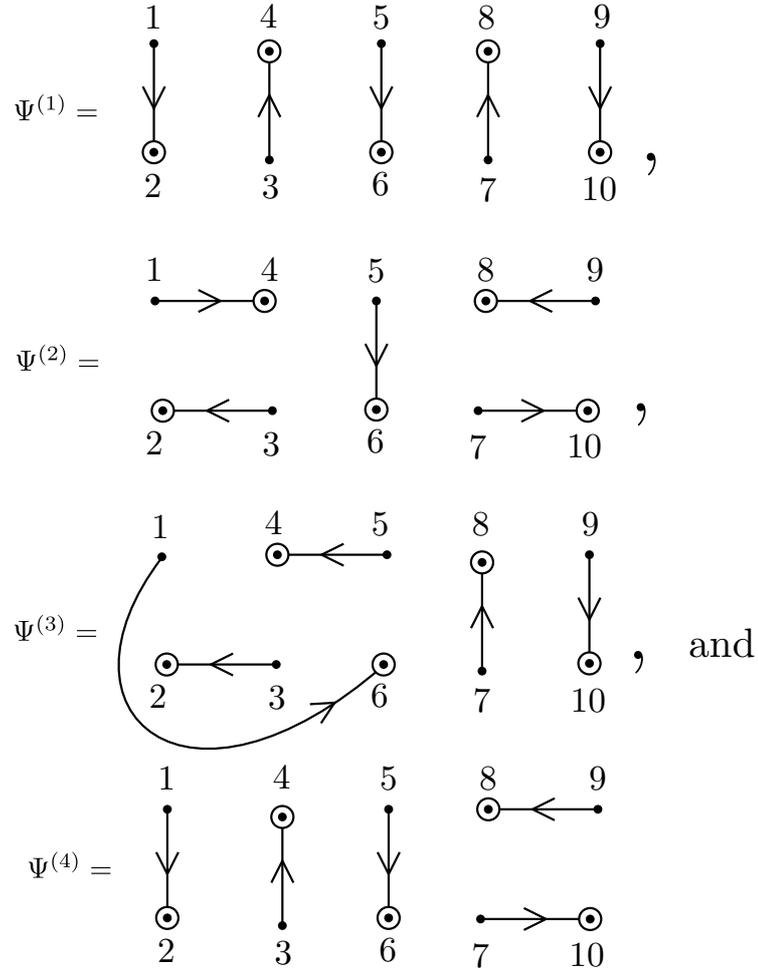}
\caption{Four coverings of ten sites. It includes a covering which is not nearest neighbour.  }\label{fig09}
\end{figure}

(b) Then 
\begin{align*}
\boldsymbol{\Psi}_{E} &= \left\{\psi^{(1)}/E , \psi^{(2)}/E, \psi^{(3)}/E\right\},\\
\boldsymbol{\Psi}_{E'} & = \left\{\psi^{(1)}/E' , \psi^{(2)}/E'\right\}.
\end{align*}

The grid is given by Fig.~\ref{fig10} which shows that $\boldsymbol{\Psi}$ is not factorable. Indeed $j' =3$, $k'=2$ whereas $s = 4$.

\begin{figure}[h]
\centering
\includegraphics[scale=1.1]{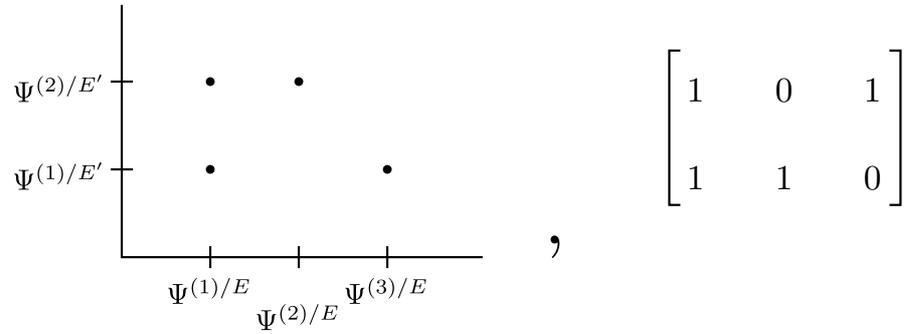}
\caption{Grid representation of coverings in Fig. \ref{fig09}.}\label{fig10}
\end{figure}


 (c) For tuples $\boldsymbol{\lambda} = (\lambda_{1}, \lambda_{2}, \lambda_{3})$, $\mu =(\mu_{1}, \mu_{2})$ of scalars, consider the polynomials $p^{\boldsymbol{\lambda}}({\boldsymbol{\x}_{E}}) = \sum\limits_{j=1}^{3} \lambda_{j} \cdot p_{j}^{0} (\boldsymbol{\x}_{E})$ and $q^{\mu}(\boldsymbol{\x}_{E'}) = \sum\limits_{k =1}^{2} \mu_{k} q_{k}^{0} (\boldsymbol{\x}_{E'})$. Then the coefficients of $\x_{1}~ \x_{4}~\x_{6} = \boldsymbol{\x}^{\{1\}} \boldsymbol{\x}^{\{4,6\}}$, $\x_{1}~ \x_{2} ~\x_{6} = \x^{\{1\}} \boldsymbol{\x}^{\{2,6\}}$, $\x_{1} \x_{2} \x_{4} = \x^{\{1\}} \boldsymbol{\x}^{\{2,4\}}$ in $p^{\boldsymbol{\lambda}}(\boldsymbol{\x}_{E})$ are $\lambda_{1}, \lambda_{2}, \lambda_{3}$ respectively. So $p^{\boldsymbol{\lambda}}(\boldsymbol{\x}_{E}) =0$ if and only if $\lambda_{1} = 0 = \lambda_{2} = \lambda_{3}$. Hence $\{p_{j}^{0}(\boldsymbol{\x}_{E}), j = 1, 2, 3\}$  is linearly independent. On the other hand, the coefficients of $\x_{7} \x_{10} = \boldsymbol{\x}^{\{7\}} \boldsymbol{\x}^{\{10\}}$ and $\x_{7} \x_{8} = \boldsymbol{\x}^{\{7\}} \boldsymbol{\x}^{\{8\}}$ in $q^{\boldsymbol{\mu}}(\boldsymbol{\x}_{E'})$ are $\mu_{1}$ and $\mu_{2}$ respectively. So $q^{\boldsymbol{\mu}} (\boldsymbol{\x}_{E'})$ is $0$ if and only if $\mu_{1} =0 =\mu_{2}$. Hence $\{q_{k}^{0}(\boldsymbol{\x}_{E'}): k = 1,2\}$ is linearly independent. 
 
  (d) Applying Theorem III.5 above, we obtain  that $| \boldsymbol{\Psi} \rangle$ is not a product vector in the cut $(E,E')$.
 
 But $\boldsymbol{\Psi}$ is not decomposable via any other bipartite cut. So $| \boldsymbol{\Psi} \rangle$ is genuinely entangled.
 \end{example}
 This motivates our next definition and result.
 
\begin{definition}\label{definition-3.3}
Let $\boldsymbol{\Psi}$ be a set of coverings of $\Gamma_{n}$ with $\# \boldsymbol{\Psi} \geq 2$.
\end{definition}

\begin{enumerate}[label = (\roman*)]
 \item Suppose that $\boldsymbol{\Psi}$ is decomposable via some $E$ with $\phi  \neq E \subset_{\neq} \Gamma_{n}$. $\boldsymbol{\Psi}$ is  said to be \textbf{independent via} $E$ if $ \boldsymbol{\Psi}_{E} = \left\{\psi_{j}^{E} : 1 \leq j \leq j' \right\}$ and $\boldsymbol{\Psi}_{E'} = \left\{\psi_{k}^{E'} : 1 \leq k \leq k' \right\}$  satisfy the following conditions.
\begin{enumerate}[label = (\alph*)]
\item For each $j, 1 \leq j \leq j'$ there exist $\phi \neq A^{E}_{j} \subset_{\neq} E \cap A$, $\phi \neq B_{j}^{E} \subset_{\neq} E \cap B$ such that $\left\{t : 1 \leq t \leq j', \psi_{t}^{E} (A_{j}^{E}) = B_{j}^{E}\right\} = \left\{j\right\}$.

\item For each $k$, $1 \leq k \leq k'$, there exist $\phi \neq A_{k}^{E'} \subset_{\neq} E' \cap A$, $\phi\neq B_{k}^{E'} \subset_{\neq} E' \cap B $ such that $\left\{ u : 1 \leq u \leq k' : \psi_{u}^{E'} (A_{k}^{E'}) = B_{k}^{E'} \right\} = \{k\}$.

In other words, $\boldsymbol{\x}^{A_{j}^{E}}$ $\boldsymbol{\x}^{E \cap B \smallsetminus B^{E}_{j}}$ does occur in and only in $p_{j}^{0} (\boldsymbol{\x}_{E})$ and $\boldsymbol{\x}^{A_{k}^{E'}}$ $\boldsymbol{\x}^{E' \cap B \smallsetminus B_{k}^{E'}}$ does occur in and only in $q_{k}^{0}(\boldsymbol{\x}_{E'})$.
\end{enumerate}

\item Neither flat nor a pole $\boldsymbol{\Psi}$ can be called \textbf{independent} if it is independent via $E$ for any $E$ via which $\boldsymbol{\Psi}$  is decomposable. We note that if $\boldsymbol{\Psi}$ is not decomposable,  $\boldsymbol{\Psi}$ is independent by this definition. On the other hand, if $\boldsymbol{\Psi}$ is flat or a pole,  $\boldsymbol{\Psi}$ is not independent by this definition.

\end{enumerate}

\begin{thm}\label{thm-3.6}
Suppose that $\boldsymbol{\Psi}$ is independent. Then $| \boldsymbol{\Psi} \rangle$ is genuinely entangled if and only if $\boldsymbol{\Psi}$ is not factorable.
\end{thm}

{\it Proof.}
In view of Theorem~\ref{thm-3.3},  if $| \boldsymbol{\Psi} \rangle$ is genuinely entangled, $\boldsymbol{\Psi}$ is not factorable. Now suppose that $\boldsymbol{\Psi}$ is not factorable. Let, if possible, $| \boldsymbol{\Psi} \rangle$ be not genuinely entangled which implies that  $| \boldsymbol{\Psi} \rangle$ is a 
product vector in some bipartition, say, $(E, E')$.  By Theorem~\ref{thm-3.1} (ii), $\boldsymbol{\Psi}$ is decomposable via $E$. Since $\boldsymbol{\Psi}$ is independent,  we have that $\boldsymbol{\Psi}$ is independent via $E$. The arguments in Example~\ref{example-3.3} (c) above can be suitably modified to give that $\left\{p_{j}^{0}(\boldsymbol{\x}_{E}) : 1 \leq j \leq j' \right\}$ and $\left\{q_{k}^{0} (\boldsymbol{\x}_{E'})  : 1 \leq k \leq k'\right\}$ are both linearly independent. Indeed, let, if possible, $p^{\lambda}(\boldsymbol{\x}_{E}) = \sum\limits_{j=1}^{j'} \lambda_{j} p_{j}(\boldsymbol{\x}_{E}) = 0$ for some tuple $\boldsymbol{\lambda} = (\lambda_{j})^{j'}_{j=1}$ of scalars. For $1 \leq j \leq j', $ the coefficient of $\boldsymbol{\x}^{A_{j}^{E}}$ $\boldsymbol{\x}^{E \cap B \smallsetminus B_{j}^{E}}$ in $p^{\boldsymbol{\lambda}} (\boldsymbol{\x}_{E})$ is $(-1)^{\# {A_{j}^{E}}} \lambda_{j}$. So $\lambda_{j} = 0$ for $1 \leq j \leq j'$. Similarly, we can prove that $\left\{q_{k}^{0}(\boldsymbol{\x}_{E'}) : 1 \leq k \leq k' \right\}$ is linearly independent. By Theorem~\ref{thm-3.5}, $\boldsymbol{\Psi}$ is factorable via $E$, a contradiction.


\vspace{0.5cm}
\textbf{(7) Master equations.}

We continue our discussion 
with inputs from other parts of the paper in the spirit of studying quantum entanglement of $| \boldsymbol{\Psi} \rangle$ in terms of properties of $\boldsymbol{\Psi}$.

Let $\boldsymbol{\Psi}$ be a set of coverings of $\Gamma_{n}$ that is neither flat nor a pole.

We begin with the case that $\boldsymbol{\Psi}$ is decomposable via some $\phi \neq E \subset_{\neq} \Gamma_{n}$ and proceed with an attempt to characterize Master Equation~\eqref{eq-3.34}.

\begin{enumerate}[label=(\alph*)]

\item Terms on the L.H.S or R.H.S of \eqref{eq-3.34} can only be of the form $\boldsymbol{\x}^{A_{3}}~\boldsymbol{\x}^{B \smallsetminus B_{3}} = \boldsymbol{\x}^{A_{1} \cup A_{2}}~~\boldsymbol{\x}^{B\smallsetminus (B_{1} \cup B_{2})}$ with $\phi \neq A_{1} \subset_{\neq} E \cap A$, $\phi \neq A_{2} \subset_{\neq} E' \cap A$, $\phi \neq B_{1} \subset_{\neq} E \cap B$, $\phi \neq B_{2} \subset_{\neq} E' \cap B$, $\# A_{1} = \# B_{1}$, $\# A_{2} = \# B_{2}$. Indeed,such a term occurs in the L.H.S of~\eqref{eq-3.34} if and only if there exist $\phi_{1}, \phi_{2} \in \boldsymbol{\Psi}$ with  $\phi_{1}(A_{1})= B_{1}$ and $\phi_{2}(A_{2}) = B_{2}$, and on the other hand, it occurs in R.H.S of \eqref{eq-3.34} if and only if there exists $\phi_{3} \in \boldsymbol{\Psi}$ with $\phi_{3}(A_{1}) = B_{1}$, $\phi_{3}(A_{2}) = B_{2}$ or equivalently, $\phi_{3}(A_{3}) = B_{3}$.

Set $U = \left\{\psi \in \boldsymbol{\Psi} : \psi(A_{1}) = B_{1}\right\} = \left\{\psi \in \boldsymbol{\Psi} : \psi /E (A_{1}) = B_{1}\right\}$,

 $V= \left\{\psi \in \boldsymbol{\Psi} : \psi(A_{2}) = B_{2}\right\} = \left\{\psi \in \boldsymbol{\Psi} : \psi/E' (A_{2}) =B_{2} \right\}$, 
 
 $W = \left\{\psi \in \boldsymbol{\Psi} : \psi(A_{1}) =B_{1}, \psi(A_{2}) = B_{2}\right\} = \left\{\psi \in \bold{\Psi} : \psi (A_{3}) = B_{3}\right\}$, 
 
 $u = \# U$, $v=\# V$, $w=\# W$,
 
  $Z= \left\{\psi \in \boldsymbol{\Psi} : \psi (A_{1}) \neq B_{1}, \psi(A_{2}) \neq B_{2}\right\} = \boldsymbol{\Psi} \smallsetminus (U \cup V)$, $z=\# Z$.
  
  The co-efficient of $\boldsymbol{\x}^{A_{3}}~~ \boldsymbol{\x}^{B\smallsetminus B_{3}}$ in L.H.S. of \eqref{eq-3.34} is $(-1)^{\# A_{3}} uv$ whereas in R.H.S. of \eqref{eq-3.34}, it is $(-1)^{\# A_{3}} sw$. Therefore, the two are equal if and only if 
  \begin{equation}
  uv = sw.
  \tag{3.41}\label{eq-3.41}
  \end{equation}
  Indeed, $uv-sw =(u-w)(v-w)-zw$. Hence, \eqref{eq-3.41} is equivalent to
  \begin{equation}
  (u-w)(v-w)= zw.
  \tag{3.42}\label{eq-3.42}
  \end{equation}
  
\item  If $U$ or $V$ is $\phi$ ,  so is $W$ and therefore, $uv=0 =w$. So \eqref{eq-3.41} is satisfied.

If $U = \boldsymbol{\Psi}$, then $u=s$, $V=W$ and therefore, $v=w$. So \eqref{eq-3.41} is satisfied.

Similarly, we obtain that if $V = \boldsymbol{\Psi}$, then \eqref{eq-3.41} is satisfied.

\item Let $A_{j}$, $B_{j} (j=1,2,3)$, $U,V,W,Z$, $u,v,w,z$, be as in (a) above with $\phi \neq U \neq \boldsymbol{\Psi}$, $\phi \neq V \neq \boldsymbol{\Psi}$, i.e., with $0 < u$, $v < s$. This is the only non-trivial class and an equivalent way of specifying it is that $A_{j} \in \mathcal{A}_{2}$ and $B_{j} \in \left\{\psi(A_{j}) : \psi \in \boldsymbol{\Psi}\right\}$ for $j=1,2$ and in this case $A_{3} \in \mathcal{A}_{2}$. Such $A_{j}$'s and $B_{j}$'s do exist as explained  
before and also in the proof of Theorem~\ref{thm-3.4}, viz., $A_{1} = \{a\}$, $A_{2} = \{a'\}$, $B_{1} = \{b\}$, $B_{2} = \{b'\}$ with $a, a', b, b'$ as chosen there. We have $0< uv < \min \{sv, us\}$. 

Now suppose that Eq. \eqref{eq-3.41} is satisfied. Then so is \eqref{eq-3.42}. So we have $uv=sw$ and $(u-w)(v-w) =zw$. This gives that $0< sw < \min \{sv, su\}$, which, in turn, gives that $0 < w < \min\{ u,v\}$, i.e., $w, u-w, v-w$ are all $> 0$. This forces $z>0$ as well. Thus \eqref{eq-3.42} 
 is equivalent to the following for this special class.
\begin{equation}
w, u-w, v-w, z\quad \text{all}\quad  > 0, (u-w)(v-w) =zw \tag{3.43}\label{eq-3.43}
\end{equation} 

We may call \eqref{eq-3.43} \textbf{criss-cross} equation in view of the pictorial representation of $\boldsymbol{\Psi}$ given in Fig. \ref{fig11} with concrete situations displayed in Example~III.4 and Figs.~\ref{fig12} to \ref{fig15} below.

Furthermore, $s$ divides $uv$ and $0 < u$, $v < s$. So $s$ cannot be a prime number. Therefore, $s$ is a composite number, say, $s = t' t''$ with $2 \leq t'$, a factor of $u$ and $2 \leq t''$, a factor of $v$. Hence, we may write as follows. 
For some 
\begin{equation}
t', t''\geq 2, u', v''\geq 1, s=t't'', u=t'u', v=t''v'', w=u'v''.\tag{3.44}\label{eq-3.44}
\end{equation}
This representation is not unique as can be seen for the case $s=16, u=12, v=8, w=6$.

\item \textbf{Pictorial representation of (c) above}. 

Let $J = \{j : 1 \leq j \leq j' \}$, $K = \{k : 1 \leq k \leq k'\}$, $J_{1} = \{j \in  J : \psi_{j}^{E} (A_{1}) = B_{1}\}$, $J_{2} = J \smallsetminus J_{1} = \{ j\in J : \psi_{j}^{E}(A_{1}) \neq B_{1}\}$, $K_{1} = \{k \in K  : \psi_{k}^{E'} (A_{2}) = B_{2}\}$, $K_{2} = K \smallsetminus K_{1}=\{k \in K : \psi_{k}^{E'}(A_{2}) \neq B_{2}\}$, $j_{1} = \# J_{1}$, $j_{2} = \# J_{2} = j'-j_{1}$, $k_{1} = \# K_{1}$, $k_{2}= \# K_{2} = k' - k_{1}$.

Then $U = \left\{\psi\in \boldsymbol{\Psi} : \psi /E  \in \{\psi^{E}_{j} : j \in J_{1} \}\right\}$, 

$V = \left\{\psi \in \boldsymbol{\Psi} : \psi/E' \in \{\psi_{k}^{E'} : k \in K_{1} \} \right\}$,

$W = \Big\{\psi \in \boldsymbol{\Psi} : \psi/E = \psi^{E}_{j}, \psi/E' = \psi^{E'}_{k} \quad \text{for some} \quad j \in J_{1}, k \in K_{1}\Big\}$,
 $Z = \Big\{\psi \in \boldsymbol{\Psi} : \psi /E = \psi^{E}_{j}, \psi/E' = \psi^{E'}_{k} \quad \text{for some}\quad j\in J_{2}, k \in K_{2} \Big\}$.

In case $j_{1}, k_{1}, j_{2}, k_{2}$ are all $> 0$, we have the pictorial representation as in Fig.~\ref{fig11} which will be called $(A_{j}, B_{j})_{j=1,2}$ \textbf{configuration}. Clearly, $w \leq j_{1} k_{1}$, $u-w \leq j_{1}k_{2}$, $v-w \leq j_{2} k_{1}$ and $z \leq j_{2} k_{2}$. 

\begin{figure}[h]
\centering
\includegraphics[scale=1.2]{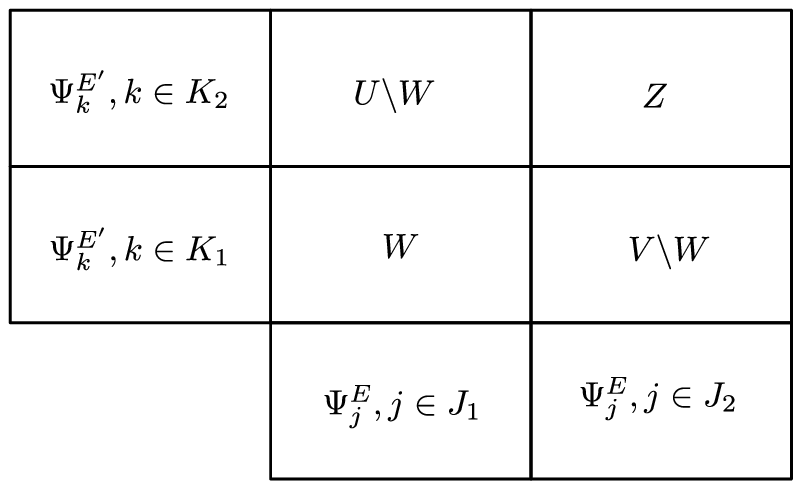}
\caption{
A configuration of \(\boldsymbol{\Psi}\).
}\label{fig11}
\end{figure}
 
\item Let us call the equations $u-w = j_{1}k_{2}$ and $v-w = j_{2} k_{1}$ by \textbf{criss}, and $w =j_{1}k_{1}$ and $z=j_{2} k_{2}$ by \textbf{cross}, Then \eqref{eq-3.43} together with criss holds if and only if \eqref{eq-3.43} with cross holds. Thus, either of this  becomes equivalent to $\boldsymbol{\Psi}$ is factorable, which, in turn, becomes equivalent to \eqref{eq-3.43} together with $U$ and $V$ are full, i.e., $u=j_{1} k'$, and $v=j'k_{1}$, i.e., cross-roads $U$ and $V$ are full.

\item For notational convenience, we write $L$ for the triple $(A_{j}, B_{j})_{1 \leq j \leq 3}$ or $(A_{j}, B_{j})_{j=1,2}$ for that matter, denote the corresponding sets $U, V, W, Z$ by $U_{L}, V_{L}, W_{L}, Z_{L}$ respectively; further the numbers $u, v, w, t', t'', u' v''$ will be written as $u_{L}, v_{L}, w_{L}, z_{L}, t_{L}', t_{L}'', u_{L}', v_{L}''$ respectively. Then the Eqs.~\eqref{eq-3.41} to \eqref{eq-3.44} 
take the forms of~\eqref{eq-3.45} to \eqref{eq-3.48} respectively given below.
\begin{align*}
u_{L} v_{L} &= sw_{L},\tag{$(3.45)_L$}\label{eq-3.45}\\
(u_{L}- w_{L}) (v_{L} - w_{L}) &= z_{L} w_{L},\tag{$(3.46)_L$}\label{eq-3.46}\\
u_{L}, u_{L}-w_{L}, v_{L}-w_{L}, z_{L}~~\text{all} > 0, (u_{L}- w_{L})(v_{L}-w_{L}) &= z_{L}w_{L}.\tag{$(3.47)_L$}\label{eq-3.47} \end{align*}

For some 
\begin{equation*}
t_{L}', t_{L}'' \geq 2, u_{L}', v_{L}'' \geq 1, s=t_{L}'~t_{L}'', u_{L}= t_{L}' u_{L}', v_{L} = t_{L}'' v_{L}'', w_{L} = u_{L}' v_{L}''. \tag{$(3.48)_L$}\label{eq-3.48}
\end{equation*}

For $L$ as above, let $\widetilde{L} = (\widetilde{A}_{j}, \widetilde{B}_{j})_{1 \leq j \leq 3}$, where $\widetilde{A}_{1} = E \cap A \smallsetminus A_{1}$, $\widetilde{A}_{2} = E' \cap A \smallsetminus A_{2}$. $\widetilde{B}_{1} = E \cap B \smallsetminus B_{1}$, $\widetilde{B}_{2} = E' \cap B \smallsetminus B_{2}$, $\widetilde{A}_{3} = \widetilde{A}_{1} \cup \widetilde{A}_{2}$, $\widetilde{B}_{3} = \widetilde{B}_{1} \cup \widetilde{B}_{2}$.

Then $U_{\widetilde{L}} = U_{L}$, $V_{\widetilde{L}} = V_{L}$, $W_{\widetilde{L}} =W_{L}$, $Z_{\widetilde{L}} =Z_{L}$, and, therefore, $u_{\widetilde{L}} = u_{L}$, $v_{\widetilde{L}} = v_{L}$, $w_{\widetilde{L}} = w_{L}$, $z_{\widetilde{L}} =z_{L}$. So the Eqs.~\eqref{eq-3.45} to \eqref{eq-3.48}
undergo no change, when we change $L$ to $\widetilde{L}$. Indeed, the same is true for $\widetilde{\widetilde{L}} = (\widetilde{A}_{1}, A_{2}, \widetilde{B}_{1}, B_{2}, \widetilde{A}_{1} \cup A_{2}, B_{1} \cup B_{2}))$ and $\widetilde{\widetilde{\widetilde{L}}} = (A_{1}, \widetilde{A}_{2}, B_{1}, \widetilde{B}_{2}, A_{1} \cup \widetilde{A}_{2}, B \cup \widetilde{B}_{2})$ as well.

\item Let $\mathcal{L}$ be the set of all $L$ as above.

Let
\begin{align*}
\hat{\mathcal{L}} & = \left\{L \in  \mathcal{L} : \# A_{1} \leq \frac{\nu_{1}}{2}\} = \{L \in \mathcal{L} : \# B_{1} \leq \frac{\nu_{1}}{2}\right\},\\
\tilde{\mathcal{L}} & = \left\{L \in \mathcal{L}  : \# A_{2} \leq \frac{\nu_{2}}{2}\} = \{L \in \mathcal{L} : \# B_{2} \leq \frac{\nu_{2}}{2}\right\}.
\end{align*}
For $a_{0} \in E \cap A$, $b_{0} \in E \cap B$, $a_{0}' \in E' \cap A$, $b_{0}' \in E' \cap B$, let
\begin{align*}
\hat{\mathcal{L}}_{a_{0}} & = \left\{L \in  \hat{\mathcal{L}} : \text{whenever}~\#~ A_{1} = \frac{v_{1}}{2}~~ \text{we have}~~ a_{0} \in A_{1} \right\},\\
\hat{\mathcal{L}}_{b_{0}} & = \left\{L \in  \hat{\mathcal{L}} : \text{whenever} ~\#~ B_{1} = \frac{v_{1}}{2}~~ \text{we have}~~ b_{0} \in B_{1} \right\},\\
\widetilde{\mathcal{L}}_{a_{0}'} & = \left\{L \in  \widetilde{\mathcal{L}} : \text{whenever} ~\#~ A_{2} = \frac{v_{2}}{2}~~ \text{we have}~~ a_{0}' \in A_{2} \right\},\quad \text{and}\\
\widetilde{\mathcal{L}}_{b_{0}'} & = \left\{L \in  \widetilde{\mathcal{L}} : \text{whenever} ~\#~ B_{2} = \frac{v_{2}}{2}~~ \text{we have}~~ b_{0}' \in B_{2} \right\}.
\end{align*}
 
Master Eq.~\eqref{eq-3.34} is equivalent to any of the criss-cross systems $S_{l, \mathcal{L}^{\ast}}$ with $l \in \{45,46, 47, 48 \}$,
{\fontsize{10.5}{12.5}\selectfont
\begin{equation*}
\mathcal{L}^{\ast} \in \left\{\mathcal{L}, \widetilde{\mathcal{L}}, \hat{\mathcal{L}}, \hat{\mathcal{L}}_{a_{0}}, \hat{\mathcal{L}}_{b_{0}}, \widetilde{\mathcal{L}}_{a_{0}'}, \widetilde{\mathcal{L}}_{b_{0}'} : a_{0} \in E \cap A, b_{0} \in E \cap B, a_{0}' \in E' \cap A, b_{0}' \in E' \cap B \right\},
\end{equation*} }
where
\begin{equation}
S_{l, \mathcal{L}^{\ast}} = \left\{(3.l)_{L} : L \in \mathcal{L}^{*} \right\}.
\tag{3.49}\label{eq-3.49}
\end{equation}
We note two of them with details as we will use them in applying our discussion.
\begin{equation}
w_{L}, u_{L}-w_{L}, v_{L}-w_{L}, z_{L}~~\text{are all} > 0, (u_{L} -w_{L})(v_{L}-w_{L}) =z_{L} w_{L}~~\text{for}~~ L \in \hat{\mathcal{L}}.
\tag{3.50}\label{eq-3.50}
\end{equation}
  
For any $L \in \hat{\mathcal{L}}$, there exist $t_{L}'$, $t_{L}'' \geq 2$, and $u_{L}', v_{L}'' \geq 1$ that satisfy
\begin{equation}
s = t_{L}' t_{L}'',~~u_{L} = t_{L}' u_{L}',~~ v_{L} = t_{L}'' v_{L}'',~~ w_{L} =u_{L}' v_{L}''.
\tag{3.51}\label{eq-3.51}
\end{equation}   
  These motivate criss-cross concepts for $\boldsymbol{\Psi}$ and $s$ and we proceed with them and their use.
  
\end{enumerate}

\begin{definition}\label{definition-3.4}
Let $\boldsymbol{\Psi}$ be a set of coverings of $\Gamma_{n}$ which is neither flat nor a pole.  $\phi  \neq E \subsetneq \Gamma_{n}$. $\boldsymbol{\Psi}$ is said to be \textbf{criss-cross via} $E$ if it is decomposable via $E$ and satisfies the criss-cross system~\eqref{eq-3.50} or equivalently,~\eqref{eq-3.51}.
$\boldsymbol{\Psi}$ is called \textbf{criss-cross} if it is criss-cross via some $E$ with $\phi \neq E \subset_{\ne} \Gamma_{n}$.

\end{definition}

\begin{thm}\label{thm-3.7}
Let $\boldsymbol{\Psi}$ be a set of coverings of $\Gamma_{n}$ such that it is neither flat nor a pole.
\begin{enumerate}[label=(\roman*)]
\item Let $\phi \neq E \subset_{\neq} \Gamma_{n}. \, | \boldsymbol{\Psi}\rangle$ is a product vector in the bipartite cut $(E, E')$ if and only if $\boldsymbol{\Psi}$ is criss-cross via $E$.
\item $| \boldsymbol{\Psi} \rangle$ is genuinely entangled if and only if $\boldsymbol{\Psi}$ is not criss-cross.
\end{enumerate}

\end{thm}

{\it Proof.}
(i) follows from entanglement properties via polynomial representation. 

(ii) It is immediate from (i). 

\begin{definition}\label{definition-3.5}
Let $x\in \mathbb{N}$ with $x \geq 4$. The number $x$ is said to be \textbf{criss-cross decomposable} if there exist $x_{1}, x_{2}, x_{3}, x_{4}$ in $\mathbb{N}$ that satisfy
\begin{equation}
x= x_{1} + x_{2} + x_{3} + x_{4}, \qquad x_{2} x_{3} = x_{4} x_{1},\tag{3.52}\label{eq-3.52}
\end{equation}
and the arrangement 
\begin{equation}
\begin{tabular}{c|c}
$x_{2}$ & $x_{4}$\\\hline\tag{3.53}\label{eq-3.53}
$x_{1}$ & $x_{3}$
\end{tabular}
\end{equation}
is called the \textbf{corresponding configuration}.
\end{definition}

\textbf{(8) Criss-cross decompositions and their use}. 


\begin{enumerate}[label=(\alph*)]
\item The property is relevant in view of $s = \# \boldsymbol{\Psi}$ satisfying $s=s_{1} + s_{2} + s_{3} + s_{4}$ with $s_{1} = w$, $s_{2} = u-w$, $s_{3}=v-w$, $s_{4} =z$ in 
(c) of Master equations and specifying the criss-cross equation~\eqref{eq-3.43} there, the configuration given in Definition~\ref{definition-3.5} expresses the cardinalities of the sets in Fig.~\ref{fig11}. 

\item At most eight rearrangements of $x_{1}, x_{2}, x_{3}, x_{4}$ satisfy the same requirements as in Definition~\ref{definition-3.5} and one and only one is non-decreasing in the sense that $x_{1} \leq x_{2} \leq x_{3} \leq x_{4}$. The eight rearrangements can be listed via corresponding configurations viz.,

\begin{equation}
\begin{array}{c|c}
x_{2} & x_{4}\nonumber\\\hline
x_{1} & x_{3}
\end{array}\\~~,~~
\begin{array}{c|c}
x_{2} & x_{1}\nonumber\\\hline
x_{4} & x_{3}
\end{array}\\~~,~~
\begin{array}{c|c}
x_{3} & x_{4}\nonumber\\\hline
x_{1} & x_{2}
\end{array}\\~~,~~
\begin{array}{c|c}
x_{3} & x_{1}\nonumber\\\hline
x_{4} & x_{2}
\end{array}
\end{equation}\\[-8pt]
\begin{equation}
\begin{array}{c|c}
x_{4} & x_{2}\nonumber\\\hline
x_{3} & x_{1}
\end{array}\\~~,~~
\begin{array}{c|c}
x_{4} & x_{3}\nonumber\\\hline
x_{2} & x_{1}
\end{array}\\~~,~~
\begin{array}{c|c}
x_{1} & x_{2}\nonumber\\\hline
x_{3} & x_{4}
\end{array}\\~~,~~
\begin{array}{c|c}
x_{1} & x_{3}\tag{3.54}\label{eq-3.54}\\\hline
x_{2} & x_{4}
\end{array}
\end{equation}

\item Suppose $x$ is criss-cross decomposable and $x_{1}, x_{2}, x_{3}, x_{4}$ are as in  \eqref{eq-3.52}. Set $y_{2} =x_{1} + x_{2}$, $y_{3} =x_{1} +x_{3}$, Then $y_{2}~y_{3} = \left(x_{1} + x_{2}\right)~\left(x_{1} + x_{3}\right)= x_{1}^{2} + x_{2}x_{1} + x_{1} x_{3} + x_{2}x_{3} = x_{1}^{2} + x_{2}x_{1} +x_{1} x_{3} + x_{1} x_{4} = x_{1} \left(x_{1} + x_{2} + x_{3} + x_{4} \right)= x_{1} x$.  

Hence, $x$ divides $y_{2}~y_{3}$. But $2 \leq y_{2}$, $y_{3} < x$. So $x$ cannot divide any one out of $y_{2}, y_{3}$. So $x$ cannot be a prime number.

On the other hand, suppose that $x$ is a composite number, say, $x=pq$ with $2 \leq p \leq q$. Consider any $1 \leq p_{1} < p$, $1 \leq q_{1} < q$. Set $x_{1} = p_{1}q_{1}$, $x_{2} = p_{1}(q-q_{1})$, $x_{3} =(p-p_{1})q_{1}$, $x_{4} = (p-p_{1})(q-q_{1})$. Then $x_{1} + x_{2} + x_{3} + x_{4} = pq =x$ and $x_{1}x_{4} = x_{2}x_{3}$ which implies $x$ is criss-cross decomposable. 
 
\item Now consider any composite number $x$ and any criss-cross decomposition of $x$. We proceed as in the first part of (c) except the last sentence. Comparing the factorization of the two sides of $y_{2} y_{3} =x_{1} x$, we can express $y_{2}= y^{'}_{2} y_{2}^{''}$, $y_{3} = y_{3}' y_{3}''$, $x=y_{2}'y_{3}'$, $x_{1} =y_{2}'' y_{3}''$  for some $y_{2}', y_{3}' \geq 2$, $y_{2}'', y_{3}'' \geq 1$. This gives us $x_{2} = y_{2}-x_{1} =\left(y_{2}'-y_{3}''\right)y_{2}''$, $x_{3} =y_{3} -x_{1} =\left(y_{3}'-y_{2}''\right)y_{3}''$, $x_{4}=x-y_{2}-x_{3} = (y_{2}'-y_{3}'')(y_{3}'-y_{2}'')$.
 We know that $x_{1}, x_{2}, x_{3}, x_{4} > 0$. Thus $1 \leq y_{3}'' < y_{2}''$ and $1 \leq y_{2}'' < y_{3}'$. Hence $x_{1}, x_{2}, x_{3}, x_{4}$ have the same form as in the second paragraph of (c) with $p,q$ replaced by $y_{2}', y_{3}'$ respectively and $p_{1}, q_{1}$ by $y_{3}'', y_{2}''$ respectively.
 
 Hence all criss-cross decompositions of $x$ are given by the method in the second paragraph of (c) by varying factorizations of $x$ that are available.
 
\item We now give some numerical illustrations of criss-cross decompositions and configurations.

For $s=4$ : $1 \leq 1 \leq 1 \leq 1$  ; $\begin{pmatrix}1 & 1\\ 1 & 1\end{pmatrix}$.

For $s=6$ : $1 \leq 1 \leq 2 \leq 2$  ; $\begin{pmatrix}1 & 1\\ 2 & 2\end{pmatrix}$,~ $\begin{pmatrix}1 & 2\\ 1 & 2\end{pmatrix}$,~ $\begin{pmatrix}2 & 1\\ 2 & 1\end{pmatrix}$,~ $\begin{pmatrix}2 & 2\\ 1 & 1\end{pmatrix}$.

{\fontsize{11}{14}\selectfont For $s=8$ : $1 \leq 1 \leq 3 \leq 3$, $2 \leq 2 \leq 2 \leq 2$~; $\begin{pmatrix}1 & 1\\ 3 & 3\end{pmatrix}$,~$\begin{pmatrix}1 & 3\\ 1 & 3\end{pmatrix}$,~$\begin{pmatrix}3 & 1\\ 3 & 1\end{pmatrix}$,~ $\begin{pmatrix}3 & 3\\ 1 & 1\end{pmatrix}$;~ $\begin{pmatrix}2 & 2\\ 2 & 2\end{pmatrix}$.} 
\end{enumerate}

For $s=9$, $1 \leq 2 \leq 2 \leq 4$~; $\begin{pmatrix}1 & 2\\ 2 & 4\end{pmatrix}$,~ $\begin{pmatrix}4 & 2\\ 2 & 1\end{pmatrix}$,~ $\begin{pmatrix}2 & 1\\ 4 & 2\end{pmatrix}$,~ $\begin{pmatrix}2 & 4\\ 1 & 2\end{pmatrix}$.


\begin{example}\label{example-3.4}
Let $\nu=6$, $E = \{1,2,3,4,5,6\}$. Next, let $\psi_{j}^{E}, \psi_{k}^{E'} : 1 \leq j, k\leq 3$ be as given in Fig.~\ref{fig12} and $\boldsymbol{\Psi} = \left\{\psi_{j}^{E} \times \psi_{k}^{E'} : j=2~~\text{or}~~ k=1~~\text{or}~~(j,k)=(3,2)\right\}$ as displayed in Fig.~\ref{fig13}.


\begin{figure}[h]
\centering
\includegraphics[scale=1]{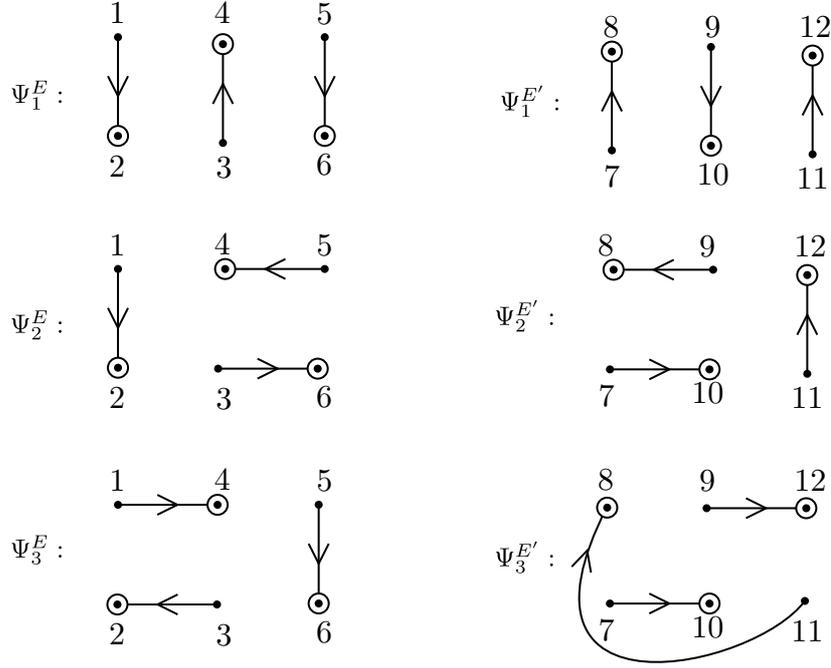}
\caption{
Example of coverings of \(E\) and \(E'\) with \(\nu=6\), \(E=\{1,2,3,4,5,6\}\).
}\label{fig12}
\end{figure}
\begin{figure}[h]
\centering
\includegraphics[scale=.9]{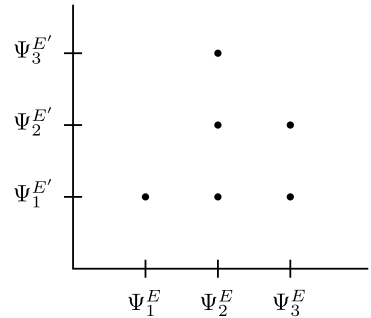}
\caption{ 
Grid of $\boldsymbol{\Psi} = \left\{\psi_{j}^{E} \times \psi_{k}^{E'} : j=2~~\text{or}~~ k=1~~\text{or}~~(j,k)=(3,2)\right\}$.}\label{fig13}
\end{figure}


\begin{enumerate}[label=(\alph*)]

\item  We consider the case $A_{1} = \{1\}$, $B_{1}=\{2\}$, $A_{2} = \{7\}$, $B_{2} = \{8\}$, i.e., $A_{3}= \{1,7\}$, $B_{3} = \{2,8\}$.

\begin{figure}[h]
\centering
\includegraphics[scale=.9]{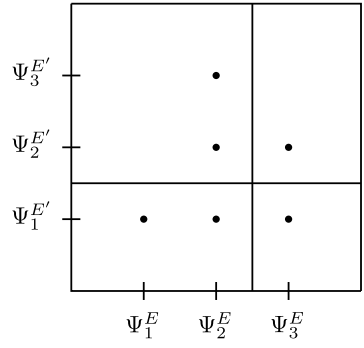}
\caption{$A_{1} = \{1\}$, $B_{1}=\{2\}$, $A_{2} = \{7\}$, $B_{2} = \{8\}$, i.e., $A_{3}= \{1,7\}$, $B_{3} = \{2,8\}$. Here $L=(A_{j}, B_{j})_{1 \leq j \leq 3}$ satisfies the criss-cross  Eq.~\eqref{eq-3.46}.}\label{fig14}
\end{figure} 

We note from Fig.~\ref{fig14} that $L=(A_{j}, B_{j})_{1 \leq j \leq 3}$ satisfies the criss-cross  Eq.~\eqref{eq-3.46}$_{L}$.


\item We consider the case $A_{1} = \{3\}$, $B_{1} = \{4\}$, $A_{2} = \{7\}$, $B_{2}=\{8\}$, i.e., $A_{3} = \{3,7\}$, $B_{3}= \{4,8\}$.
\begin{figure}[h]
\centering
\includegraphics[scale=.9]{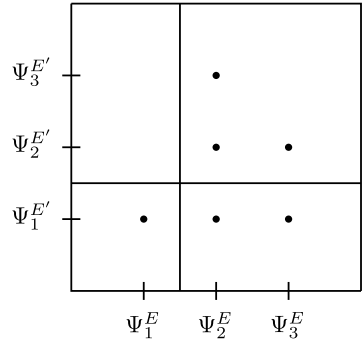}
\caption{\(A_1 = \{3\}\), \(B_1 = \{4\}\), \(A_2 = \{7\}\), \(B_2 = \{8\}\), i.e., \(A_3 = \{3, 7\}\), \(B_3 = \{4, 8\}\). Here \(L = (A_j, B_j)\) \(1\le j \le 3\) does not  satisfy the
criss-cross Eq.~\eqref{eq-3.46}.
}\label{fig15}
\end{figure} 
$\boldsymbol{\Psi}$ does not satisfy the Criss-cross Eq.~\eqref{eq-3.46} for this triple $L = \left(A_{j}, B_{j}\right)_{1 \leq j \leq 3}$. One can see Figure~\ref{fig15}.

\item Hence by Theorem~\ref{thm-3.7} $| \boldsymbol{\Psi} \rangle$ is genuinely entangled.

\item (a) and (b) show that~\eqref{eq-3.46}$_{L_1}$ and \eqref{eq-3.46}$_{L_{2}}$ can be in different situations of holding or not holding for certain $L_{1}, L_{2} \in \mathcal{L}$ in 
(f) of Master equations.
\end{enumerate}
\end{example}

We can reword Theorem~\ref{thm-3.7} in terms of criss-cross decompositions of $s$ in place of~\eqref{eq-3.50} and \eqref{eq-3.51} as done below.

\begin{thm}\label{thm-3.8}
(i) (a) If $\boldsymbol{\Psi}$ is such that every configuration as in 
(7)(d) of Master equations and Fig.~\ref{fig11} corresponds to some criss-cross decomposition of $s$, then $|\boldsymbol{\Psi}\rangle$ is a product vector in the bipartite cut $(E, E')$.

(b) The converse of (a) is also true. 

(ii) (i)(a) does happen if  $\bold{\Psi}$ is factorable.

(iii) If (i) (a) does not happen for any $\phi  \neq E \subsetneq \Gamma_{n}$, then $| \boldsymbol{\Psi} \rangle$ is genuinely entangled.

\end{thm}


{\bfseries(9) Alternative.}

We now look at the same situation as above from another point of view. We continue with the notation and terminology above, particularly that used in
Master equations and Criss-cross representation. 
We consider the case that $j_{1}, k_{1}, j_{2}, k_{2} > 0$ but $\boldsymbol{\Psi}$ is not factorable. 

\begin{enumerate}[label =(\alph*)]

\item Let $\sigma = j' k'-s$, $y_{1} = j_{1}k_{1}-w$, $y_{2}= j_{1}k_{2}-(u-w)$, $y_{3} = j_{2}k_{1}-(v-w)$, $y_{4} = j_{2} k_{2}-z$. Then 
\begin{equation}
y_{1}, y_{2}, y_{3}, y_{4} \geq 0,\quad y_{1} + y_{2} + y_{3} + y_{4} = \sigma > 0\tag{3.55}\label{eq-3.55}
\end{equation}
Equation~\eqref{eq-3.43} can be rewritten as
$$
y_{1} < j_{1} k_{1},~ y_{2} < j_{1}k_{2},~ y_{3} < j_{2}k_{1},~ y_{4} < j_{2}k_{2},
$$
\begin{equation}
(j_{1}k_{2} -y_{2})(j_{2} k_{1} - y_{3}) = (j_{1}k_{1}-y_{1})(j_{2}k_{2}-y_{4})\tag{3.56}\label{eq-3.56}
\end{equation}

\item We combine~\eqref{eq-3.55} and \eqref{eq-3.56} and obtain that \eqref{eq-3.56} is equivalent to 

$0 \leq y_{1} < j_{1} k_{1}$, $0 \leq y_{2} < j_{1} k_{2}$, $0 \leq y_{3} < j_{2} k_{1}$, $ 0 \leq y_{4} < j_{2} k_{2}$, $\left(y_{1}, y_{4}\right) \neq (0,0) \neq \left(y_{2}, y_{3}\right)$, $\sigma = y_{1} + y_{2} + y_{3} + y_{4} \geq 2$, 
\begin{equation}
j_{2}k_{2}y_{1} + j_{1} k_{1} y_{4} - y_{1}y_{4} = j_{2} k_{1} y_{2} + j_{1} k_{2} y_{3} - y_{2}y_{3}\tag{3.57}\label{eq-3.57}
\end{equation}

We may proceed as in 
(7)(f) and (7)(g) in Master equations, change $j_{1}, j_{2}, k_{1}, k_{2}$, $y_{1}, y_{2}, y_{3}, y_{4}$ to $j_{1}^{L}, j_{2}^{L}, k_{1}^{L}, k_{2}^{L}, y_{1}^{L}, y_{2}^{L}, y_{3}^{L}, y_{4}^{L}$ respectively in \eqref{eq-3.57} and call the new equations~\eqref{eq-3.57}$_L$.

We readily obtain that $\boldsymbol{\Psi}$ is criss-cross via $E$ if and only if \eqref{eq-3.57}$_L$ holds for $L \in \hat{\mathcal{L}}$.

\item We note that even though $s$ depends only on $\boldsymbol{\Psi}$, $j'$ and $k'$ depend on $E$ and $\boldsymbol{\Psi}$. So we can denote $\sigma$ by $\sigma_{E}$, if the need be.

\item We note different  ways of writing~\eqref{eq-3.57} to facilitate the process of finding the  solutions:
\begin{align*}
\left(j_{2}k_{2} - y_{4} \right)y_{1} + j_{1}k_{1} y_{4} & =\left(j_{2}k_{1}-y_{3}\right)y_{2} + j_{1}k_{2}y_{3},
\tag{3.58}\label{eq-3.58}\\
\left(j_{2}k_{2} - y_{4} \right)y_{1} + j_{1} k_{1} y_{4} &= j_{2}k_{1}y_{2} + \left(j_{1} k_{2} -y_{2} \right)y_{3}, \tag{3.59}\label{eq-3.59}\\
j_{2}k_{1}y_{1} + \left(j_{1}k_{1}-y_{1}\right)y_{4} &= \left(j_{2}k_{1} -y_{3}\right)y_{2} + j_{1}k_{2}y_{3},
\tag{3.60}\label{eq-3.60}\\
j_{2}k_{1}y_{1} + \left(j_{1}k_{1}-y_{1} \right)y_{4} &= j_{2}k_{1}y_{2} + \left(j_{1}k_{2}-y_{2}\right)y_{3}.
\tag{3.61}\label{eq-3.61}
\end{align*}
\end{enumerate} 

\begin{example}\label{example-3.5}
\textbf{Solutions of \eqref{eq-3.57} and consequences}. We follow the notation and terminology as in
alternative criss-cross decomposition and book for solutions $\mathbf{y}=\left(y_{1} y_{2}, y_{3}, y_{4}\right)$ for~\eqref{eq-3.57}.

\begin{enumerate}[label=(\alph*)]

\item Trivially, there is no solution for $\sigma=1$.

\item Let $\sigma=2$. Then $\mathbf{y} = (1,1,0,0)$ is a solution if and only if $k_{1}=k_{2}$ if and only if $\mathbf{y}= (0,0,1,1)$ is a solution. In this case $k'$ is even. Next, $\mathbf{y} = (1,0,1,0)$ is a solution. if and only if $j_{1}=j_{2}$ if and only if $\mathbf{y}= (0,1,0,1)$ is a solution. In this case $j'$ is even.

Hence if both $j'$ and $k'$ are odd then there is no solution for~\eqref{eq-3.57}.

\item Let $\sigma = 3$. Then $\mathbf{y} = (1,1,1,0)$ is a solution if and only if $1 = j_{2}k_{1} + (j_{1}-j_{2})k_{2} = j_{1}k_{2} + j_{2}(k_{1}-k_{2})$.

In this case $j_{1}\leq j_{2}$ and $k_{1} \leq k_{2}$.

But $\mathbf{y} = (1,1,0,1)$ is a solution if and only if $1=j_{2}k_{2} + (j_{1}-j_{2})k_{1}=j_{1}k_{1} + j_{2}\left(k_{2}-k_{1}\right)$. In this case, $j_{1} \leq j_{2}$ and $k_{2} \leq k_{1}$.

Next, $\mathbf{y} = (1,0,1,1)$ is a solution if and only if $1=\left(j_{2}-j_{1}\right)k_{2} + j_{1}k_{1} \equiv j_{1}(k_{1}-k_{2}) + j_{2} k_{2}$. In this case $j_{2} \leq j_{1}$ and $k_{1}\leq k_{2}$.

Finally, $\mathbf{y}=(0,1,,1,1)$ is a solution if and only if\\ $1=\left(j_{2}-j_{1}\right) k_{1} + j_{1}k_{2}\equiv \left(k_{2}-k_{1}\right)j_{1} + j_{2}k_{1} $. In this case $j_{2} \leq j_{1}$ and $k_{2}\leq k_{1}$.

We now, come to 8 possible solutions of the type $(2,1,0,0)$. We first note that $\mathbf{y}=(2,1,0,0)$ is a solution if and only if $2k_{2}=k_{1}$ if and only if $\mathbf{y}=(0,0,2,1)$ is a solution. But $\mathbf{y}=(1,2,0,0)$ is a solution if and only if $k_{2}=2k_{1}$ if and only if $(0,0,1,2)$ is a solution.

Similarly we obtain conditions for other four like  $(2,0,1,0)$ in terms of $j_{1}$ and $j_{2}$. 

\item Let $\sigma = 4$. Then $\mathbf{y} = (1,1,1,1)$ is a solution if and only if $\left(j_{2} -j_{1} \right)\left(k_{2}-k_{1}\right) = 0$, i.e., either $j_{1} =j_{2}$ or $k_{1}=k_{2}$.

In this case, either $j'$ or $k'$ is even. 

Under respective condition in (b), a solution $\mathbf{y}$ of $\sigma=2$ renders $2\mathbf{y}$ as a solution for $\sigma=4$. 

The remaining possible solutions are 12 of the type $(2,1,1,0)$ with conditions of the type $1 = j_{2}k_{1} + (j_{1} -2j_{2})k_{2} \equiv (k_{1}-2k_{2})j_{2} + j_{1}k_{2}$ (this can happen only if $j_{1}-2j_{2} \leq 0$ and $k_{1} - 2k_{2} \leq 0$) on the one hand, and of the type $(3,1,0,0)$ with conditions similar to those for $(2,1,0, 0)$  etc. in (c) above., on the other hand. 

\end{enumerate}
\end{example}

\textbf{(10) Modification of $\boldsymbol{\Psi}$.} 

Let $\boldsymbol{\Psi}$ be as in Theorem~\ref{thm-3.8} (i) above. We work in the setup of~
(7)(c) and (7)(d) of Master equations above.

\begin{enumerate}[label =(\alph*)]
\item Consider any set $\boldsymbol{\widetilde{\Psi}}$ of coverings of $\Gamma_{n}$ for which either $\boldsymbol{\widetilde{\Psi}} \subset_{\neq} \boldsymbol{\Psi}$ or $\boldsymbol{\widetilde{\Psi}} \supset\neq \boldsymbol{\Psi}$. We call $\boldsymbol{\widetilde{\Psi}}$ \textbf{compatible with $\boldsymbol{{\Psi}}$} if $\boldsymbol{\widetilde{\Psi}}_{E}= \boldsymbol{\Psi}_{\boldsymbol{E}}$ and $\boldsymbol{\widetilde{\Psi}}_{E'} = \boldsymbol{\Psi}_{E'}$. Such a compatible $\boldsymbol{\widetilde{\Psi}}$ will be called a \textbf{reduction of $\boldsymbol{\Psi}$} if $\boldsymbol{\widetilde{\Psi}} \subset \boldsymbol{\Psi}$ and \textbf{extension of} $\boldsymbol{\Psi}$ if $\boldsymbol{\Psi} \subset \boldsymbol{\widetilde{\Psi}}$. For any such modification of $\boldsymbol{\Psi}$, the \textbf{modification size} is the number $m_{\boldsymbol{\widetilde{\Psi}}} = | \# \boldsymbol{\widetilde{\Psi}} - \# \boldsymbol{\Psi}|$.

\item If modification $\boldsymbol{\widetilde{\Psi}}$ takes place in any of the four cells $W, U \smallsetminus W, V\smallsetminus W, Z$ as in Fig.~\ref{fig11}, then the analogue of the Criss-Cross Eq.~\eqref{eq-3.43} for $\boldsymbol{\Psi}$ can not be satisfied. So $| \boldsymbol{\widetilde{\Psi}} \rangle$ is not a product vector in the bipartite cut $(E, E')$.

\item The situation in (b) does occur if $m_{\boldsymbol{\widetilde{\Psi}}} =1$. In particular, if we take $\boldsymbol{\Psi}$ to be factorable so that $s=j'k'$, then for $\boldsymbol{\widetilde{\Psi}} \subset \boldsymbol{\Psi}$ with $\# \boldsymbol{\widetilde{\Psi}} = s-1$, $| \boldsymbol{\Psi} \rangle$ is not a product vector in the bipartite cut $(E, E')$. This is in line with~\eqref{eq-3.57} 
in (a) of (9) Alternative criss-cross decomposition and Example III.5(a) above.
\end{enumerate}

\textbf{(11) Further refinement of Criss-Cross and pictorial representation.}

Let $\phi\neq A_{1} \subset_{\neq} E \cap A$, $\phi \neq A_{2} \subset_{\neq} E' \cap A$ with $A_{1}, A_{2} \in \mathcal{A}_{2}$.

\begin{enumerate}[label=(\alph*)]
\item Let $\left\{\psi(A_{1}) : \psi \in \boldsymbol{\Psi}\right\} = \left\{B(A_{1}, p) : 1 \leq p \leq j'_{A_{1}}\right\}$ and\\ $\left\{\psi (A_{2}) : \psi \in \boldsymbol{\Psi}\right\} = \left\{B(A_{2}, q) : 1 \leq q \leq k'_{A_{2}}\right\}$, say with $B(A_{1}, p)$s being all distinct and $B(A_{2}, q)$s being all distinct. Since  $A_{1}, A_{2} \in \mathcal{A}_{2}$,  we have $j_{A_{1}}' \geq 2$ and $k_{A_{2}}' \geq 2$. Let $1 \leq p \leq j_{A_{1}}'$, $1 \leq q \leq k_{A_{2}}'$. Set 
\begin{align*}
J_{p} &= \left\{j \in J  : \psi_{j}^{E} (A_{1}) = B(A_{1}, p)\right\}, j_{p} = \# J_{p},\\
K_{q} & = \left\{k \in K : \psi_{k}^{E'} (A_{2}) = B(A_{2}, q)\right\}, k_{q} = \# K_{q},\\
U_{p} & = \left\{\psi \in \boldsymbol{\Psi} : \psi / E \in \left\{\psi_{j}^{E} : j \in J_{p}\right\}\right\}, u_{p} = \# U_{p},\\
V_{q} & = \left\{\psi \in \boldsymbol{\Psi} : \psi/E' \in \left\{\psi_{k}^{E'} : k \in K_{q}\right\} \right\}, v_{q} = \# V_{q},\\
W_{p,q} & = U_{p} \cap V_{q},\quad w_{p, q} = \# W_{p,q},\\
Z_{p,q} & = \left\{\psi \in \boldsymbol{\Psi} : \psi /E \notin \{ \psi_{j}^{E} : j \in J_{p}\}, \psi/E' \notin \{\psi_{k}^{E'} : k \in K_{q}\} \right\}\\
 & = \boldsymbol{\Psi} \smallsetminus (U_{p} \cup V_{q}), z_{p,q} = \# Z_{p,q}.
\end{align*}

Then $J_{p} \neq \phi \neq K_{q}$ and $U_{p} \neq \phi \neq V_{q}$. Further, $J$ is the disjoint union of $\left\{J_{p} : 1 \leq p \leq j'_{A_{1}}\right\}$ whereas $K$ is the disjoint union of $\left\{K_{q} : 1 \leq q \leq k'_{A_{2}}\right\}$. Moreover, $\boldsymbol{\Psi}$ is the disjoint union of $\left\{ U_{p}: 1 \leq p \leq j'_{A_{1}}\right\}$, and also of $\left\{V_{q}: 1 \leq q \leq k'_{A_2} \right\},$ for $1 \leq p \leq j'_{A_{1}}$, $U_{p}$ is the disjoint union of $\{W_{p,q} : 1 \leq q \leq k'_{A_{2}} \}$ whereas for $1 \leq q \leq k'_{A_{2}}$, $V_{q}$ is the disjoint union of $\{W_{p_{1},q} : 1 \leq p_{1} \leq j'_{A_{1}}\}$.

\item It follows from (a) that 
\begin{equation}
j' = \sum_{p=1}^{j'_{A_{1}}} j_{p}, ~~ k' = \sum_{q=1}^{k'_{A_{2}}} k_{q},~~ s= \sum_{p=1}^{j'_{A_{1}}} u_{p}= \sum_{q=1}^{{k'_{A_{2}}}}v_{q}. 
\tag{3.62}\label{eq-3.62}
\end{equation}
Also, for $1 \leq p \leq j'_{A_{1}}$, $1 \leq q \leq k'_{A_{2}}$, we have
\begin{equation}
0 \leq w_{p,q} \leq j_{p}k_{q},~~1 \leq j_{p} \leq u_{p} \leq j_{p}k',~~ 1 \leq k_{q} \leq v_{q} \leq k_{q}j',
\tag{3.63}\label{eq-3.63}
\end{equation}
\begin{equation}
\phi \neq U_{p} \subsetneq \boldsymbol{\Psi},~~ \phi \neq V_{q} \subsetneq \boldsymbol{\Psi},
\tag{3.64}\label{eq-3.64}
\end{equation}
\begin{equation}
u_{p}= \sum_{q_{1}=1}^{k'_{A_{2}}} w_{p,q_{1}}, v_{q}=\sum_{p_{1}=1}^{j'_{A_{1}}}w_{p_{1}, q}.\tag{3.65}\label{eq-3.35}
\end{equation}
As a consequence,
\begin{equation}
s =\sum \left\{w_{p,q} : 1 \leq p \leq j'_{A_{1}}, 1 \leq q \leq k'_{A_{2}}\right\}\tag{3.66}\label{eq-3.66}
\end{equation}

Hence $\boldsymbol{\Psi}$ is factorable if and only if for $1 \leq p \leq j'_{A_{1}}$, $1 \leq q \leq k'_{A_{2}}$,
\begin{equation}
w_{p,q} = j_{p}k_{q}.
\tag{3.67}\label{eq-3.67}
\end{equation}

\item Now, suppose that the Master Eq.~\eqref{eq-3.34} is satisfied. In view of (3.64) in (b), we may apply the discussion in 
(7)(c) of Master equations to $U_{p}, V_{q}, W_{p,q}$ etc. (in place of $U, V, W$ etc.) and obtain respective collective versions of~\eqref{eq-3.41} and (3.43) as follows

Let $1 \leq p \leq j'_{A_{1}}$, $1 \leq q \leq k'_{A_{2}}$. Then
\begin{equation}
u_{p}v_{q} = sw_{p,q}, \tag{3.68}\label{eq-3.68}
\end{equation} 
$w_{p,q}>0$, $u_{p}-w_{p,q} > 0$, $v_{q}-w_{p,q} > 0$, $z_{p,q}>0$ and 
\begin{equation}
(u_{p}-w_{p,q})(v_{q}-w_{p,q}) = z_{p,q}w_{p,q}.
\tag{3.69}\label{eq-3.69}
\end{equation}
In view of \eqref{eq-3.68}, we do obtain $u_{p}-w_{p,q}>0$, $v_{q}-w_{p,q}>0$ and $z_{p,q} >0$ for $1 \leq p \leq j'_{A_{1}}$, $1 \leq q \leq k'_{A_{2}}$ right from $w_{p,q}>0$ for $1 \leq p \leq j'_{A_{1}}$, $1 \leq p \leq k'_{A_{2}}$ simply because $z_{p,q} = \sum \left\{w_{p_{1},q_{1}} : 1 \leq p_{1} \leq j'_{A_{1}}, 1 \leq q_{1} \leq k'_{A_{2}}, p_{1} \neq p, q_{1} \neq q \right\}$.


We may infer that
\begin{equation}
s \geq j'_{A_{1}} k'_{A_{2}} \tag{3.71}\label{eq-3.71}
\end{equation} 
We show the situation in $(A_{1}, A_{2})$-grid displayed in Fig.~\ref{fig16}.
\begin{figure}[h]
\centering
\includegraphics[scale=1]{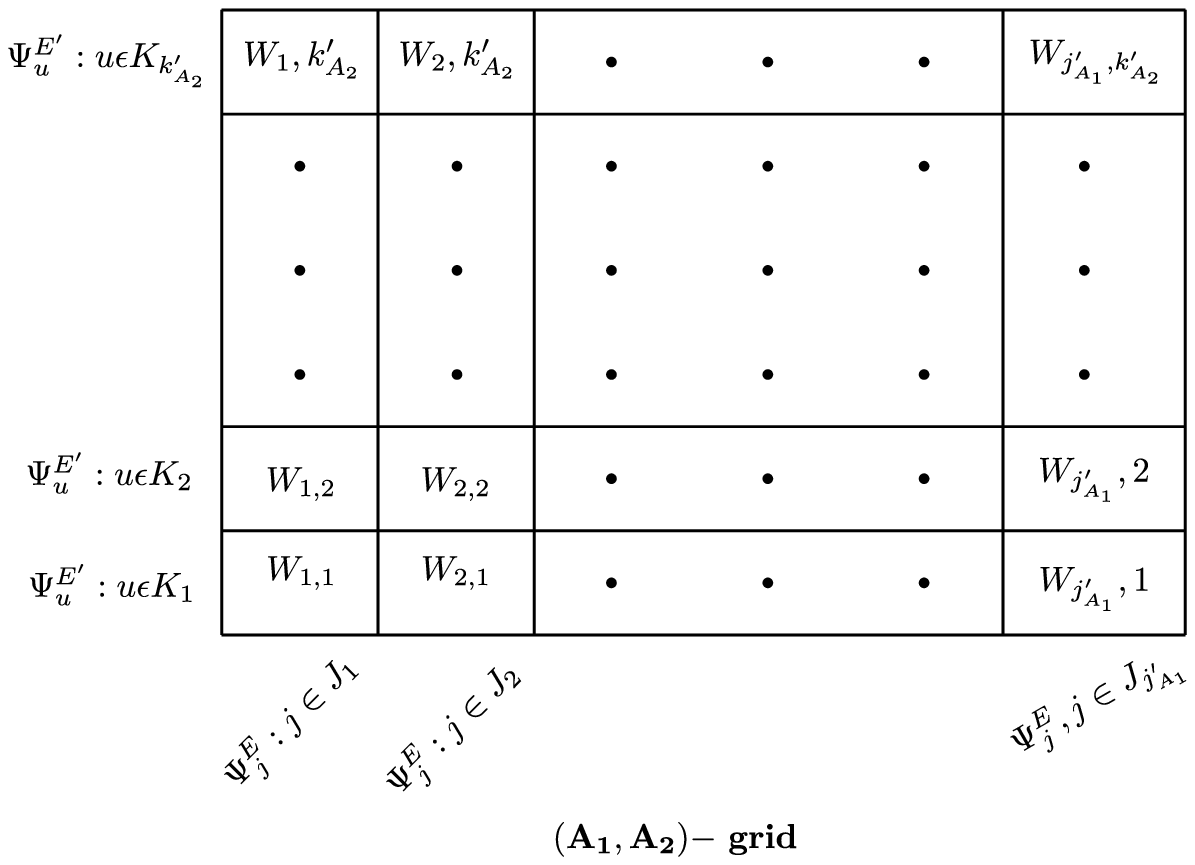}
\caption{
}
\label{fig16}
\end{figure}

We may rewrite~\eqref{eq-3.69} in terms of $w_{p,q}$'s alone as:
For $1 \leq p \leq p'_{A_{1}}$, $1 \leq q \leq k'_{A_{2}}$,
\begin{align*}
w_{p,q}>0,& \left(\sum\{w_{p,q_{1}} : 1 \leq q_{1} \leq k'_{A_{2}}, q\neq q_1 \}\right) \left(\sum \{w_{p_{1},q} : 1 \leq p_{1} \leq j'_{A_{1}}, p_{1} \neq p\} \right)\\
& = w_{p,q} \left(\sum \{w_{p_{1}, q_{1}} : 1 \leq p_{1} \leq j'_{A_{1}}, 1 \leq q_{1} \leq k'_{A_{2}}, p_{1} \neq p, q_{1} \neq q \} \right).\tag{3.72}\label{eq-3.72}
\end{align*}
We call \eqref{eq-3.72} ($\mathbf{A_{1}, A_{2}}$)-\textbf{criss-cross equations}.

\item Steps in (c) above can be reversed. Hence the Master Equation \eqref{eq-3.34} is equivalent to: 
For $\phi \neq A_1 \subsetneq E \cap A, \phi \neq A_2 \subsetneq E' \cap A,$ $A_1, A_2 \in \mathcal{A}_2$, the $(A_1, A_2)$-criss-cross equations (3.72) are satisfied.

This leads to the following Theorem in analogy with Theorem III.7.

\end{enumerate}

\begin{thm}\label{thm-3.9}

Let $\boldsymbol{\Psi}$ be a set of coverings of $\Gamma_{n}$ which is neither flat nor a pole.

\end{thm}

\begin{enumerate}[label=(\roman*)]
\item Let $(E, E')$ be a bipartite cut with $\phi \neq E \subset_{\neq} \Gamma_{n}$. Then the following are equivalent.
 \begin{enumerate}[label=(\alph*)]
\item $| \boldsymbol{\Psi} \rangle$ is a product vector in the bipartite cut $(E, E')$.

\item $\boldsymbol{\Psi}$ is decomposable via $E$ and for $\phi \neq A_{1} \subset_{\neq} E \cap A$, $\phi \neq A_{2} \subset_{\neq}  E' \cap A$, $A_{1}, A_{2} \in \mathcal{A}_{2}$, $(A_{1}, A_{2})$ criss-cross Eq.~\eqref{eq-3.72} are satisfied.

\item $\boldsymbol{\Psi}$ is criss-cross via $E$.
 \end{enumerate}

\item In case (i) (b) we have $s \geq j'_{A_{1}} \cdot k'_{A_{2}}$ and therefore,\\ $s \geq s_{E} = \max\left\{j'_{A_{1}} \cdot k'_{A_{2}} : \phi \neq A_{1} \subset_{\neq} E \cap A, \phi\neq A_{2} \subset_{\neq} E' \cap A, A_{1}, A_{2} \in \mathcal{A}_{2} \right\}$. As a consequence, if $s < s_{E}$ then $| \bold{\Psi} \rangle$ is not a product vector in the bipartite cut $(E, E')$. 

\item Suppose $| \boldsymbol{\Psi} \rangle$ is a product vector in the bipartite cut $(E, E')$ but $\boldsymbol{\Psi}$ is not factorable via $E$. Then for $\phi \neq A_{1} \subset_{\neq} E \cap A$, $\phi \neq A_{2} \subset_{\neq} E' \cap A$, $A_{1}, A_{2} \in \mathcal{A}_{2}$, no crossroad $\left(U_{p}, V_{q}\right), 1 \leq p \leq j'_{A_{1}}, 1 \leq q \leq k'_{A_{2}}$ is full in the terminology of
(7)(e) of Master equations.

\item Let $\boldsymbol{\Psi}$ be decomposable. Set
$$
s_{0} = \min \left\{s_{E} : \boldsymbol{\Psi}~ \text{is decomposable via}~E \right\}.
$$
 
 If $s < s_{0}$, then  $| \boldsymbol{\Psi} \rangle $ is genuinely entangled. 
 
\end{enumerate}

\subsection{Polynomial representation of doped RVB states, i.e., RVB states with holes and their quantum entanglement}
\label{subsection-3.2}

Doped RVB states are genuinely entangled and the proof is rather easy using their polynomial representation.

Let $1 \leq \gamma <v$ and $\mu=v-\gamma$ so that $1 \leq \mu \leq v-1$. We will study RVB states with $\gamma$ holes unless otherwise stated.

\begin{enumerate}[label=(\alph*)]
\item  For $a \in A$, $b \in B$, a hole or a dope at $a, b$ is the unit vector in $\mathcal{H}_{a,b} = \mathcal{H}_{a} \bigotimes \mathcal{H}_{b}$ given by 
$$
[a, b]_{h} = |\uparrow \rangle_{a} | \uparrow \rangle_{b}.
$$
It is orthogonal to $[a,b]$. Its polynomial representation is  
\begin{equation}
h_{a,b} (\x_{a}, \x_{b})=1, \,\text{the constant function 1 in variables} \x_{a}, \x_{b}.
\tag{3.73}\label{eq-3.73}
\end{equation}

\item Let $C$ be a covering of $\Gamma_{n}$, i.e., a bijective map $\psi$ on $A$ to $B$. We introduce $\gamma$ holes in $C$ or $\psi$ in all possible ways, alternatively we introduce $\mu$ resonants in $C$ in all possible ways. To elaborate, let $D\subset A$ with $\# D = \gamma$, the corresponding $R = A \smallsetminus D \subset A$ with $\# R = \mu$, say, $D = \{a'_{k}: 1 \leq k \leq \gamma \}$, $R = \{a_{j} : 1 \leq j \leq \mu \}$. The corresponding holes and resonants are given by $(D, \psi) = \left\{\left(a'_{k}, \psi(a'_{k})\right) : 1 \leq k \leq \gamma \right\}$, $(R, \psi) = \left\{\left(a_{j}, \psi(a_{j})\right) : 1 \leq j \leq \mu \right\}$ respectively. Let $\mathcal{D} = \left\{D \subset A : \# D=\gamma \right\}$, $\mathcal{R}=\left\{R \subset A : \# R = \mu\right\}$, $\mathcal{S}=\left\{S \subset B: \# S=\mu\right\}$, $(\mathcal{D}, \psi) = \{(D, \psi) : D \in  \mathcal{D}\}$, $(\mathcal{R},\psi) = \{(R, \psi) : R \in \mathcal{R}\};$ all these five sets have the same cardinality $v_{C_{\mu}} = v_{C_{\gamma}} = \frac{v !}{\gamma ! \mu !}$.

\item For $D \in \mathcal{D}$, the corresponding $R = A \smallsetminus D \in \mathcal{R}$, we put $| (D , \psi) \rangle_{h} = | (R, \psi) \rangle = \left(\bigotimes\limits_{j=1}^{\mu} [a_{j}, \psi (a_{j})] \right) \bigotimes \left(\bigotimes\limits_{k=1}^{\gamma} [a'_{k}, \psi (a'_{k})]_{h}\right)$ considered as an element of $\mathcal{H}$ identified with $\left(\bigotimes\limits_{j=1}^{\mu} \mathcal{H}_{a_{j}, \psi(a_{j})}\right) \bigotimes \left(\bigotimes\limits_{k=1}^{\gamma} \mathcal{H}_{a'_{k}, \psi (a'_{k})} \right)$. It is a unit vector and its polynomial representation is
\begin{align*}
F_{D, \psi}^{h} (\boldsymbol{\x}) &= F_{R, \psi}^{h}(\boldsymbol{\x}) = 2^{-\mu/2} \prod_{j=1}^{\mu} \left(\x_{\psi(a_{j}) - \x_{a_{j}}}\right)\\
 & = 2^{-\mu/2} \left(\boldsymbol{\x}^{\psi(R)} + (-1)^{\mu} \boldsymbol{\x}^{R} + \sum_{1 \leq \mu_{1} \leq \mu-1} \sum_{\substack{A_{1} \subset R \\ \#A_{1} = \mu_{1}}} (-1)^{\mu_1}\boldsymbol{\x}^{A_{1}} \boldsymbol{\x}^{\psi(R \smallsetminus A_{1})}\right)\tag{3.74}\label{eq-3.74}
\end{align*} 
This is a homogeneous polynomial of degree $\mu$ and $X_{l}$ occurs in it for each $l \in R \cup \psi (R)$.
The third part is absent and by our convention, zero if and only if $\mu=1$.

\item $\left\{| (D, \psi) \rangle_{h} : D \in \mathcal{D} \right\} = \left\{ | (R, \psi) \rangle : R \in \mathcal{R}\right\}$ is an orthonormal subset of $\mathcal{H}$.

\item The doped state $| C \rangle_{h} = | \psi \rangle_{h}$ with $\gamma$ holes in $C$ is the element of $\mathcal{H}$ given by
$$
| C \rangle_{h} = | \psi \rangle_{h} = \frac{1}{\sqrt{\# \mathcal{D}}} \sum_{D \in \mathcal{D}} | (D, \psi) \rangle_{h} = \frac{1}{\sqrt{\# R}} \sum_{R \in \mathcal{R}} | (R, \psi) \rangle.
$$

\item The polynomial representation of $| \psi \rangle_{h}$ is given by 
{\fontsize{8}{10}\selectfont
\begin{align*}
F_{\psi}^{h}(\boldsymbol{\x}) &= \frac{1}{\sqrt{\# \mathcal{R}}} \sum_{R \in \mathcal{R}} F_{R, \psi}^{h}(\boldsymbol{\x})\\
 & = \frac{1}{\sqrt{\# \mathcal{R}}} 2^{-\mu/2}\left(\sum_{R \in \mathcal{R}} \boldsymbol{\x}^{\psi(R)} + (-1)^{\mu}\sum_{R \in \mathcal{R}} \boldsymbol{\x}^{R} + \sum_{1 \leq \mu_{1} \leq \mu-1} \sum_{\substack{A_{1} \subset A,\\ \# A_{1}= \mu_{1}}}(-1)^{\mu_{1}} \boldsymbol{\x}^{A_{1}} \left(\sum_{\substack{A_{2} \subset A \smallsetminus A_{1},\\ \# A_{2}= \mu-\mu_{1}}}  \boldsymbol{\x}^{\psi(A_{2})}\right)\right)\\
 & = \frac{1}{\sqrt{\# \mathcal{R}}} 2^{-\mu/2} \left(\sum_{S \in \mathcal{S}} \boldsymbol{\x}^{S} + (-1)^{\mu} \sum_{R \in \mathcal{R}} \boldsymbol{\x}^{R}\right) + \frac{1}{\sqrt{\# \mathcal{R}}} 2^{-\mu/2} \left(\sum_{1 \leq \mu_{1} \leq \mu-1} \left( \sum_{\substack{A_{1} \subset A,\\
 \# A_{1}=\mu_{1}}} (-1)^{\mu_{1}} \boldsymbol{\x}^{A_{1}} \left( \sum_{\substack{A_{2} \subset A \smallsetminus A_{1},\\ \# A_{2} = \mu-\mu_{1}}} \boldsymbol{\x}^{\psi(A_{2})}\right)\right)\right)\\
 & = F^{hi} (\boldsymbol{\x}) + F_{\psi}^{hd} (\boldsymbol{\x}), \quad \text{say}.
 \tag{3.75}\label{eq-3.75}
\end{align*}}
 
\item We note that $F^{hi}(\boldsymbol{\x})$ is independent of $\psi$, it has full support, i.e., $\Gamma_{n}$ and has no $X_{j}$ as a factor. But $F_{\psi}^{hd}(\boldsymbol{\x})$ does depend on $\psi$, indeed, it is absent and thus, by convention, is zero if and only if $\mu=1$.

In case $\mu =1$, by Theorem~\ref{thm-2.1} (iii)(b), $| \psi \rangle_{h} $ is genuinely entangled.

\item For $\mu \geq 2$, $F_{\psi}^{hd}(\boldsymbol{\x})$ has an alternative form, viz., 
\begin{equation}
F_{\psi}^{hd} (\boldsymbol{\x}) = \frac{1}{ \sqrt{\#} \mathcal{R}} 2^{-\mu/2} \sum_{1 \leq \mu_{2} \leq \mu-1} \sum_{\substack{B_{2} \subset B,\\
\# B_{2} = \mu_{2}}} \boldsymbol{\x}^{B_{2}} \left(\sum_{\substack{B_{1} \subset B \smallsetminus B_{2} \\ \# B_{1} = \mu-\mu_{2}}} (-1)^{\mu-\mu_{2}} \boldsymbol{\x}^{\psi^{-1} (B_{1})}\right).
\tag{3.76}\label{eq-3.76}
\end{equation}

\item Let $\boldsymbol{\Psi}$ be a set of coverings of $\Gamma_{n}$ with $s=\# \boldsymbol{\Psi} \geq 1$. Then the \textbf{doped vector} $| \boldsymbol{\Psi} \rangle_{h}$ in $\mathcal{H}$ is given by $\frac{1}{\# \boldsymbol{\Psi}} \sum_{\psi \in \boldsymbol{\Psi}} | \psi \rangle_{h}$. Then $|| | \boldsymbol{\Psi} \rangle_{h}|| \leq 1$.

The polynomial representation of $| \boldsymbol{\Psi}\rangle_{h}$ is given by
\begin{align*}
F_{\boldsymbol{\Psi}}^{h}(\boldsymbol{\x}) &= F^{hi}(\boldsymbol{\x}) + \frac{1}{\# \bold{\Psi}} \sum_{\boldsymbol{\boldsymbol{\psi} \in \boldsymbol{\Psi}}} F_{\psi}^{hd}(\boldsymbol{\x})\\
& = F^{hi}(\boldsymbol{\x}) + F_{\bold{\Psi}}^{hd} (\boldsymbol{\x}),\quad \text{say}.
\tag{3.77}\label{eq-3.77}
\end{align*} 
Thus as in (g) above, $F_{\boldsymbol{\Psi}}^h (\boldsymbol{\x})$ is a homogeneous polynomial of degree $\mu$, it has full support, i.e., $\Gamma_{n}$ and has no  $X_{j}$ as a factor.

Furthermore, for $\mu=1$, $F_{\bold{\Psi}}^{h} (\boldsymbol{\x})= F^{hi}(\boldsymbol{\x})$ and $| \boldsymbol{\Psi} \rangle_{h}$ is genuinely entangled.
 
\item Let $\mu\geq 2$. For $\phi \neq A_{1} \subset_{\neq} A$, $\phi \neq B_{2} \subset_{\neq}B$ with $\# A_{1} + \# B_{2}= \mu$, let $\boldsymbol{\Psi}_{A_{1}, B_{2}}^{h} = \{\psi \in \boldsymbol{\Psi} : \psi(A_{1})\cap B_{2} = \phi\}$ and $w(A_{1}, B_{2}) = \frac{1}{\# \boldsymbol{\Psi}} \# \boldsymbol{\Psi}^{h}_{A_{1}, B_{2}}$. We write $A_{1}wB_{2}$ to mean that $w(A_{1}, B_{2}) \neq 0$. Using \eqref{eq-3.75}, we have
{\fontsize{10}{12}\selectfont
\begin{align*}
F_{\boldsymbol{\Psi}}^{hd} (\boldsymbol{\x}) &= \frac{1}{\# \boldsymbol{\Psi}} \frac{1}{\sqrt{\#\mathcal{R}}} 2^{-\mu/2} \sum_{1 \leq \mu_{1} \leq \mu-1} \left(\sum_{\substack{A_{1}\subset A,\\\# A_{1}=\mu_{1}}} (-1)^{\mu_{1}} \boldsymbol{\x}^{A_{1}}\left( \sum_{\psi \in \boldsymbol{\Psi}} \left( \sum_{\substack{A_{2} \subset A \smallsetminus A_{1},\\ \# A_{2} = \mu-\mu_{1}}} \boldsymbol{\x}^{\psi(A_{2})}\right) \right) \right)\\
& = \frac{1}{\sqrt{\#\mathcal{R}}} 2^{-\mu/2} \sum_{1 \leq \mu_{1}\leq \mu-1} \sum_{\substack{A_{1}\subset A, B_{2}\subset B, \\ \# A_{1}=\mu_{1}, \# B_{2}= \mu-\mu_{1}}} (-1)^{\mu_{1}} w(A_{1}, B_{2}) \boldsymbol{\x}^{A_{1}} \boldsymbol{\x}^{B_{2}}\\
& = \frac{1}{\sqrt{\# \mathcal{R}}} 2^{-\mu/2} \sum_{1 \leq \mu_{1} \leq \mu-1} (-1)^{\mu_{1}} \sum \left\{w(A_{1}, B_{2})\boldsymbol{\x}^{A_{1}} \boldsymbol{\x}^{B_{2}}: \# A_{1} =\mu_{1}, \# B_{2} = \mu-\mu_{1}, A_{1} w B_{2}\right\}.
\tag{3.78}\label{eq-3.78}
\end{align*}} 

The last specification is only to specify the non-zero co-efficient representation of $F_{\boldsymbol{\Psi}}^{hd}(\boldsymbol{\x})$ as a sum of mutually orthogonal elements in $\mathcal{H}$. Thus $F_{\boldsymbol{\Psi}}^{h}(\boldsymbol{\x})$ is the sum of mutually orthogonal (non-zero) elements as follows:
\begin{equation}
\begin{multlined}
F_{\boldsymbol{\Psi}}^{h}(\boldsymbol{\x}) = \frac{1}{\sqrt{\# \mathcal{R}}} 2^{-\mu/2} \Bigg(\sum_{S \in \mathcal{S}} \boldsymbol{\x}^{S} + \sum_{R \in \mathcal{R}} (-1)^{\mu} \boldsymbol{\x}^{R}  + \sum \Big\{(-1)^{\mu_{1}} w(A_{1}, B_{2}) \boldsymbol{\x}^{A_{1}} \boldsymbol{\x}^{B_{2}} :\\ \# A_{1} = \mu_{1}, \# B_{2} = \mu-\mu_{1}, A_{1}wB_{2},~~ 1 \leq \mu_{1} \leq \mu-1 \Big\}\Bigg).
\tag{3.79}\label{eq-3.79}
\end{multlined}
\end{equation} 
 
\item Using~\eqref{eq-3.79}, 
\begin{align*}
|| | \boldsymbol{\Psi} \rangle_{h} ||^{2} &= \frac{1}{\# \mathcal{R}} 2^{-\mu} \bigg(\# \mathcal{S} + \# \mathcal{R} + \sum\bigg\{ (w(A_{1}, B_{2}))^{2}: \# A_{1}=\mu_{1}, \#B_{2}=\mu -\mu_{1},  A_{1} wB_{2},  \\
&\hspace{7cm} 1 \leq \mu_{1} \leq \mu-1\bigg\}\bigg)\\
& = 2^{-\mu} \bigg(2 + \sum \bigg\{(w(A_{1}, B_{2}))^{2} /\# \mathcal{R}: \# A_{1}=\mu_{1}, \# B_{2}=\mu-\mu_{1}, A_{1}wB_{2},\\&\hspace{7cm} 1 \leq \mu_{1} \leq \mu-1 \bigg\} \bigg).
\tag{3.80}\label{eq-3.80}
\end{align*}

Thus, we may normalize $| \boldsymbol{\Psi} \rangle_{h}$ by replacing it by $| \boldsymbol{\Psi} \rangle_{h}^{\tilde{}} = \frac{1}{|| | \boldsymbol{\Psi}\rangle_{h}||} ~~ | \boldsymbol{\Psi} \rangle_{h}$ to obtain the corresponding doped state.
The entanglement properties of $| \boldsymbol{\Psi} \rangle_{h}$ and $| \boldsymbol{\Psi}\rangle_{h}^{\tilde{}}$ are the same and indeed all we need is observations made in (i) above and a good look at $F^{hi}(\boldsymbol{\x})$ for that purpose as we show in the theorem that follows.
\end{enumerate}

\begin{thm}\label{thm-3.9}
The doped state $| \boldsymbol{\Psi}\rangle_{h}^{\tilde{}}$ is genuinely entangled.
\end{thm}

{\it Proof.}
Let, if possible, $| \boldsymbol{\Psi}\rangle_{h}^{\tilde{}}$ be not genuinely entangled. Then the same is true for $|\boldsymbol{\Psi} \rangle_{h}$. Thus it is a product vector in some bipartite cut $(E, E')$. Then $F_{\boldsymbol{\Psi}}^{h}(\boldsymbol{\x}) = p(\boldsymbol{\x}_{E}) q(\boldsymbol{\x}_{E'})$ for some polynomials $p$ and $q$. Now $F_{\boldsymbol{\Psi}}^{h} (\boldsymbol{\x})$ is a homogeneous polynomial of degree $\mu$. So by (2)(e) in~\ref{subsection-2.2},
$p$ and $q$ are homogeneous polynomials of degrees, say $\mu_{1}$ and $\mu_{2}$ respectively with $\mu_{1} + \mu_{2} =\mu$. By~\ref{subsection-3.2}(i), support of $F_{\boldsymbol{\Psi}}^{h}(\boldsymbol{\x})$ is full. So $\mu_{1} \geq 1$ and $\mu_{2}\geq 1$. This forces $\mu \geq 2$. We note in passing that this does give that for $\mu=1$, $| \boldsymbol{\Psi} \rangle_{h}^{~}$ is genuinely entangled, a fact already noted in \ref{subsection-3.2}(i). Let $R \in \mathcal{R}$. Then by~\eqref{eq-3.79}, $\boldsymbol{\x}^{R}$ occurs in $F_{\boldsymbol{\Psi}}^{h}(\boldsymbol{\x})$, The only way that it can occur in $p(\boldsymbol{\x}_{E})$$q(\boldsymbol{\x}_{E'})$ is that $\boldsymbol{\x}^{E \cap R}$ occurs in $p(\boldsymbol{\x}_{E})$ and $\boldsymbol{\x}^{E' \cap R}$ occurs in $q(\boldsymbol{\x}_{E'})$. Let $A_{1}= E \cap R$, $A_{2} = E'\cap R$. Then $\# A_{1}= \mu_{1}$, $\# A_{2}=\mu_{2}$ and therefore, $A_{1}, A_{2}$ are both non-empty subsets of $A$. Now $\# A_{1} \cup A_{2} = \mu < v = \# A$. So there exists an $a \in A \smallsetminus A_{1} \cup A_{2}$. But $A=(E \cap A)\cup (E' \cap A)$. So either $a \in E \cap A$ or $a \in E' \cap A$. Consider the case when $a \in E \cap A$. There exists $a_{2} \in A_{2}$. Let $R_{1} = A_{1} \cup \{a\} \cup (A_{2} \smallsetminus \{a_{2}\})$. Then $R_{1} \subset A$, $\# R_{1} =\mu$. So $R_{1} \in \mathcal{R}$. By~\ref{eq-3.79}, $\boldsymbol{\x}^{R_{1}}$ occurs in $F_{\boldsymbol{\Psi}}^{h}(\boldsymbol{\x})$.
So as argued before for $R$, we have $\boldsymbol{\x}^{E \cap R_{1}}$ occurs in $p(\boldsymbol{\x}_{E})$. $\# E \cap R_{1} = \# A_{1} \cup \{a\} = \mu_1 + 1$, a contradiction. Similarly, we can deal with the case, $a \in E' \cap A$ to arrive at a contradiction. Hence $| \boldsymbol{\Psi} \rangle_{h}^{~}$ is genuinely entangled.

\section{Generalities, Analogies and Differences}
\label{sec-4}

The basic idea of this section is to take $v \geq 2$ and a set $\boldsymbol{\Psi}$ of coverings of $\Gamma_{n} = \Gamma_{2v}$ with $s=\# \boldsymbol{\Psi} \geq 2$ as in \ref{subsection-3.1} 2(c) although take a general convex combination or even a balanced convex combination of $| \psi \rangle$'s with $\psi$ in $\boldsymbol{\Psi}$ instead of just the arithmetic mean $| \boldsymbol{\Psi} \rangle = \sum_{\psi \in \boldsymbol{\Psi}} \frac{1}{\# \boldsymbol{\Psi}}| \psi \rangle$ and study its  entanglement properties via its polynomial representation. We define the corresponding concepts and obtain some analogies and some differences for the discussion and results in Section 3, which we shall freely use.

\subsection{Basic definitions and results}
\label{subsection-4.1}

We begin with a definition.

\begin{definition}\label{definition-4.1}
Let $\alpha$ be a function on $\boldsymbol{\Psi}$ to $\vertchar[.08ex]{C} \smallsetminus \{0\}$ with $\sum\limits_{\psi \in \boldsymbol{\Psi}} |\alpha({\psi})| =1$.
\end{definition}
\begin{enumerate}[label=(\roman*)]

\item $| \alpha \rangle  = \sum\limits_{\psi \in \boldsymbol{\Psi}} \alpha (\psi) | \psi \rangle$ is called a \textbf{generalized RVB vector} in $\mathcal{H}$.

\item If $\alpha$ is such that $| \alpha \rangle \neq 0$, then $| \widetilde{\alpha} \rangle = \frac{1}{|| |\alpha \rangle||} | \alpha \rangle$ is said to be a \textbf{generalized RVB (pure) state}.

\item For $S \subset \boldsymbol{\Psi}, | \alpha \rangle_{s} = \sum\limits_{\psi \in S} \alpha(\psi) | \psi \rangle$ is called \textbf{S-generalized RVB vector} in $\mathcal{H}$.

\item For $S \subset \boldsymbol{\Psi}$ let $\alpha^{\#}(S)= \sum_{\psi \epsilon S} \alpha(\psi)$, we shall denote $\alpha \# (S)$ by  $s_{\alpha}$ or $ \alpha_{S}$ and $\alpha\#(\bold{\Psi})$ by $\hat{\alpha}$.

\item $\alpha(\psi) > 0$ for all $\psi \in \boldsymbol{\Psi}$ if and only if $\hat{\alpha} =1$, and in this case, $| \alpha \rangle$ can be called a \textbf{convex RVB Vector}.

\item If $\hat{\alpha} =0$ then $| \alpha \rangle$ is called a \textbf{conditional RVB Vector}, otherwise $| \alpha \rangle$ is said to be \textbf{non-conditional RVB}.

\end{enumerate}


\textbf{(1) Properties of \(\alpha^\#\).}

We note a few facts about $\alpha^{\#}$ and set up useful notation for further use.

\begin{enumerate}[label=(\alph*)]
	
\item For any disjoint collection $\{S_{j} : 1 \leq j \leq j_{0} \}$ of subsets of $\boldsymbol{\Psi}$,
\begin{equation}
\alpha^{\#} \left(\bigcup\limits_{j=1}^{j_{0}} S_{j} \right) = \sum\limits_{j=1}^{j_{0}} \alpha^{\#}(S_{j}).
\tag{4.1}\label{eq-4.1}
\end{equation}

\item For $\phi \neq S \subset \boldsymbol{\Psi}$, $\alpha^{\#}(S)=0$ implies that 
\begin{equation}
\# S \geq 2.
\tag{4.2}\label{eq-4.2}
\end{equation}
By convention (that empty sums are zero), $\alpha^{\#}(\phi)=0$ and $| \alpha \rangle_{\phi} =0$.

\item Let $A_{1} \subset A$, $B_{1} \subset B$. Set $\boldsymbol{\Psi}_{A_{1}, B_{1}} = \{\psi \in \boldsymbol{\Psi} : \psi(A_{1}) = B_{1}\}$, $\{\psi(A_{1}) : \psi \in \boldsymbol{\Psi} \} = \{B(A_{1}, j) : 1 \leq j \leq j_{A_{1}} \}$ with $B(A_{1},j)$'s all distinct and $\{\psi^{-1}(B_{1}) : \psi \in  \boldsymbol{\Psi}\} = \{A(B_{1}, k) : 1\leq k \leq k_{B_{1}}\}$ with $A(B_{1}, k)$'s all distinct. Then $\boldsymbol{\Psi}_{A_{1}, B_{1}} \neq \phi$ if and only if $B_{1} = B(A_{1}, j)$ for some $j$ with $1 \leq j \leq j_{A_{1}}$ if and only if $A_{1} = A(B_{1}, k)$ for some $k$ with $1 \leq k \leq k_{B_{1}}$. Moreover, $\boldsymbol{\Psi}_{A_{1}, B_{1}} = \boldsymbol{\Psi}_{A \smallsetminus A_{1}, B \smallsetminus B_{1}}$ and $\boldsymbol{\Psi}$ is the disjoint union of non-empty sets $\boldsymbol{\Psi}_{A_{1}, B(A_{1,j})}, $ $ 1 \leq j \leq j_{A_{1}}$ and also of $\boldsymbol{\Psi}_{A(B_{1}, k), B_{1}}, 1 \leq k \leq k_{B_{1}}$.


For notational convenience, we write $\alpha(A_1, B_1)$ for $\alpha\# (\boldsymbol{\Psi}_{A_1, B_1}).$ We have  
\begin{equation}
\alpha(A_{1}, B_{1}) = \alpha(A \smallsetminus A_{1}, B \smallsetminus B_{1}).
\tag{4.3}\label{eq-4.3}
\end{equation}
\begin{equation}
\hat{\alpha} = \alpha \# (\boldsymbol{\Psi}) = \sum_{j=1}^{j_{A_{1}}} \alpha (A_{1}, B(A_{1}, j)) = \sum_{k=1}^{k_{B_{1}}} \alpha (A(B_{1}, k), B_{1}).\tag{4.4}\label{eq-4.4}
\end{equation}

\item For $A_{1} \in \mathcal{A}_{1}, B_{1} \in \mathcal{B}_{1}$, we have $j_{A_{1}}=1=k_{B_{1}}$ and therefore,

\begin{equation}
\widehat{\alpha}  = \alpha (A_{1}, B(A_{1}, 1)= \alpha(A(B_{1}, 1), B_{1})\tag{4.5}\label{eq-4.5}
\end{equation}
 On the other hand, for $A_{1} \in \mathcal{A}_{2}$, $B_{1} \in \mathcal{B}_{2}$ we have $j_{A_{1}} \geq 2$, $k_{B_{1}}\geq 2$ and 
\begin{equation}
s \geq \max\{j_{A_{1}}, k_{B_{1}}\}.
\tag{4.6}\label{eq-4.6}
\end{equation}
\end{enumerate}


\textbf{(2) Polynomial representation of $| \alpha \rangle $.} 

\begin{enumerate}[label=(\alph*)]
\item The polynomial representation $F_{\alpha}(\boldsymbol{\x})$ of $| \alpha \rangle$ is given by
\begin{align*}
F_{\alpha}(\boldsymbol{\x}) &= \sum_{\psi \in \boldsymbol{\Psi}} \alpha(\psi) F_{\psi} (\boldsymbol{\x}) = \sum_{\psi \in \boldsymbol{\Psi}} \alpha(\psi) 2^{-v/2} \prod_{a \in A} \left(\x_{\psi(a)} - \x_{a}\right)\\
& = 2^{-v/2} \sum_{\psi \in \boldsymbol{\Psi}} \alpha(\psi)\left(\sum_{A_{1}\subset A} (-1)^{\#A_{1}} \boldsymbol{\x}^{A_{1}} \boldsymbol{\x}^{B \smallsetminus \psi (A_{1})} \right)\\
& = 2^{-v/2} \sum_{\psi \in \boldsymbol{\Psi}} \alpha(\psi) \left(\sum_{B_{1}\subset B} (-1)^{\# B_{1}} \boldsymbol{\x}^{\psi^{-1}(B_{1})} \boldsymbol{\x}^{B \smallsetminus B_{1}} \right).
\tag{4.7}\label{eq-4.7}
\end{align*}
Thus, $F_{\alpha}$ is a homogeneous polynomial in $\boldsymbol{\x}$ of degree $v$.

\item We combine (a) above with~\eqref{subsection-4.1}(1) (c) and (d) and have, in analogy with~\eqref{subsection-3.1}(3), particularly~\eqref{eq-3.10},
\begin{align*}
F_{\alpha}(\boldsymbol{\x}) &= 2^{-v/2} \sum_{A_{1}\subset A}(-1)^{\# A_{1}} \boldsymbol{\x}^{A_{1}} \sum_{j=1}^{j_{A_{1}}} \alpha(A_{1}, B(A_{1}, j)) \boldsymbol{\x}^{B\smallsetminus B(A_{1},j)}\\
& = 2^{-v/2} \hat{\alpha} \sum_{A_{1}\in \mathcal{A}_{1}} (-1)^{\# A_{1}} \boldsymbol{\x}^{A_{1}} \boldsymbol{\x}^{B \smallsetminus B(A_{1},1)} +\\
    & \qquad\qquad2^{-v/2} \sum_{A_{1}\in \mathcal{A}_{2}} \sum_{j=1}^{j_{A_{1}}} (-1)^{\# A_{1}} \alpha(A_{1}, B(A_{1}, j)) \boldsymbol{\x}^{A_{1}} \boldsymbol{\x}^{B \smallsetminus B(A_{1}, j)}\\
 & = \hat{\alpha} F_{\boldsymbol{\Psi}}^{1} (\boldsymbol{\x}) + F_{\alpha}^{2}(\boldsymbol{\x}) = F_{\alpha}^{1}(\boldsymbol{\x}) + F_{\alpha}^{2} (\boldsymbol{\x}),~~ \text{say}.
 \tag{4.8}\label{eq-4.8}    
\end{align*}

Here $F_{\boldsymbol{\Psi}}^{1}(\boldsymbol{\x})$ is independent of $\alpha$, is non-zero and has support $\Gamma_{n}$, but $F_{\alpha}^{2}(\boldsymbol{\x})$ depends on $\alpha$, it is either zero or has support $\Gamma_{n}$ in view of (4.3) 
above. In the latter case, no $\x_{j}$ is a factor of $F_{\alpha}^{2}(\boldsymbol{\x})$.
An alternative form for $F_{\alpha}^{2} (\boldsymbol{\x})$ is 
\begin{equation}
F^{2}_{\alpha}(\boldsymbol{\x}) = 2^{-\nu/2} \sum_{B_{1} \in \mathcal{B}_{2}} \sum_{k=1}^{k_{B_{1}}} (-1)^{\# B_{1}} \alpha(A(B_{1}, k), B_{1}) \boldsymbol{\x}^{A{(B_{1}, k)}} \boldsymbol{\x}^{B \smallsetminus B_{1}}.
\tag{4.9}\label{eq-4.9} 
\end{equation}

\item $\hat{\alpha}~ F_{\boldsymbol{\Psi}}^{1} (\boldsymbol{\x}) \neq 0$ if and only if $\hat{\alpha} \neq 0$ if and only if $\boldsymbol{\x}^{B}$ occurs in $F_{\alpha}(\boldsymbol{\x})$ if and only if $\boldsymbol{\x}^{A}$ occurs in $F_{\alpha}(\boldsymbol{\x})$. In this case, $F_{\alpha}^{1}(\boldsymbol{\x})$ in support $\Gamma_{n}$ and no $\x_{j}$ is a factor of $F_{\alpha}^{1}(\boldsymbol{\x})$.

\item Now suppose $\hat{\alpha} \neq 0$. Then by (4.4) 
above, for $A_{1} \in \mathcal{A}_{2}$, $B_{1} \in \mathcal{B}_{2}$, we have that $\alpha(A_{1}, B(A_{1}, j)) \neq 0$ for some $j$ with $1 \leq j \leq j_{A_{1}}$, and 
\begin{equation}
\alpha {(A(B_{1}, k), B_{1})} \neq 0~\text{for some}~k~\text{with}~ 1 \leq k \leq k_{B_{1}}\tag{4.10}\label{eq-4.10} 
\end{equation}

Therefore, $F_{\alpha}^{2}(\boldsymbol{\x})$ has at least $\# \mathcal{A}_{2}$ (non-zero) terms and is, thus, non-zero. Moreover, using (4.3)  
above,
we have that $\alpha\left(A \smallsetminus A_{1},  B\smallsetminus B (A_{1}, j) \right) \neq 0 \neq \alpha \left(A\smallsetminus A (B_{1}, k), B \smallsetminus B_{1} \right)$ with $j$ and $k$ as in \eqref{eq-4.10} above, So $F_{\alpha}^{2}(\boldsymbol{\x})$ has support $\Gamma_{n}$ and no $\x_{j}$ is a factor of $F_{\alpha}^{2}(\boldsymbol{\x})$.

\item We now come to the case $\hat{\alpha}=0$.

Then $F_{\alpha}^{2} (\boldsymbol{\x}) \neq 0$ if and only if for some $A_{1} \in \mathcal{A}_{2}$, 
\begin{equation}
\alpha\left(A_{1}, B(A_{1}, j)\right)\neq 0 \,\,\text{for some} \,j \,\text{with}\,\, 1 \leq j \leq j_{A_{1}},
\tag{4.11}\label{eq-4.11}
\end{equation}
if and only if for some $B_{1} \in \mathcal{B}_{2}$, 
\begin{equation}
\alpha\left(A(B_{1}, k), B_{1}\right)  \neq 0 \,\,\text{ for some}\,\, k\,\,\text{ with}\, 1 \leq k \leq k_{B_{1}}.
\tag{4.12}\label{eq-4.12}
\end{equation}
But $\hat{\alpha} = 0$ and therefore, by \eqref{eq-4.4}, if \eqref{eq-4.11} holds for $j$, then it has to hold for some $1 \leq \tilde{j} \neq j \leq j_{A_{1}}$ as well and as a consequence, 
\begin{align*}
&\alpha\left(A_{1}, B(A_{1}, j)\right),~~\alpha \left(A_{1}, B(A_{1}, \tilde{j}) \right)~, \alpha\left(A\smallsetminus A_{1}, B\smallsetminus B(A_{1}, j)\right)~,\\
  &~\alpha\left(A \smallsetminus A_{1}, B \smallsetminus B(A_{1}, \tilde{j})\right)~ \text{are all non-zero}.
  ~\tag{4.13}\label{eq-4.13}
\end{align*}
Similarly, if~\eqref{eq-4.12} holds for $k$, for some 
\begin{align*}
&1 \leq \tilde{k} \neq k \leq k_{B_{1}}, \alpha(A (B_{1}, k), B_{1}), \alpha\left(A(B_{1}, \tilde{k}), B_{1}\right),\\
& \alpha(A \smallsetminus A(B_{1}, k), B\smallsetminus B_{1}), \alpha(A \smallsetminus A (B_{1}, \tilde{k}), B\smallsetminus B_{1})~ \text{are all non-zero}.
\tag{4.14}\label{eq-4.14} 
\end{align*}

Hence $F_{\alpha}^{2}(\boldsymbol{\x}) \neq 0$ if and only if $F_{\alpha}^{2}(\boldsymbol{\x})$ has at least four terms coming from co-efficients in \eqref{eq-4.13} and \eqref{eq-4.14} jotted in \eqref{eq-4.8} and \eqref{eq-4.9} respectively. In this case, $F^{2}_{\alpha}({\boldsymbol{\x}})$ has support $\Gamma_{n}$ and has no $\boldsymbol{\x}_{j }$ as a factor.

\item We continue with the case $\hat{\alpha} = 0$.

Now $F_{\alpha}^{2}(\boldsymbol{\x}) =0$ if and only if for 
\begin{equation}
A_{1} \in \mathcal{A}_{2}, \alpha(A_{1}, B(A_{1}, j))=0~~ \text{for}~~ 1 \leq j \leq j_{A_{1}},
\tag{4.15}\label{eq-4.15}
\end{equation} 
if and only if for 
\begin{equation}
B_{1} \in \mathcal{B}_{2},~~ \alpha(A(B_{1}, k), B_{1}) =0 ~~\text{for}~~ 1 \leq k \leq k_{B_{1}}.
\tag{4.16}\label{eq-4.16}
\end{equation}
If \eqref{eq-4.15} and \eqref{eq-4.16} hold then $\# \boldsymbol{\Psi}(A_{1}, B(A_{1}, j))\geq 2$ and $\# \boldsymbol{\Psi} (A(B_{1}, k), B_{1}) \geq 2$ and therefore, 
$$
s= \# \boldsymbol{\Psi} \geq 2 \max \left\{j_{A_{1}}, k_{B_{1}} : A_{1} \in \mathcal{A}_{2}, B_{1} \in \mathcal{B}_2\right\}=2s_{0}, \text{say}.
$$


Also by (4.6), $2 \leq s_{0} \leq s$.\\
Hence if $F_{\alpha}^{2}(\boldsymbol{\x})=0$, then $s \geq 2s_{0} \geq 4.$\\
In other words, if $s < 2~s_{0}$, then $F_{\alpha}^{2} (\boldsymbol{\x}) \neq 0$. In particular, it is so if $s=2$ or $3$.

\begin{table}
\label{tab17}
\begin{center}
\begin{tabular}{c|c||c|c|}
\cline{2-4}
$v$ & $v!$ & $n=2v$ & $2^{n}$ \\\cline{1-4}
\multicolumn{1}{|c|}{2} & 2 & 4 & 16\\\cline{1-4}
\multicolumn{1}{|c|}{3} & 6 & 6 & 64 \\\hline
\multicolumn{1}{|c|}{4} & 24 & 8 & 256\\\hline
\multicolumn{1}{|c|}{5} & 120 & 10 & 1024\\\hline
\multicolumn{1}{|c|}{6} & 720 & 12 &4,096\\\hline
\multicolumn{1}{|c|}{7} & 5,040 & 14 & 16,384\\\hline
\multicolumn{1}{|c|}{8} & 40,320 & 16 & 65,536\\\hline
\multicolumn{1}{|c|}{9} & 362,880 & 18 & 262,144\\\hline
\end{tabular}
\end{center}
\caption*{Table 1:  A comparison between factorials and exponentials.}
\end{table}

\item We note that for $s > 2^{n} = {\rm dim}~\mathcal{H}, \left\{| \psi \rangle : \psi \in \boldsymbol{\Psi} \right\}$ is linearly dependent and therefore, there exists $\phi \neq \boldsymbol{\Phi} \subset \boldsymbol{\Psi}$ and $\alpha : \boldsymbol{\Phi} \rightarrow  \vertchar[.08ex]{C} \smallsetminus \{0\}$ with $| \alpha \rangle = 0$. And such $\boldsymbol{\Psi}$s can exist if and only if $v! > 2^{n} = 2^{2v}$ if and only if $v \geq 9$, as simple computations show the situation given in Table~1.

\item In any case by~\eqref{eq-4.8} and \eqref{eq-4.9}, we have 
\begin{align*}
|| | \alpha \rangle ||^{2} &= 2^{-v} \left(| \hat{\alpha}|^{2} \# \mathcal{A}_{1} + \sum_{A_{1} \in \mathcal{A}_{2}} \sum_{j=1}^{j_{A_{1}}} 1 | \alpha(A_{1}, B(A_{1}, j))|^{2} \right)\\
&= 2^{-v} \left(| \hat{\alpha}|^{2} \# \mathcal{A}_{1} + \sum_{B_{1} \in \mathcal{B}_{2}} \sum_{k=1}^{k_{B_{1}}} | \alpha (A(B_{1}, k), B_{1}) |^{2}\right).
\tag{4.17}\label{eq-4.17}
\end{align*}

\item In view of (a) to (f) above, we may apply results or modify proofs in Secs. \ref{section-2} and \ref{section-3} above to study quantum entanglement properties of generalized RVB states $| \tilde{\alpha} \rangle$. We shall note a few straightforward analogies and give a few contrasting examples 
in the next subsections.
	
\end{enumerate}


\begin{thm}\label{thm-4.1}
Let $\xi = | \alpha \rangle$ be a non-zero generalized RVB vector.
\begin{enumerate}[label= (\roman*)]
\item $\xi$ is an entangled vector.

\item In case $\boldsymbol{\Psi}$ is flat or a pole via some $E$ with $\phi \neq E \subset_{\neq} \Gamma_{n}$,  $\xi$ is a product vector in  the bipartite cut $(E, E')$.

It does happen if $\boldsymbol{\Psi}$ is decomposable via $E$ and $v=3$.

\item Suppose $\xi$ is a convex RVB vector, i.e., $\alpha(\psi) > 0$ for $\psi \in \boldsymbol{\Psi}$, i.e., $\hat{\alpha}=1$. 
\begin{enumerate}[label=(\alph*)]
\item If $\xi$ is product vector in some bipartite cut $(E, E')$,  then $\boldsymbol{\Psi}$ is decomposable via $E$.

\item The converse of (a) is not true in general for $v > 3$.

\item If $\nu = 2$,  $\xi$ is genuinely entangled.

\item If $\boldsymbol{\Psi}$ is NN or PNN,  $\xi$ is genuinely entangled.
\end{enumerate}

 \end{enumerate}

\end{thm}

\emph{Proof.}
\begin{enumerate}[label=(\roman*)]
\item It follows from Theorem~\ref{thm-2.1}(i)(a) or (iv)(a) simply because the degree of $F_{\alpha}(\boldsymbol{\x})$ is $d=v < 2v = n$.

\item We follow the notation as in \eqref{subsection-3.1}. It is enough to consider the case that $\boldsymbol{\Psi}$ is flat via $E$ i.e., $\boldsymbol{\Psi}_{E'}= \left\{\psi^{E'}_{1}\right\}$. Let $\alpha_{E}$ be the function on $\boldsymbol{\Psi}_{E}$ given by $\alpha_{E}\left(\psi_{j}^{E}\right) = \alpha\left(\psi_{j}^{E} \times \psi_{1}^{E'}\right)$, $1 \leq j \leq j'$. Then $\xi = | \alpha \rangle = \left(\sum\limits_{j=1}^{j'} \alpha(\psi_{j}^{E}) | \psi_{j}^{E} \rangle \right) \otimes | \psi_{1}^{E'} \rangle$ and is, therefore, a product vector in the bipartite cut $(E, E')$.

For the second part, we already know from the proof of Theorem~\ref{thm-3.1}(v)(a) that, in this case, viz, $v=3$ and $\boldsymbol{\Psi}$ is decomposable via $E$,  $\boldsymbol{\Psi}$ is flat or a pole via $E$.

\item Proofs of the counterparts in Theorem~\ref{thm-3.1} can be modified in a straightway fashion to give different parts here as indicated in Table~ 2.

\begin{table}
\begin{center}
\begin{tabular}{c|c}
Theorem~\ref{thm-4.1}(iii) & Theorem~\ref{thm-3.1}\\[4pt]
\hline
(a) & (ii)(b)\\[4pt]
\hline
(b) & (v)(a), Example~\ref{example-4.1} as well\\[4pt]
\hline
(c) & (iii)\\[4pt]
\hline
(d) & (iv)
\end{tabular}
\end{center}
\caption*{Table 2: An analogy.}\label{fig018}
\end{table}

\end{enumerate}

\begin{example}\label{example-4.1}
Let $\boldsymbol{\Psi}$ be factorable via $E$ for some $\phi \neq E \subset_{\neq} \Gamma_{n}$, say, 
\begin{align*}
\boldsymbol{\Psi}_{E} &= \left\{\psi_{j}^{E} : 1 \leq j \leq j' \right\},~~~\boldsymbol{\Psi}_{E'} = \left\{\psi_{k}^{E'} : 1 \leq k \leq k' \right\},\\
\boldsymbol{\Psi} &= \left\{\psi_{j}^{E} \times \psi_{k}^{E'} : 1 \leq j \leq j', 1 \leq k \leq k' \right\}.
\end{align*}

\begin{enumerate}[label=(\roman*)]
\item Consider any matrix $T = [\alpha_{jk}]_{\substack{1 \leq j \leq j'\\ 1 \leq k \leq k'}}$ with non-zero scalar entries and set $\alpha (\psi_{j}^{E} \times \psi_{k}^{E'}) = \alpha_{jk}$ for $1 \leq j \leq j'$, $1 \leq k \leq k'$. Let us call $\boldsymbol{\alpha}$ \textbf{factorable} if $T$ is of rank one, i.e., $\alpha_{jk} =\beta_{j}\cdot \gamma_{k}$ for $1 \leq j \leq j'$, $1 \leq k \leq k'$ for some tuples $\boldsymbol{\beta} = (\beta_{j})_{j=1}^{j'}$, $\boldsymbol{\gamma} = (\gamma_{k})_{k=1}^{k'}$ of non-zero scalars. Then $\xi = | \alpha \rangle = \eta \otimes \zeta $ with $\eta = \sum\limits_{j=1}^{j'} \beta_{j} | \psi_{j}^{E} \rangle$, $\zeta = \sum\limits_{k=1}^{k'} \gamma_{k} | \psi_{k}^{E'}\rangle$  and $\xi \neq 0$ if and only if $\eta \neq 0$, $\zeta \neq 0$.

\item Thus we have a restricted analogue of Theorem~\ref{thm-3.3}, viz., If $\alpha$ is factorable via $E$ then $| \alpha \rangle$ is a product vector in the bipartite cut $(E, E')$.

\item The converse is easily seen to hold for the case $j'=2=k'$. Indeed, in this case, there exist $a \in E \cap A$, $a' \in E' \cap A$ with $\psi_{1}^{E}(a) = b_{1} \neq b_{2} = \psi_{2}^{E}(a)$, $\psi_{1}^{E'}(a') = b'_{1} \neq b'_{2} = \psi^{ E'}_2(a')$. So $\boldsymbol{\x}^{\{a,a'\}} (\alpha_{11} \boldsymbol{\x}^{B \smallsetminus \{b_{1}, b'_{1}\}} + \alpha_{1, 2} \boldsymbol{\x}^{B\smallsetminus \{b_{1}, b'_{2}\}} + \alpha_{2, 1} \boldsymbol{\x}^{B\smallsetminus \{b_{2}, b'_{1}\}} + \alpha_{2 2} \boldsymbol{\x}^{B \smallsetminus \{b_{2}, b'_{2}\}} )$  is a part of $F_{\alpha}(\boldsymbol{\x})$. But $F_{\alpha}(\boldsymbol{\x}) = p (\boldsymbol{\x}_{E})q(\boldsymbol{\x}_{E'})$ for some polynomials $p$ and $q$. So it is necessary that for some scalars $\beta_{1}, \beta_{2}, \gamma_{1}, \gamma_{2} $, $\boldsymbol{\x}^{a} \left(\beta_{1}\boldsymbol{\x}^{E \cap B \smallsetminus \{b_{1}\}} + \beta_{2} \boldsymbol{\x}^{E \cap B \smallsetminus\{b_{2}\}} \right)$ is a part of $p(\boldsymbol{\x}_{E})$ and\\ $\boldsymbol{\x}^{a'}(\gamma_{1} \x^{E'\cap B \smallsetminus \{b'_{1}\}} + \gamma_{2} \boldsymbol{\x}^{E'\cap B\smallsetminus \{b_{2}'\}})$ is a part of $q(\x_{E'})$. But that forces $\alpha_{11} = \beta_{1}\gamma_{1}$, $\alpha_{12}= \beta_{1} \gamma_{2},$ $\alpha_{21}=\beta_{2} \gamma_1$ and $\alpha_{22}= \beta_{2}\gamma_{2}$. This gives $\alpha_{11} \alpha_{22} = \alpha_{12} \alpha_{21}$. In other words $[\alpha_{jk}]_{1 \leq j \leq j', 1 \leq k \leq k'}$ has rank one.

\item Several 2 $\times$ 2 matrices like $\begin{bmatrix}0.1 & 0.2\\0.3 & 0.4 \end{bmatrix}$ give that  there is no general analogue of Theorem~\ref{thm-3.3} and also that Theorem~\ref{thm-4.1}(iii)(b) is true even for $v=4$.

\item Indeed, if $\{| \psi_{j}^{E} \rangle : 1 \leq j \leq j' \}$ and $\{| \psi_{k}^{E'} \rangle : 1 \leq k \leq k'\}$ are linearly independent in $\mathcal{H}_{E}$ and $\mathcal{H}_{E'}$ respectively, we can extend them to form bases $\mathcal{B}$ and $\mathcal{B}'$ of $\mathcal{H}_{E}$ and $\mathcal{H}_{E'}$ respectively. Then $\sum_{\substack{\beta \in \mathcal{B} \\ \beta' \in \mathcal{B}'}} \gamma_{\beta \beta'} |\beta \otimes \beta' \rangle$ is a product vector in $\mathcal{H}_{E}\otimes \mathcal{H}_{E'}$ if and only if the matrix $[\gamma_{\beta \beta'}]_{\substack{\beta \in \mathcal{B} \\ \beta' \in \mathcal{B}'}}$, has rank one. As a consequence $| \alpha \rangle$ is a product vector in $\mathcal{H}_{E} \otimes \mathcal{H}_{E'}$ if and only if $[\alpha_{jk}]_{\substack{1 \leq j \leq j'\\ 1 \leq k \leq k'}}$ has rank one. 

\end{enumerate}
\end{example}

\begin{example}\label{example-4.2}
Let $v=2$ and $\boldsymbol{\Psi}$ the NN covering of $\Gamma_{4}$, i.e., $\boldsymbol{\Psi} = \{\psi_{1}, \psi_{2}\}$ with $\psi_{1}(1)=2$, $\psi_{1}(3) =4$, $\psi_{2}(1)=4$, $\psi_{2}(3) =2$, then $\boldsymbol{\Psi}$ is not decomposable.

\begin{enumerate}[label=(\alph*)]
\item Consider $\alpha : \boldsymbol{\Psi} \rightarrow \vertchar[.08ex]{C} \smallsetminus \{0\}$ given by $\alpha(\psi_{1}) = \frac{1}{2}$, $ \alpha(\psi_{2}) = \frac{-1}{2}$. Then $\hat{\alpha} =0$. Further,
\begin{align*}
F_{\alpha}(\boldsymbol{\x}) &= \frac{1}{2} \cdot \frac{1}{2} \left(\x_{2} - \x_{1}\right)\left(\x_{4}-\x_{3} \right) - \frac{1}{2}~\frac{1}{2} \left(\x_{4}-\x_{1}\right) \left(\x_{2}-\x_{3} \right)\\
&= -\frac{1}{4} \left(\x_{1} \x_{4} + \x_{2} \x_{3} \right) + \frac{1}{4} \left(\x_{1} \x_{2} + \x_{3} \x_{4} \right)\\
& = \frac{1}{4} \left(\x_{1}-\x_{3}\right)\left(\x_{2}-\x_{4} \right).
\end{align*}
Hence $| \alpha \rangle \neq 0$ and it is a product vector in the bipartite cut $(A, B)= (\{1,3\}, \{2,4\})$. Therefore, $| \alpha \rangle$ is not genuinely entangled.
Hence the condition $\alpha(\psi) > 0$ for $\psi \in \boldsymbol{\Psi}$, i.e., $\hat{\alpha}=1$ is important in Theorem~\ref{thm-4.1}(a), (c), (d).\\
Indeed the arguments apply to any $\alpha : \boldsymbol{\Psi} \rightarrow \vertchar[.08ex]{C} \smallsetminus\{0\}$ with $\hat{\alpha} = 0$ as in Definition~\ref{definition-4.1}.

\item Consider any $\alpha : \boldsymbol{\Psi} \rightarrow \vertchar[.08ex]{C}$ as in Definition~\ref{definition-4.1} with $\hat{\alpha} \neq 0$.
 $| \alpha \rangle$ has polynomial representation,
\begin{equation*}
\begin{split}
F_{\alpha}(\boldsymbol{\x}) &= \frac{1}{2} \Bigg(\hat{\alpha}\left(\x_{2} \x_{4} + \x_{1} \x_{3}\right)- \alpha \left(\psi_{1}\right) \left(\x_{1} \x_{4} + \x_{2} \x_{3} \right)\\
&\qquad\qquad - \alpha \left(\psi_{2}\right) \left(\x_{1} \x_{2} + \x_{3} \x_{4}\right) \Bigg),
\end{split}
\end{equation*}
which has  six terms. Let, if possible, $| \alpha \rangle$ be a product vector in a bipartite cut $(E, E')$, so that $F_{\alpha}(\boldsymbol{\x}) = p(\boldsymbol{\x}_{E})~~ q(\boldsymbol{\x}_{E'})$. Because supp $F_{\alpha}= \Gamma_{4}$ and $F_{\alpha}(\boldsymbol{\x})$ is homogeneous of degree 2 , both $p$ and $q$ have degree 1. So the number of terms in $p(\boldsymbol{\x}_{E})~~q(\boldsymbol{\x}_{E'})$ can be 3 or 4. Hence $| \alpha \rangle$ is genuinely entangled. Indeed, we have repeated the simpler version of  Theorem  II.1 (iv)(c) in this case.
\end{enumerate}
\end{example}

\subsection{Quantum entanglement of generalized RVB states via polynomial representation in matrix forms}
\label{subsection-4.2}

Let $\boldsymbol{\Psi}, \alpha$ etc. be as in Subsec.~\ref{subsection-4.1} with $| \alpha \rangle \neq 0$ and $\xi = 2^{v/2} | \alpha \rangle$. Consider any bipartite cut $(E, E')$ and $\mathcal{H}$ expressed as $\mathcal{H}(E) \otimes \mathcal(H)(E')$. The entanglement properties in this break-up are well-studied, particularly in terms of the rank of the associated  matrix $A_{\xi}$ in terms of any chosen bases $\mathcal{B}_{E}$ and $\mathcal{B}_{E'}$ for $\mathcal{H}(E)$ and $\mathcal{H}(E')$ respectively. As seen in Subsec. \ref{subsection-4.1}, the polynomial representation $F_{\xi}(\boldsymbol{\x})$ of $\xi$ is a homogeneous polynomial of degree $v$. So discussion in various sections above can be used to advantage. We proceed to work that out. Let $n_{1}=\# E$, $n_{2} = \#E'$ so that $n_{2}= n-n_{1}$. Let $\tilde{\xi} = \frac{1}{|| \xi ||} \xi $, so that $\tilde{\xi}$ is a unit vector.

\vspace{0.5cm}
\textbf{(1) Matrix representation of vectors in $\mathcal{H}$ in the bipartite cut $(E, E')$}. 
\begin{enumerate}[label = (\alph*)]
\item We arrange ${\rm P}_{E}$ and ${\rm P}_{E'}$ in any way, we like. One way that will be useful is to arrange them in blocks according to their sizes with any order within the block. To elaborate, for $0 \leq j \leq n_{1}$, $0 \leq k \leq n_{2}$, let $P_{j}^{E} = \{K \subset E : \# K =j\}$ and $P_{k}^{E'} = \{L \subset E' : \# L =k\}$. Then $P_{n}= P_{E} \times P_{E'}= \bigcup\limits_{\substack{0 \leq j \leq n_{1}\\ 0 \leq k \leq n_{2}}} {\rm P}_{j}^{E} \times {\rm P}_{k}^{E'}$ via $M \subset \Gamma_{n} \rightarrow (K=M \cap E, L=M \cap E')$. This permits us to identity the polynomial $F(\boldsymbol{\x}) = \sum\limits_{M \subset \Gamma_{n}} a_{M} \boldsymbol{\x}^{M}$ with 
\begin{equation}
\sum_{\substack{K \in P_{E},\\ L \in P_{E'}}} a_{K, L} \boldsymbol{\x}^{K} \boldsymbol{\x}^{L} =  \sum_{\substack{0 \leq j \leq n_{1} \\0 \leq k \leq n_{2}}} \left(\sum_{\substack{K \in P_{j}^{E},\\ L \in P_{k}^{E'}}} a_{K, L} \boldsymbol{\x}^{K} \boldsymbol{\x}^{L} \right),
\tag{4.18}\label{eq-4.18}
\end{equation}
which, in turn, gives rise to the corresponding matrix $G_{E} = [a_{K, L}]_{\substack{K \in P_{E}, L \in P_{E'}}}$, and the block matrix $G^{E} =[G_{jk}^{E}]_{\substack{0 \leq j \leq n_{1}\\ 0 \leq k \leq n_{2}}}$ with 
\begin{equation}
G_{jk}^{E}= [a_{K, L}]_{\substack{K \in P_{j}^{E}, L \in P_{k}^{E'}}}, 0 \leq j \leq n_{1}, 0 \leq k \leq n_{2}.
\tag{4.19}\label{eq-4.19}
\end{equation}
Clearly, $F \neq 0$ if and only if $G_{E} \neq 0$ if and only if $G_{j k}^{E} \neq 0$ for some $0 \leq j \leq n_{1}$, and $0 \leq k \leq n_{2}$. Let $\xi_{F}$ be the vector represented by $F$.

\item Suppose $\xi_{F}$ is a unit vector. Then $||G_{E}|| = \max\{| \langle \chi, \xi_{F} \rangle | : \chi~~\text{is a unit product vector in the bipartite cut}~~(E, E')\}$. Further, $||G_{E}||=1$ if and only if $\xi_{F}$ is a product vector in the bipartite cut $(E, E')$ if and only if rank $G_{E}=1$.

\item It is immediate from (b) above that unit vector $\xi_{F}$ is genuinely entangled if and only if $||G_{E}|| < 1 $ for $\phi \neq E \subset_{\neq} \Gamma_{n}$ if and only if rank $G_{E} \geq 2$ for $E$ with $\phi \neq E \subset_{\neq} \Gamma_{n}$.

In this case, the \textbf{generalized geometric measure} (GGM, in short) \cite{ggm1, ggm2} of $\xi_{F}$ is given by 
\begin{equation}
1 -\max_{\substack{\phi \neq E \subset_{\neq}\Gamma_{n}}} || G_{E}||^{2}\tag{4.20}\label{eq-4.20}
\end{equation} 

\item Now suppose that $F$ is a non-zero homogeneous polynomial of degree $d \geq 2$, supp$F = \Gamma_{n}$ and no $X_j$ is a factor of $F(\boldsymbol{X}).$ 

By Theorem~\ref{thm-2.1}(iv), $\xi_{F}$ is a product vector in the bipartite cut $(E, E')$ if and only if $F(\boldsymbol{\x}) = p(\boldsymbol{\x}_{E})q(\boldsymbol{\x}_{E'})$ for some polynomials $p$ and $q$ of degree $d_{1}$ and $d_{2}$ respectively with $1 \leq d_{1} < d$, $d_{2}=d-d_{1}$. In view of (b) above, this is true if and only if for some $d_{1}, d_{2}$ with $1 \leq d_{1} < d$, $d_{2}=d-d_{1}$, $G_{jk} =0$ for $(j,k) \neq (d_{1}, d_{2})$ and $||G_{d_{1}, d_{2}}||=|| \xi_{F}||$ if and only if for some $d_{1}, d_{2}$ with $1 \leq d_{1} < d$, $d_{2}=d-d_{1}$, $G_{jk}=0$ for $(j,k) \neq (d_{1}, d_{2})$ and rank $G_{d_{1},d_{2}}=1$.
 
 \item In case (d) happens, $G_{j~0} =0$ and $G_{0~k}=0$ for $0 \leq j \leq n_{1}$, $0 \leq k \leq n_{2}$; in other words $F(\boldsymbol{\x})$ has no terms of the type $\lambda \boldsymbol{\x}^{K}$ or $\lambda \boldsymbol{\x}^{L}$ with non-zero scalar $\lambda$ and $K \subset E$ or $L \subset E'$.
 
 \item Let $F$ be as in the first paragraph of (d) above. If it does not satisfy the conditions in (d) stated in the second paragraph of (d) for any $\phi \neq E \subset_{\neq} \Gamma_{n}$, then $\xi_{F}$ is genuinely entangled.
 
 \item It is clear from (d) and (e) above that we can limit our choice of the bipartite cuts $(E, E')$ for which to check the details. One such restriction is that for $\lambda\boldsymbol{\x}^{M}$ present in $F(\boldsymbol{\x})$ with $\lambda\neq 0$, we must have $K = M \cap E \neq \phi \neq M \cap E' =L$. 
 Moreover, it is enough to take $0 \leq j \leq \min \{n_{1}, d\}$ and \,$0 \leq k \leq \min \{n_{2}, d\}$. 

\end{enumerate}

\textbf{(2) Quantum entanglement of $\xi$ via matrix representation}. 

We apply \ref{subsection-4.2} (1) above to $\xi$ via $F_{\xi}(\boldsymbol{\x})$.
\begin{enumerate}[label=(\alph*)]
\item $F_{\xi}$ satisfies the conditions for $F$ in the first paragraph of    \ref{subsection-4.2}(1)(d) with $d=v$, $n_{1} + n_{2} =n = 2v$, so either $n_{1}\leq v \leq n_{2}$ or $n_{2} \leq v \leq n_{1}$.

In view of  \ref{subsection-4.2}(1)(g), if $0 \neq \lambda \boldsymbol{\x}^{M}$ is part of $F_{\xi}(\boldsymbol{\x})$ for $M=L \subset E'$ in the first case and $M = K \subset E$ in the second case, then $\xi$ is not a product vector in the bipartite cut $(E, E')$.

\item We refer to   \ref{subsection-4.2}(1)(e) and take $F= F_{\xi}$. For $K\subset E$, $L \subset E'$, $A_{1} \subset A$, $B_{1} \subset B$, we can have $K \cup L = A_{1} \cup (B \smallsetminus B_{1})$ if and only if $A_{1} = (K \cup L) \cap A$, $B_{1}= B \smallsetminus (K \cup L)\cap B$ if and only if $K =(E \cap A_{1}) \cup (E \cap (B\smallsetminus B_{1}))$, $L=(E' \cap A_{1})\cup (E' \cap (B \smallsetminus B_{1}))$. As a consequence, $E \smallsetminus K = (E \cap (A \smallsetminus A_{1})) \cup (E \cap B_{1})$, $E' \smallsetminus L = E' \cap (A \smallsetminus A_{1}) \cup (E' \cap B_{1})$. As a sample of applications (4.3) gives
\begin{align*}
a_{K L} &= (-1)^{\#(K \cup L)\cap A} \alpha \left((K \cup L) \cap A, B \smallsetminus (K \cup L) \cap B \right) = (-1)^{\# A_{1}} \alpha(A _{1}, B_{1})\\
 & = (-1)^{v} (-1)^{\# (A \smallsetminus (K \cup L) \cap A) }\alpha (A \smallsetminus (K \cup L) \cap A, (K \cup L )\cap B)\\
 & = (-1)^{v} a_{E \smallsetminus K, E' \smallsetminus L}.\tag{4.21}\label{eq-4.21}
\end{align*}

\item It helps to make the proofs and discussion shorter and transparent if we order $P_{j}^{E}$'s and $P_{k}^{E'}$s in a convenient manner for $0 \leq j \leq n_{1}$, $0 \leq k \leq n_{2}$. We first note that the complementation maps $S \rightarrow E \smallsetminus S$ and $T \rightarrow E' \smallsetminus T$ set up one-to-one correspondence between $P_{j}^{E}$ and $P_{n_{1}-j}^{E}$ and between $P_{k}^{E'}$ and $P_{n_{2}-k}^{E'}$ respectively for $0 \leq j \leq n_{1}$, $0 \leq k \leq n_{2}$. Indeed for $n_{1}$ even, say $n_{1}= 2v_{1}$, so is $n_{2}$ say $2v_{2}$ with $v_{2}= v-v_{1}$; and in this case, for $j=v_{1}$, $k=v_{2}$, we have $P_{j}^{E}= P_{n_{1}-j}^{E}$ and $P_{k}^{E'} = P_{n_{2}-k}^{E'}$ and the complementation maps are only permutations of order 2.

\item For $0 \leq j < \frac{n_{1}}{2}$, $0 \leq k < \frac{n_{2}}{2}$, we order $P_{j}^{E}$ and $P_{k}^{E'}$ the way we like and then follow the order in $P_{n_{1}-j}^{E}$ and $P_{n_{2}-k}^{E'}$ dictated by the complementation maps. This permits us, in view of~\eqref{eq-4.21} in (b) above, to write
\begin{equation}
G_{j,k}^{E} = (-1)^{v} G_{n_{1}-j, n_{2}-k}^{E}.\tag{4.22}\label{eq-4.22}
\end{equation}

\item If $n_{1}$ is odd (equivalently, $n_{2}$ is odd),  for $0 \leq j \leq n_{1}$ and $0 \leq k \leq n_{2}$, we have $(j,k) \neq (n_{1}-j, n_{2}-k)$ and, therefore, at least two non-zero blocks of the form $G_{j,k}$ and $G_{n_{1}-j, n_{2}-k}$ for some $0\leq j \leq n_{1}$, $0 \leq k \leq n_{2}$. 

\item By \ref{subsection-4.2}(1)(d), $\xi$ cannot be a product vector for the bipartite cut $(E, E')$ if $n_{1}$ is odd (equivalently, $n_{2}$ is odd) in view of $(e)$ above. This is in line with Theorem~\ref{thm-3.1}(ii)(a), second part.

\item Now suppose $n_{1}$ is even, say, $n_{1}=2v_{1}$ with $1 \leq v_{1} < v$.  $n_{2}=2v_{2}$ with $v_{2}=v-v_{1}$.

Consider any $K \in P_{v_{1}}^{E}$, $L \in P_{v_{2}}^{E'}$. Then $E \smallsetminus K \in P_{v_{1}}^{E}$ and $E'\smallsetminus L \in P_{v_{2}}^{E'}$. So after relabelling amongst these four, if the need be,
\begin{align*}
    M_{K,~L} = 
\begin{bmatrix}
a_{K,E' \smallsetminus L} &  a_{E \smallsetminus K, E'\smallsetminus L}\\ 
a_{K,L} & a_{E \smallsetminus K, L} 
\end{bmatrix}
\tag{4.23}
\label{eq-4.23}
\end{align*}

is a 2 $\times$ 2-minor in $G_{v_{1}, v_{2}}^{E}$. By~\eqref{eq-4.21}, we have $a_{E \smallsetminus K, ~L} = (-1)^{v} a_{K, E'\smallsetminus L}$ and, therefore, the determinant of $M_{K, L}$ has the value $m_{K,L} = (-1)^{v} \left((a_{K, E' \smallsetminus L})^{2} -\right.$ $\left.(a_{K, L})^{2} \right)$.

Now suppose $\xi$ is a product vector in the bipartite cut $(E, E')$. Because of  \ref{subsection-4.2}(1)(d) and (e), we have $G_{j~k}^{E}=0$ for $(j,k) \neq (v_{1}, v_{2}) $, $0 \leq j \leq n_{1}$, $0 \leq k \leq n_{2}$ and $G_{v_{1}, v_{2}}^{E}$ has rank one. Hence for $(K, L) \in P_{v_{1}}^{E} \times P_{v_{2}}^{E'}$, we have $m_{K_{1}~L} = 0$, i.e., $a_{K, E'\smallsetminus L} \pm a_{K, L} $ and as a consequences, $M_{KL}= a_{KL}\begin{pmatrix}\pm 1 & (-1)^{v} \\ 1 & \pm (-1)^{v}\end{pmatrix}$. Moreover, at least one such minor has all entries non-zero. Part (b) tells us how to shift the discussion in terms of ($A_{1}, B_{1}$) etc..

\item We continue with the case that $n_{1}$ is even as in the item (g) above.

Fix any $x\in E$. Let $P_{v_{1}, x}^{E} = \{K \in P_{{\nu_{1}}}^{E} : x\in K\}$. Then $\# P_{v_{1}, x}^{E}  = \,^{2\nu_{1}-1}C_{ \nu_1-1}=\,{\left(\frac{1}{2}\right)\, ^{ 2\nu_1}}C_{\nu_1}=c_{{\nu_{1}}}$, say. Order $P_{v_{1}, x}^{E}$ in any manner that we like.

Then $P_{v_{1}}^{E} \smallsetminus P_{v_{1}, x}^{E} = \{K \in P_{v_{1}}^{E} : x \notin K\} = \{E \smallsetminus K : K \in P_{v_{1}, x}^{E}\}$.

We order $P_{v}^{E} \smallsetminus P_{v_{1}, x}^{E}$ in the opposite order to that of $P_{v_{1}, x}^{E}$ dictated by the complementation map.

Similarly, we fix a $y \in E'$ and follow the procedure above  for $(E, x)$ for $(E', y)$. We put $c_{\nu_{2}}' = \,{\left(\frac{1}{2}\right)\, ^{ 2\nu_2}}C_{\nu_2}$. 

Now consider the matrix $G_{v_{1}, v_{2}}^{E}$. Then the 2 $\times$ 2 minors $M_{K,L}$ as in (g) above are placed symmetrically above the horizontal and vertical lines through the point $(c_{v_{1}} + \frac{1}{2}, c_{v_{2}}' + \frac{1}{2})$ in the matrix. We call them symmetrical minors.

A geometrical interpretation for (g) is that if $\xi$ is a product vector  in the bipartite cut $(E, E')$,  at least one such minor has non-zero entries, all of them have the form
\begin{align*}
    M_{K, L} = a_{K, L}\begin{bmatrix}\pm 1 & (-1)^{v}\\ 1 & \pm (-1)^{v} \end{bmatrix}, K \in P_{v_{1}}^{E}, L \in P_{v_{2}}^{E'},
    \tag{4.24}
    \label{eq-4.24}
\end{align*}
and $G_{j k}^{E}=0$ for $(j,k) \neq (v_{1}, v_{2})$.

\item We consider the case, when $E = A$ and $E'=B$ motivated by Example~\ref{example-4.2}. By \ref{subsection-4.2}(1)(g), it needs  consideration if and only if $\boldsymbol{\x}^{E}$ and $\boldsymbol{\x}^{E'}$ do not occur in $F_{\xi}(\boldsymbol{\x})$ if and only if $\hat{\alpha} =0$. Hence, in Subsection IV A, we consider the case when $\hat{\alpha}=0$.

Next, by  (f) above, $\xi$ is not a product vector in the bipartite cut $(E, E')$ if $\nu$ is odd. So we confine our attention to the case that $\nu$ is even, say, $v=2v_{1}$. Then $\xi$ is a product vector in the cut $(E, E')$ if and only if $G_{jk}^{E}=0$ for $(j, k)\neq (v_{1}, v_{1})$ and $G_{v_{1}, v_{1}}^E$ has rank one. We note that $G^{E}_{\nu_{1}, \nu_1}=(-1)^{\nu_{1}}[\alpha (A_{1}, B_{1})]_{ A_{1} \in P^{A}_{\nu_1}, B_1 \epsilon P^{B}_{\nu_{1}}}$ and $G^{E}_{jk} =0$ for $(j,k) \neq (\nu_{1}, \nu_{1})$.

We are now ready to consolidate the discussion as our next theorem.

\end{enumerate}


\begin{thm}\label{thm-4.2}
Let $\alpha, \xi, F_{\xi}, E , G^{E}$ etc. be as above.
\begin{enumerate}[label=(\roman*)]
 \item If $\xi$ is a product vector in the bipartite cut $(E, E')$,  the followings hold.
\begin{enumerate}[label= (\alph*)]
\item $n_{1}=\# E$ is even, say, $2v_{1}$, $n_{2}= \# E'$ is even, say=$2v_{2}$ with $1 \leq v_{1} < v$, $v_{2}= v-v_{1}$.

\item If $0 \neq \lambda {\boldsymbol{\x}}^{M}$ is a part of $F_{\xi}(\boldsymbol{\x})$, $K = M \cap E \neq  \phi \neq M \cap E' = L$, say.

\item $G_{jk}^E =0$ for $(j,k) \neq (v_{1}, v_{2})$.

\item $||G_{v_{1}, v_{2}}^{E}||= |\xi |||$.

\item $G_{v_{1}, v_{2}}^{E}$ has rank one.

\item All symmetrical 2 $\times$ 2 minors have the form 
$$
M_{K, L} = a_{K L} \begin{bmatrix}\pm 1 & (-1)^{v}\\ 1  & \pm(-1)^{v} \end{bmatrix},
$$
and at least one $M_{K, L}$ is non-zero.
\end{enumerate}

\item Suppose $E$ is as in (i) (a) above. If (i) (c) together with (i) (d) or with (i) (e) are satisfied then $\xi$ is a product vector in the bipartite cut $(E, E')$.

\item If for each $E$ as in (i) (a) above, (i) (c) or (i) (d) or (i) (e) or (i) (f) is not satisfied,  $\xi$ is genuinely entangled.

\item $\xi$ is a product vector in the cut $(A, B)$ if and only if the following hold.

\begin{enumerate}[label= (\alph*)]
\item $\hat{\alpha} = 0$.

\item $v$ is even, say $v=2v_{1}$.

\item For $0 \leq j \leq v$, $j \neq v_{1}$, $[\alpha(A_{1}, B_{1})]_{A_{1} \in P_{j}^{A}, B_{1}\in P_{j}^{B}}$ is zero.

\item $[\alpha(A_{1}, B_{1})]_{A_{1} \in P_{\nu_1}^{A}, B_{1}\in P_{\nu_1}^{B}}$  has rank one or has norm equal to that of $\xi.$

\end{enumerate}

\end{enumerate}

\end{thm}

\subsection{Quantum entanglement of generalized RVB states for decomposable sets of coverings via polynomial representations}
\label{subsection-4.3}

Let $\boldsymbol{\Psi}$ be a set of coverings with $\# \boldsymbol{\Psi} \geq 2$ such that $\boldsymbol{\Psi}$ is decomposable  via $E$ for some $\phi \neq E \subset_{\neq} \Gamma_{n}$. In view of Theorem~\ref{thm-4.1} it is enough to confine our attention to $v \geq 4$ and $\boldsymbol{\Psi}$ neither flat nor a pole. We consider any $\alpha : \boldsymbol{\Psi} \rightarrow \vertchar[.08ex]{C} \smallsetminus \{0\}$ with $| \alpha \rangle \neq  0$ and try to find conditions under which $| \alpha \rangle$ is a product vector in the bipartite cut $(E, E')$ using the notation, terminology and results in previous sections particularly Section III.

It will be convenient to consider $\xi = 2^{v/2} | \alpha \rangle$ which has same entanglement properties as $| \alpha \rangle $ and its polynomial representation $F_{\xi}(\boldsymbol{\x}) =2^{v/2}F_{\alpha}(\boldsymbol{\x})$. 

\vspace{0.5cm}

\textbf{Master equations for \(\xi\).}

\begin{enumerate}[label=(\alph*)]
\item Part (d) of \ref{subsection-3.1}(6) undergoes obvious changes to give a modified version using~ Subsec. \ref{subsection-4.1}, particularly \eqref{eq-4.8} above to give
\begin{align*}
F_{\xi}(\boldsymbol{\x}) &= \hat{\alpha}\left(\boldsymbol{\x}^{B} + (-1)^{\nu} \boldsymbol{\x}^{ A}\right)\\
 & + \hat{\alpha}\left((-1)^{v_{1}} \boldsymbol{\x}^{E \cap A} \boldsymbol{\x}^{E' \cap B} + (-1)^{v_{2}} \boldsymbol{\x}^{E \cap B} \boldsymbol{\x}^{E' \cap A}\right)\\
 & + \left(\boldsymbol{\x}^{E \cap B} + (-1)^{v_{1}} \boldsymbol{\x}^{E \cap A}\right) \left(\sum_{\psi \in \boldsymbol{\Psi}} \alpha(\psi) q_{\psi}^{0} (\boldsymbol{\x}_{E'}) \right)\\
 & + \left(\sum_{\psi \in \boldsymbol{\Psi}} \alpha (\psi) p_{\psi}^{0} (\boldsymbol{\x}_{E}) \right) \left(\boldsymbol{\x}^{E' \cap B} + (-1)^{v_{2}} \boldsymbol{\x}^{E' \cap A} \right)\\
 & + \sum_{\psi \in \boldsymbol{\Psi}} \alpha(\psi) p_{\psi}^{0} (\boldsymbol{\x}_{E}) q_{\psi}^{0}(\boldsymbol{\x}_{E'}).
 \tag{4.25}\label{eq-4.25}
\end{align*}
We will utilize \eqref{eq-3.29} in Subsec. \eqref{subsection-3.1}  and compare it with \eqref{eq-4.25} above to have that $F_{\xi}(\boldsymbol{\x}) = p(\boldsymbol{\x}_{E}) q(\boldsymbol{\x}_{E'})$ if and only if 
\begin{equation}
\beta\beta' = \hat{\alpha} = (-1)^{v} \delta \delta' = (-1)^{v_{1}} \delta\beta' = (-1)^{v_{2}} \delta' \beta,
\tag{4.26}\label{eq-4.26}
\end{equation}
\begin{equation}
\beta q_{0} (\boldsymbol{\x}_{E'}) = \sum_{\psi \in \boldsymbol{\Psi}} \alpha (\psi) q_{\psi}^{0} (\boldsymbol{\x}_{E'}) = (-1)^{v_{1}} \delta q_{0} (\boldsymbol{\x}_{E'}),
\tag{4.27}\label{eq-4.27}
\end{equation}
\begin{equation}
\beta' p_{0} (\boldsymbol{\x}_{E}) = \sum_{\psi \in \boldsymbol{\Psi}} \alpha (\psi) p_{\psi}^{0} (\boldsymbol{\x}_{E}) = (-1)^{v_{2}} \delta' p_{0} (\boldsymbol{\x}_{E}),
\tag{4.28}\label{eq-4.28}
\end{equation}
\begin{equation}
\text{and}\quad p_{0}(\boldsymbol{\x}_{E}) q_{0}(\boldsymbol{\x}_{E'}) = \sum_{\psi \in \boldsymbol{\Psi}} \alpha (\psi) p_{\psi}^{0} (\boldsymbol{\x}_{E}) q_{\psi}^{0} (\boldsymbol{\x}_{E'}).
\tag{4.29}\label{eq-4.29}
\end{equation}
\item Let us first consider the case when 
the condition in \(F_\xi(\boldsymbol{\x}) = p(\boldsymbol{\x}_{E})q(\boldsymbol{\x}_{E'})\)
does happen together with $\hat{\alpha} \neq 0$.

\eqref{eq-4.26}, $\delta = (-1)^{v_{1}}\beta$, $\delta' =(-1)^{v_{2}}\beta'$ are all non-zero. So we may take $\beta = \hat{\alpha}$, $\delta = (-1)^{v_{1}} \hat{\alpha}$, $\beta' =1$, $\delta' = (-1)^{v_{2}}$ and multiply $p_{0}(\boldsymbol{\x}_{E})$ and $q_{0}(\boldsymbol{\x}_{E'})$ by suitable scalars if the need be and retain the same notation for them. Then \eqref{eq-4.27} and \eqref{eq-4.28}  give
$$
\hat{\alpha} q_{0} (\boldsymbol{\x}_{E'}) = \sum_{\psi \in \boldsymbol{\Psi}} \alpha(\psi) q_{\psi}^{0}(\boldsymbol{\x}_{E'}),
$$
$$
\text{and}~~ p_{0} (\boldsymbol{\x}_{E}) = \sum_{\psi \in \boldsymbol{\Psi}} \alpha(\psi) p_{\psi}^{0}(\boldsymbol{\x}_{E}).
$$
This turns \eqref{eq-4.29} into
\begin{equation}
\left(\sum_{\psi \in \boldsymbol{\Psi}} \alpha(\psi) p_{\psi}^{0} (\boldsymbol{\x}_{E}) \right) \left(\sum_{\psi \in \boldsymbol{\Psi}} \alpha(\psi) q_{\psi}^{0} (\boldsymbol{\x}_{E'}) \right) = \hat{\alpha} \sum_{\psi \in \boldsymbol{\Psi}} \alpha(\psi) p_{\psi}^{0} (\boldsymbol{\x}_{E}) q_{\psi}^{0} (\boldsymbol{\x}_{E'}).
\tag{4.30}\label{eq-4.30}
\end{equation}

\item Arguments in (b) can be reversed and, therefore, for $\hat{\alpha} \neq 0$, we call \eqref{eq-4.30},  the \textbf{Master equation} for $F_{\xi}(\boldsymbol{\x})$ to be expressible as product of some polynomials $p(\boldsymbol{\x}_{E})$ and $q(\boldsymbol{\x}_{E'})$ in analogy with \eqref{eq-3.34}. We emphasize just as in Subsec. 
\ref{subsection-3.1}, that for $\hat{\alpha} \neq 0$, Master Equation does not involve $p(\boldsymbol{\x}_{E})$ or $q(\boldsymbol{\x}_{E'})$ and, in fact, they are determined in an explicit way as explained in (a) above. Moreover, it is an equivalent condition for $\xi$ to be a product vector in the bipartite cut $(E, E')$.

 We could also have analogues of \eqref{eq-3.37} and \eqref{eq-3.38}  by replacing $\# T_{j}$ by $\alpha^{\#}(T_{j}) =\sum_{k \in T_{j}} \alpha (\psi_{j}^{E} \times \psi_{k}^{E'})$, $\# S_{k}$ by $\alpha^{\#}(S_{k}) =\sum_{j\in S_{k}} \alpha (\psi_{j}^{E} \times \psi_{k}^{E'})$, $s$ by $\hat{\alpha}$. However, these Master Equations have restricted applications compared to \eqref{eq-4.30}, when it comes to developing analogues of Theorems depending on these Master Equations.

\item We note the simplified versions of certain polynomials in \eqref{eq-4.27} to \eqref{eq-4.30} in terms of $p_{j}^{0}(\boldsymbol{\x}_{E})=p_{\psi^{E}_{j}}^{0}(\boldsymbol{\x}_{E})$, $q_{k}^{0}(\boldsymbol{\x}_{E'}) = q_{\psi^{E'}_{k}}^{0}(\boldsymbol{\x}_{E'})$, $1 \leq j \leq j'$, $1 \leq k \leq k'$ for record and further use regardless of $\hat{\alpha} \neq  0$ or $\hat{\alpha}=0$. 
\begin{equation}
P_{\alpha}^{0}(\boldsymbol{\x}_{E}) = \sum_{\psi \in \boldsymbol{\Psi}} \alpha (\psi) p_{\psi}^{0}(\boldsymbol{\x}_{E}) = \sum_{j=1}^{j'} \alpha^{\#} (T_{j}) p_{j}^{0}(\boldsymbol{\x}_{E}),\tag{4.31}\label{eq-4.31}
\end{equation}
\begin{equation}
q_{\alpha}^{0}(\boldsymbol{\x}_{E'}) = \sum_{\psi \in \boldsymbol{\Psi}} \alpha(\psi) q_{\psi}^{0}(\boldsymbol{\x}_{E'}) =\sum_{k=1}^{k'} \alpha^{\#}(S_{k}) q_{k}^{0} (\boldsymbol{\x}_{E'}),\tag{4.32}\label{eq-4.32}
\end{equation}
\begin{align*}
F_{\alpha}^{0}(\boldsymbol{\x}) &= \sum_{\psi \in \boldsymbol{\Psi}}\alpha (\psi) p_{\psi}^{0}(\boldsymbol{\x}_{E}) q_{\psi}^{0}(\boldsymbol{\x}_{E'}) = \sum_{j=1}^{j'} p_{j}^{0}(\boldsymbol{\x}_{E}) \left(\sum_{k \in T_{j}} \alpha(\psi_{j}^{E} \times \psi_{k}^{E'}) q_{k}^{0}(\boldsymbol{\x}_{E'})\right)\\
& = \sum_{k=1}^{k'} \left(\sum_{j \in S_{k}}  \alpha (\psi_{j}^{E} \times \psi_{k}^{E'}) p_{j}^{0} (\boldsymbol{\x}_{E})\right) q_{k}^{0}(\boldsymbol{\x}_{E'}),
\tag{4.33}\label{eq-4.33}
\end{align*} 
and also note that
\begin{equation}
\hat{\alpha} = \sum_{j=1}^{j'} \alpha^{\#}(T_{j}) = \sum_{k=1}^{k'} \alpha^{\#}(S_{k}).\tag{4.34}\label{eq-4.34}
\end{equation}
Hence, \eqref{eq-4.30} takes two equivalent forms analogous to \eqref{eq-3.37} and \eqref{eq-3.38}.

\item We now come to the case that 
the condition in \(F_\xi(\boldsymbol{\x}) = p(\boldsymbol{\x}_E)q(\boldsymbol{\x}_{E'})\)
does happen together with $\hat{\alpha} =0$. 
We see that
\begin{align*}
F_{\xi}(\boldsymbol{\x})& = \left(\boldsymbol{\x}^{E \cap B}  + (-1)^{v_{1}} \boldsymbol{\x}^{E \cap A}\right) q_{\alpha}^{0} (\boldsymbol{\x}_{E})\\
 &\qquad + p_{\alpha}^{0} (\boldsymbol{\x}_{E}) (\boldsymbol{\x}^{E' \cap B} + (-1)^{v_{2}} \boldsymbol{\x}^{E' \cap A}) + F_{\alpha}^{0}(\boldsymbol{\x}).\tag{4.35}\label{eq-4.35}
\end{align*}
 Next, \eqref{eq-4.26} takes the form
 \begin{equation}
 \beta\beta' = 0 = \delta\delta' = \delta \beta' = \delta'\beta.\tag{4.36}\label{eq-4.36}
 \end{equation}
The trivial solution $\beta = \delta = \beta' = \delta' = 0$ turns  \eqref{eq-4.27} and \eqref{eq-4.28} into $q_{\alpha}^{0} (\boldsymbol{\x}_{E'}) = 0$ and $p_{\alpha}^{0}(\boldsymbol{\x}_{E})=0$ respectively. This forces $F_{\xi}(\boldsymbol{\x})$ as in \eqref{eq-4.28} reduce to $F_{\alpha}^{0}(\boldsymbol{\x})$ which has to be non-zero. So \eqref{eq-4.29} persists as
\begin{equation}
0 \neq p_{0}(\boldsymbol{\x}_{E}) q_{0}(\boldsymbol{\x}_{E'}) = F_{\alpha}^{0}(\boldsymbol{\x}),
\tag{4.37}\label{eq-4.37}
\end{equation}
which is not really helpful to reduce the task.

So we have two subcases to look into, viz., $(\beta, \delta) =(0,0) \neq (\beta', \delta')$ and $(\beta', \delta') = (0,0) \neq (\beta, \delta)$. They are similar in the sense that we have essentially to interchange $p$'s with $q$'s to get from one to the other. Let us consider the subcase $(\beta, \delta) =(0,0) \neq (\beta', \delta')$. Because of $(\beta, \delta) = (0,0)$, \eqref{eq-4.27} becomes 
\begin{equation}
q_{\alpha}^{0}(\boldsymbol{\x}_{E'})= 0.
\tag{4.38}\label{eq-4.38}
\end{equation}  

Let, if possible, $p_{0}(\boldsymbol{\x}_{E})=0$. Then \eqref{eq-4.28} gives that $p_{0}(\boldsymbol{\x}_{E})=0 = p_{\alpha}^{0}(\boldsymbol{\x}_{E})$. But that forces $F_{\alpha}^{0}(\boldsymbol{\x})=0$ in view of \eqref{eq-4.29}, which in turn gives $F_{\xi}(\boldsymbol{\x})=0$ using \eqref{eq-4.35}. But that is not so. As a consequence, $p_{0}(\boldsymbol{\x}_{E}) \neq  0$. This combined with $(\beta', \delta') \neq (0,0)$ and \eqref{eq-4.28} renders 
\begin{equation}
\beta'=(-1)^{v_{2}} \delta' \neq 0, ~~p_{0}(\boldsymbol{\x}_{E}) = \frac{1}{\beta'} p_{\alpha}^{0}(\boldsymbol{\x}_{E}) \neq 0.\tag{4.39}\label{eq-4.39}
\end{equation} 
This turns 
\eqref{eq-4.29} 
into
\begin{equation}
p_{\alpha}^{0} (\boldsymbol{\x}_{E}) \left(\frac{1}{\beta'} q_{0} (\boldsymbol{\x}_{E'}) \right) = F_{\alpha}^{0}(\boldsymbol{\x}).\tag{4.40}\label{eq-4.40}
\end{equation}
On the other hand, for $(\beta', \delta') = (0,0) \neq (\beta, \delta)$, we obtain the following counter parts of \eqref{eq-4.38} to \eqref{eq-4.40}:
\begin{align*}
&p_{\alpha}^{0} (\boldsymbol{\x}_{E}) = 0,\tag{4.41}\label{eq-4.41}\\
\beta &= (-1)^{v_{1}} \delta \neq 0, \quad q_{0}(\boldsymbol{\x}_{E'}) = \frac{1}{\beta} ~~q_{\alpha}^{0} (\boldsymbol{\x}_{E'}) \neq 0,\tag{4.42}\label{eq-4.42}\\
&\left(\frac{1}{\beta} p_{0}(\boldsymbol{\x}_{E}) \right)~~q_{\alpha}^{0} (\boldsymbol{\x}_{E'}) = F_{\alpha}^{0}(\boldsymbol{\x}).\tag{4.43}\label{eq-4.43}
\end{align*} 

\item The arguments in (e) can be reversed, but the Eqs. \eqref{eq-4.40} and \eqref{eq-4.43} do leave the task of determining $q_{0}(\boldsymbol{\x}_{E'})$ and $p_{0}(\boldsymbol{\x}_{E})$  respectively while checking their validity. We can simplify them a bit by incorporating the constant $\frac{1}{\beta'}$ and $\frac{1}{\beta}$ with $q_{0}(\boldsymbol{\x}_{E'})$ and $p_{0}(\boldsymbol{\x}_{E})$ respectively.

We write the final outcome in a consolidated form as a Theorem and leave it to the reader to formulate further analogues of Theorems and discussion etc in Subsec. \ref{subsection-3.1}.
\end{enumerate}

\begin{thm}\label{thm-4.3}
Let $\boldsymbol{\Psi}$, $\alpha$, $\xi$ etc. be as in the opening paragraph of Subsec.~\ref{subsection-4.3}.

$\xi$ is a product vector in the bipartite cut $(E, E')$ if and only if the Master Equations listed below case-wise hold.
\begin{enumerate}[label = (\roman*)]
\item \textbf{ME} for ${\boldsymbol{\hat{\alpha} \neq 0}}$.
$p_{\alpha}^{0}(\boldsymbol{\x}_{E}) ~~q_{\alpha}^{0} (\boldsymbol{\x}_{E'}) = \hat{\alpha} F_{\alpha}^{0} (\boldsymbol{\x})$, i.e.,
\begin{equation}
\left(\sum_{\psi \in \boldsymbol{\Psi}} \alpha(\psi)p_{\psi}^{0} (\boldsymbol{\x}_{E})\right) \left(\sum_{\psi \in \boldsymbol{\Psi}}\alpha(\psi)  q_{\psi}^{0} (\boldsymbol{\x}_{E'})\right) = \hat{\alpha} \sum_{\psi \in \boldsymbol{\Psi}} \alpha(\psi) p_{\psi}^{0} (\boldsymbol{\x}_{E}) q_{\psi}^{0}(\boldsymbol{\x}_{E'}).\tag{4.44}\label{eq-4.44}
\end{equation}
In this case, $F_{\xi}(\boldsymbol{\x}) = p(\boldsymbol{\x}_{E}) q(\boldsymbol{\x}_{E'})$ with
\begin{align*}
p(\boldsymbol{\x}_{E}) &= \hat{\alpha} \left(\boldsymbol{\x}^{E \cap B} + (-1)^{v_{1}} \boldsymbol{\x}^{E \cap A} \right) + p_{\alpha}^{0}(\boldsymbol{\x}_{E}),\\
q(\boldsymbol{\x}_{E'}) &= \boldsymbol{\x}^{E' \cap B} + (-1)^{v_{2}} \boldsymbol{\x}^{E' \cap A} + \frac{1}{\hat{\alpha}} q_{\alpha}^{0} (\boldsymbol{\x}_{E'}).\tag{4.45}\label{eq-4.45}
\end{align*}

\item \textbf{Trichotomy} for ME for $\hat{\alpha} = 0$. One and only one of the following hold.

\textbf{ME(a)} $p_{\alpha}^{0} (\boldsymbol{\x}_{E}) = 0$, $q_{\alpha}^{0} (\boldsymbol{\x}_{E'}) = 0$,
\begin{equation}
0 \neq p_{0} (\boldsymbol{\x}_{E}) q_{0}(\boldsymbol{\x}_{E'}) = F_{\alpha}^{0} (\boldsymbol{\x}) = F_{\xi}(\boldsymbol{\x}),\tag{4.46}\label{eq-4.46}
\end{equation}
for some polynomials $p_{0}$ and $q_{0}$ (with $\boldsymbol{\x}^{E \cap A}$ and $\boldsymbol{\x}^{E \cap B}$ not occurring in $p_{0}(\boldsymbol{\x}_{E})$ and $\boldsymbol{\x}^{E \cap A}$ and $\boldsymbol{\x}^{E' \cap B}$ not occurring in $q_{0} \boldsymbol{\x}_{E'})$.

\textbf{ME(b)} $q_{\alpha}^{0} (\boldsymbol{\x}_{E'}) = 0$, $p_{\alpha}^{0}(\boldsymbol{\x}_{E}) \neq 0$,
\begin{equation}
p_{\alpha}^{0}(\boldsymbol{\x}_{E}) q_{0}(\boldsymbol{\x}_{E'}) = F_{\alpha}^{0} (\boldsymbol{\x}), \tag{4.47}\label{eq-4.47}
\end{equation}
for some polynomial $q_{0}$ (with $\boldsymbol{\x}^{E' \cap B}$ and $\boldsymbol{\x}^{E' \cap A}$ not occurring in $q_{0}(\boldsymbol{\x}_{E'})$).
In this case, \begin{equation}F_{\xi}(\boldsymbol{\x}) = p_{\alpha}^{0} (\boldsymbol{\x}_{E'}) \left(\boldsymbol{\x}^{E' \cap B} + (-1)^{v_{2}} \boldsymbol{\x}^{E' \cap A}  + q_{0} (\boldsymbol{\x}_{E'})\right).\tag{4.48}\label{eq-4.48}\end{equation}

\textbf{ME(c)} $p_{\alpha}^{0} (\boldsymbol{\x}_{E})= 0$, $q_{\alpha}^{0} (\boldsymbol{\x}_{E'}) \neq 0$,
\begin{equation}
p_{0}(\boldsymbol{\x}_{E}) q_{\alpha}^{0}(\boldsymbol{\x}_{E'}) = F_{\alpha}^{0}(\boldsymbol{\x}),\tag{4.49}\label{eq-4.49}
\end{equation} 
for some polynomial $p_{0}$ (with $\boldsymbol{\x}^{E \cap B}$ and $\boldsymbol{\x}^{E \cap A}$) not occurring in $p_{0}(\boldsymbol{\x}_{E})$.

In this case, 
\begin{equation}
F_{\xi}(\boldsymbol{\x}) = \left(\boldsymbol{\x}^{E \cap B} + (-1)^{v_{1}} \boldsymbol{\x}^{E \cap A} + p_{0} (\boldsymbol{\x}_{E}) \right) q_{\alpha}^{0}(\boldsymbol{\x}_{E'}).\tag{4.50} \label{eq-4.50}
\end{equation}

\end{enumerate}

\end{thm}

\subsection{Quantum entanglement of generalized doped RVB states}\label{subsection-4.4}

We proceed as in Subsec.~\ref{subsection-3.2} 
and $\# \boldsymbol{\Psi} \geq 2$.

\textbf{(1) The concept.} 

Let $\boldsymbol{\Psi}$ be a set of coverings of $\Gamma_{n}$ with $\# \boldsymbol{\Psi} \geq 2$, $\alpha : \boldsymbol{\Psi} \rightarrow \vertchar[.08ex]{C} \smallsetminus \{0\}$ with $\sum_{\psi \in \boldsymbol{\Psi}} |\alpha (\psi)|= 1$ as in Subsec.~\ref{subsection-4.1}. Let $1 \leq \gamma \leq v$ and $\mu = \nu - \gamma$ as in Subsec.~\ref{subsection-3.2}.
\begin{enumerate}[label = (\alph*)]
\item The \textbf{generalized RVB vector with $\gamma$ holes} is $| \alpha \rangle_{h} = \sum_{\psi \in \boldsymbol{\Psi}} \alpha (\psi) | \psi \rangle_{h}$, where $| \psi \rangle_{h}$ is as in~\ref{subsection-3.2} (e) with polynomial representation $F_{\psi}^{h}(\boldsymbol{\x})$ as in~\ref{subsection-3.2}  (f), viz., \eqref{eq-3.75}.

\item The polynomial representation of $| \alpha \rangle_{h}$ is given by modifying \eqref{eq-3.77} to \eqref{eq-3.80} accordingly. To begin with,
\begin{equation}
F_{\alpha}^{h}(\boldsymbol{\x}) = \hat{\alpha} F^{hi}(\boldsymbol{\x}) + \sum_{\psi \in \boldsymbol{\Psi}} \alpha(\psi) F_{\psi}^{hd}(\boldsymbol{\x}).\tag{4.51}\label{eq-4.51}
\end{equation}
Thus $F_{\alpha}^{h}(\boldsymbol{\x})$ is a homogeneous polynomial of degree $\mu$.
 
\item Let $\mu=1$. We have 
\begin{equation}
F_{\alpha}^{h}(\boldsymbol{\x}) = \hat{\alpha} F^{hc}(\boldsymbol{\x})=\frac{1}{\sqrt{\#\mathcal{R}}} 2^{-\mu/2} \hat{\alpha} \left(\sum_{S \in \mathcal{S}} \boldsymbol{\x}^{S} + (-1)^{\mu} \sum_{R \in \mathcal{R}} \boldsymbol{\x}^{R} \right).\tag{4.52}\label{eq-4.52} 
\end{equation}
Hence $F_{\alpha}^{h}(\boldsymbol{\x}) \neq 0$ if and only if $\hat{\alpha} \neq 0$.  For $\hat{\alpha} \neq 0$, we have that $F_{\alpha}^{h}(\boldsymbol{\x})$ is a homogeneous polynomial of degree $1$ with supp equal to $\Gamma_{n}$. So, for this case, by Theorem \ref{thm-2.1}(iii)(b), $| \alpha \rangle_h$ is genuinely entangled.
We note that for Example 4.2 (a), $| \alpha \rangle_{h}=0$. 

\item Now suppose that $\mu \geq 2$. This forces $v \geq 3$. As in Subsec.~\ref{subsection-3.2} (j), let for $\phi \neq A_{1} \subset_{\neq} A$, $\phi \neq B_{2} \subset_{\neq} B$ with $\# A_{1} + \# B_{2} = \mu$, $\boldsymbol{\Psi}_{A_{1}, B_{2}}^{h} = \{\psi \in \boldsymbol{\Psi} : \psi (A_{1}) \cap B_{2} = \phi\}$. We now set $\varpi^{\alpha} (A_{1}, B_{2}) = \alpha^{\#}\left(\boldsymbol{\Psi}_{A_{1}, B_{2}}^{h}\right)$. We write $A_{1} \omega^{\alpha} B_{2}$ to mean that $\omega^{\alpha}(A_{1}, B_{2}) \neq 0$. Then we obtain an analogue of \eqref{eq-3.78} with $F_{\boldsymbol{\Psi}}^{hd}(\boldsymbol{\x})$ replaced by $F_{\alpha}^{hd}(\boldsymbol{\x}) = \sum_{\psi \in \boldsymbol{\Psi}} \alpha (\psi) F_{\psi}^{hd}(\boldsymbol{\x})$ occurring in \eqref{eq-4.51} above, $\omega (A_{1}, B_{2})$ by $\omega^{\alpha}(A_{1}, B_{2})$ and $A_{1} \omega B_{2}$ by $A_{1} \omega^{\alpha} B_{2}$.

We write the analogue of \eqref{eq-3.79} in full as
\begin{align*}
F_{\alpha}^{h} (\boldsymbol{\x}) &= \frac{1}{\sqrt{\# \mathcal{R}}} 2^{-\mu/2} \left(\hat{\alpha} \left(\sum_{S \in \mathcal{S}} \boldsymbol{\x}^{S} + \sum_{R \in \mathcal{R}} (-1)^{\mu} \boldsymbol{\x}^{R} \right) +\right.\\
&\quad\sum \left\{(-1)^{\mu_{1}} \omega^{\alpha} (A_{1}, B_{2}) \boldsymbol{\x}^{A_{1}} \boldsymbol{\x}^{B_{2}} : \# A_{1}= \mu_1, \# B_{2} = \mu-\mu_{1}, A_{1} \omega^{\alpha} B_{2},\right.\\
 &\quad\left.\left. 1 \leq \mu_{1} \leq \mu-1\right\} \right).\tag{4.53}\label{eq-4.53}\\ 
\end{align*}
So $| \alpha \rangle_{h} =0$ if and only if $F_{\alpha}^{h} (\boldsymbol{\x})=0$ if and only if $\hat{\alpha}=0$ and $\omega^{\alpha}(A_{1}, B_{2}) =0$ for $A_{1} \subset A$, $B_{1} \subset B$ with $\# A_{1}= \mu_{1}$, $\# B_{2}=\mu-\mu_{1}$, $ 1 \leq \mu_{1} \leq \mu-1$. 
In other words, $| \alpha \rangle_{h} \neq 0$ if and only if either $\hat{\alpha} \neq 0$ or $\omega^{\alpha}(A_{1}, B_{2}) \neq 0$ for some $A_{1}\subset A$, $B_{2}\subset B$ with $\# A_{1} = \mu_{1}$, $\# B_{2} = \mu-\mu_{1}$ for some $1 \leq \mu_{1} \leq \mu-1$.

\item For the case $\mu \geq 2$ as in (d) above,
\begin{align*}
|| | \alpha \rangle_{h} ||^{2} &= 2^{-\mu} \left(2| \hat{\alpha} |^{2} \right.+ \sum \left\{ | \omega^{\alpha}(A_{1}, B_{2}) |^{2}/\#\mathcal{R} : \# A_{1} = \mu_{1},\right.\\ 
&\left. \left.\# B_{2} = \mu-\mu_{1}, A_{1} \omega^{\alpha} B_{2}, 1 \leq \mu_{1} \leq \mu-1 \right\}\right).\tag{4.54}\label{eq-4.54}
\end{align*}
So in case $| \alpha \rangle_{h} \neq 0$ as specified in (d) above, we may normalize $| \alpha \rangle_{h}$ by replacing it by  $| \alpha \rangle_{h}^{\sim} = \frac{1}{|| | \alpha \rangle_{h} ||} | \alpha \rangle_{h}$ to obtain the corresponding generalized doped RVB state, if we like. 
\end{enumerate}


\begin{thm}\label{thm-4.4}
If $\hat{\alpha} \neq 0$, then $| \alpha \rangle_{h}^{\sim}$ is genuinely entangled.
\end{thm}

\emph{Proof.}
It follows on the lines of that of Theorem~\ref{thm-3.9}.

\vspace{0.5cm}
\textbf{(2) Case \(\hat{\alpha} =0\).}

We have already noted in  \ref{subsection-4.4}(1)(c) that $| \alpha \rangle_{h} = 0$ for Example 4.2(a) and indeed for the subcase $\mu=1$ for any general situation as well.

Now suppose $\mu \geq 2$ and $| \alpha \rangle_{h} \neq 0$. This foces $v \geq 3$.


\begin{enumerate}[label=(\alph*)]

\item By \eqref{eq-4.53}, 
\begin{align*}
F_{\alpha}^{h}(\boldsymbol{\x}) &= \sum \left\{(-1)^{\mu_1} \omega^{\alpha}(A_{1}, B_{2}) \boldsymbol{\x}^{A_{1}} \boldsymbol{\x}^{B_{2}} : \# A_{1} = \mu_{1}, \# B_{2} = \mu-\mu_{1},\right.\\
&\left.\quad A_{1}\omega^{\alpha} B_{2}, 1 \leq \mu_{1} \leq \mu-1\right\}.
\tag{4.55}\label{eq-4.55} 
\end{align*}

\item If $S_{F^{h}_{\alpha}} \neq \Gamma_{n}$ or some $\x_{j}$ is a factor of $F_{\alpha}^{h}(\boldsymbol{\x})$,  $F_{h}^{\alpha}$ is a product vector in some bipartite cut and, therefore, $| \alpha \rangle_{h}$ is not genuinely entangled. 

\item Now suppose $S_{F_{\alpha}^{h}}= \Gamma_{n}$ and no $\x_{j}$ is a factor of $F_{\alpha}^{h}(\boldsymbol{\x})$. We can carry out the discussion as in previous sections to study quantum entanglement properties of $| \alpha \rangle_{h}$.

Because $\# A_{1} + \# B_{2} = \mu$ for $A_{1}$, $B_{2}$ occurring in (4.55) and $2\mu \leq 2v-2 < n = \# \Gamma_{n}$, we have at least three terms in \eqref{eq-4.55}, i.e., there exist three distinct $(A_{1}^{u}, B_{2}^{u})$, $u= 1,2,3$. that satisfy $A^{u}_{1} \omega^{\alpha} B^{u}_{2}$ for $u=1,2,3$. 
\end{enumerate}

\vspace{0.5cm}

\textbf{(3) The case $\hat{\alpha} = 0$, $\mu = 2$, $v \geq 3$.}

This case is rather simple. We consider the case $| \alpha \rangle_{h} \neq 0$ and let $t$ be the number of terms in $F_{\alpha}^{h}(\boldsymbol{\x})$. For notational convenience, for $z \in \Gamma_n,$ we write $\boldsymbol{X}^z$ for $X_z=\boldsymbol{X}^{\{z\}}.$
\begin{enumerate}[label=(\alph*)]
\item By \eqref{eq-4.53}, we have 
\begin{equation}
F_{\alpha}^{h}(\boldsymbol{\x}) = - \sum_{\substack{a \in A}, b \in B} \omega^{\alpha} (\{a\}, \{b\}) \boldsymbol{\x}^{a} \boldsymbol{\x}^{b}.
\tag{4.56}\label{eq-4.56}
\end{equation}

\item For $a \in A$, $b \in B$, $a \in S_{F_{\alpha}^{h}}$ if and only if for some $b_{a} \in B$, $\varpi^{\alpha}(\{a\}, \{b_{a}\})\neq 0$ whereas $b \in S_{F_{\alpha}^{h}}$ if and only if for some $a_{b} \in B$, $\omega^{\alpha}(\{a_{b}\}, \{b\}) \neq 0$.

As a consequence, $S_{F_{\alpha}^{h}} = \Gamma_{n}$ only if $t \geq v$ on the one hand, and on the other hand, it is so if $t \geq v(v-1) + 1$.

\item For $a \in A$, $b \in B$, $\boldsymbol{\x}^{a}$ is a factor of $F_{\alpha}^{h}(\boldsymbol{\x})$ if and only if $S_{F_{\alpha}^{h}} \cap A = \{a\}$ while $\boldsymbol{\x}^{b}$ is a factor of $F_{\alpha}^{h}(\boldsymbol{\x})$ if and only if $S_{F^h_\alpha} \cap B=\{b\}.$ As a consequence, for  $K \subset \Gamma_n$ with $\#K=2,$ $\boldsymbol{X}^K$ is a factor of $F^h_\alpha(\boldsymbol{X})$ in case $K \subset A$ or $K \subset B$ whereas for $K =\{a, b\}$ for some $a \in A$, $b \in B$, $\boldsymbol{\x}^{K}$ is a factor of $F_{\alpha}^{h}(\boldsymbol{\x})$ if and only if $t=1$ and $F_{\alpha}^{h}(\boldsymbol{\x}) = -\varpi^{\alpha} (\{a\}, \{b\}) \boldsymbol{\x^{K}}$.

\item It follows from (b) and (c) that if $S_{F_{\alpha}^{h}} = \Gamma_{n}$, no $\boldsymbol{\x}^{K}$ is a factor of $F_{\alpha}^{h}(\boldsymbol{\x})$ for $\phi \neq K \subset_{\neq } \Gamma_{n}$.

\item Suppose $S_{F_{\alpha}^{h}} = \Gamma_{n}$. Then $| \alpha \rangle_{h}$ is a product vector in some bipartite cut $(E, E')$ if and only if $(E, E') = (A, B)$ or $(B, A)$ and for  some tuples $(\alpha_{a})_{a \in A} $ and $(\beta_{b})_{b \in B}$ of non-zero scalars, 
\begin{equation}
F_{\alpha}^{h}(\boldsymbol{\x}) = \left(\sum_{a \in A} \alpha_{a} \boldsymbol{\x}^{a} \right) \left(\sum_{b \in B} \beta_{b} \boldsymbol{\x}^{b}\right)~~\text{i.e.,}~~ \omega^{\alpha} (\{a\}, \{b\}) = -\alpha_{a}\beta_{b}~~\text{for}~~a\in A , b \in B.
\tag{4.57}\label{eq-4.57}
\end{equation}
In this case,
\begin{equation}
t = \# A \# B =v^{2}. \tag{4.58}\label{eq-4.58}
\end{equation}

\item $| \alpha \rangle_{h}$ is genuinely entangled if and only if $S_{F_{\alpha}^{h}} = \Gamma_{n}$ and (4.57) does not hold. 

In particular, if $S_{F_{\alpha}^{h}} = \Gamma_{n}$ and $t < v^{2}$, then $| \alpha \rangle_{h}$ is genuinely entangled, whereas if $t < v$,  $| \alpha \rangle_{h}$ is not genuinely entangled.
\end{enumerate}

\section{Conclusion and outlook}
\label{sec-conclu}

Resonating valence bond states have been proposed as a means for explaining a variety of phenomena, including high-Tc superconductivity, and as a 
substrate for quantum devices, including measurement-based quantum computation, although these proposals remain contested. Multipartite entanglement in RVB states are potentially the fundamental reason for the possible success of the states in these applications. Moreover, multiparty entanglement is known to be useful in a variety of applications, and therefore, it is plausible that a robust understanding of the multiparty entanglement in RVB states will lead to interesting future applications of these states. 

Polynomial representations of entanglement has been in use in the community, and is known to be useful in understanding a variety of physical phenomena. We have used a polynomial representation of RVB states on ladder lattices, possibly with doping, and possibly including unequal amplitudes of the dimer covering involved, to show the existence of genuine multiparty entanglement in the states. 

The technique developed can potentially be applied to more general classes of quantum states and in lattices of more involved geometries, and possibly also for the quantification of genuine multisite entanglement. We also hope to extend the technique to noisy situations, that would typically require us to handle mixed multiparty quantum states. 

\begin{acknowledgments}

The work began when AIS visited the Quantum Information and Computation Group at the Harish-Chandra Research Institute (HRI), Allahabad (Prayagraj) in 2015, and it has been continued through subsequent visits and emails. AIS thanks HRI 
for their kind hospitality. She would also like to thank G. Baskaran for his insightful discussion on the topic at The Institute of Mathematical Sciences in  Chennai and his encouragement. It is a pleasure for her to thank Richard Josza, Gerardo Adesso, and their research groups at Cambridge and Nottingham respectively for their kind hospitality and interesting discussion sessions. She thanks The Indian National Science Academy for its continuous support. The authors thank Sudipto Singha Roy for his interest in the topic. The authors also thank Vishvesh Kumar and Savita Rani for reading the paper and unusual contribution to corrections. They also thank M/s. Sriranga Digital Software Technologies for the latex typing of the paper. 
ASD and US acknowledge partial support from the Interdisciplinary Cyber-Physical Systems (ICPS) program of the Department of Science and Technology (DST), Government of India, Grant Nos.~DST/ICPS/QuST/Theme-1/2019/23 and DST/ICPS/QuST/Theme-3/2019/120.

\end{acknowledgments}

\end{document}